\documentclass[oneside,english]{book}
\usepackage[T1]{fontenc}
\usepackage[latin9]{inputenc}
\usepackage{babel}
\usepackage{amsmath}
\usepackage{amssymb}
\usepackage{graphicx}
\usepackage{setspace}
\usepackage{esint}
\usepackage[unicode=true,pdfusetitle,
 bookmarks=true,bookmarksnumbered=false,bookmarksopen=false,
 breaklinks=false,pdfborder={0 0 1},backref=false,colorlinks=false]
 {hyperref}

\makeatletter
\pdfoutput=1

\numberwithin{equation}{section}
\numberwithin{figure}{section}

\usepackage{braket}

\makeatother

\begin{document}
\begin{titlepage} 
\begin{center}
\textbf{\Huge{}{}Quantum Magnetism,}{\Huge\par}
\par\end{center}

\begin{center}
 
\par\end{center}

\begin{center}
\textbf{\Huge{}{}Spin Waves, }{\Huge\par}
\par\end{center}

\begin{center}
 
\par\end{center}

\begin{center}
\textbf{\Huge{}{}and Optical Cavities}{\Huge\par}
\par\end{center}

\begin{center}
 
\par\end{center}

\begin{doublespace}
\begin{center}
{\huge{}{}Silvia Viola Kusminskiy}{\huge\par}
\par\end{center}
\end{doublespace}

\begin{center}
{\LARGE{}{}Max Planck Institute for the Science of Light}{\LARGE\par}
\par\end{center}

\begin{center}
 
\par\end{center}

\begin{center}
{\large{}{}Staudtstraße 2, 91058 Erlangen, Germany}{\large\par}
\par\end{center}

\begin{center}
 {\large{}{}and}{\large\par}
\par\end{center}

\begin{center}
{\LARGE{}{}Friedrich-Alexander University Erlangen-Nuremberg}{\LARGE\par}
\par\end{center}

\begin{center}
{\large{}{}Staudtstraße 7, 91058 Erlangen, Germany}{\large\par}
\par\end{center}

\end{titlepage}

\chapter*{Abstract}

Both magnetic materials and light have always played a predominant
role in information technologies, and continue to do so as we move
into the realm of quantum technologies. In this course we review the
basics of magnetism and quantum mechanics, before going into more
advanced subjects. Magnetism is intrinsically quantum mechanical in
nature, and magnetic ordering can only be explained by use of quantum
theory. We will go over the interactions and the resulting Hamiltonian
that governs magnetic phenomena, and discuss its elementary excitations,
denominated magnons. After that we will study magneto-optical effects
and derive the classical Faraday effect. We will then move on to the
quantization of the electric field and the basics of optical cavities.
This will allow us to understand a topic of current research denominated
\emph{Cavity Optomagnonics}.

These notes were written as the accompanying material to the course
I taught in the Summer Semesters of 2018 and 2019 at the Friedrich-Alexander Universität
in Erlangen. The course is intended for Master or advanced Bachelor
students. Basic knowledge of quantum mechanics, electromagnetism,
and solid state at the bachelor level is assumed. Each section is
followed by a couple of simple exercises which should serve as to
" fill in the blanks" of what has been
derived and a couple of check-points for the main concepts developed.

\tableofcontents{}

\chapter{\label{chap:Electromagnetism}Electromagnetism}

The history of magnetism is ancient: just to give an example, the
magnetic compass was invented in China more that 2000 years ago. The
fact that magnetism is intrinsically connected to moving electric
charges (and not to " magnetic charges"
) however was not discovered until much later. In the year 1820, Oersted
experimentally demonstrated that a current-carrying wire had an effect
on the orientation of a magnetic compass needle placed in its proximity.
In the following few years, Ampere realized that a small current loop
generates a magnetic field which is equivalent to that of a small
magnet, and speculated that all magnetic fields are caused by charges
in motion. In the next few sections we will review these concepts
and the basics of \emph{magnetostatics}.

\section{\label{sec:Basic-magnetostatics}Basic magnetostatics}

As the name indicates, magnetostatics deals with magnetic fields that
are constant in time. The condition for that is a steady-state current,
in which both the charge density $\rho$ and the current density $j=I/A_{{\rm s}}$
($A_{{\rm s}}$ cross-sectional area) are independent of time 
\begin{eqnarray}
\frac{\partial\rho}{\partial t} & =0\\
\frac{\partial\mathbf{j}}{\partial t} & =0\,.
\end{eqnarray}
From the continuity equation 
\begin{equation}
\nabla\cdot\mathbf{j}+\frac{\partial\rho}{\partial t}=0
\end{equation}
we moreover obtain 
\begin{equation}
\nabla\cdot\mathbf{j}=0\,.
\end{equation}

In these notes we will call \emph{magnetic induction} to $\mathbf{B}$
and \emph{magnetic field} to $\mathbf{H}$.\footnote{Some authors call instead $\mathbf{B}$ the magnetic field and $\mathbf{H}$
the \emph{auxiliary field. }} In free space, these two fields are related by 
\begin{equation}
\mathbf{B}=\mu_{0}\mathbf{H}\label{eq:B free}
\end{equation}
being $\mu_{0}=4\pi\times10^{-7}{\rm NA}^{-2}$ the permeability of
free space. We will use the SI units system throughout these notes,
therefore \textbf{$\mathbf{B}$ }is measured in Teslas (${\rm T}={\rm V}.{\rm s}.{\rm m}^{-2}$)
and $\mathbf{H}$ in Amperes per meter (${\rm A}.{\rm m}^{-1}$).

\begin{figure}
\includegraphics[width=1\textwidth]{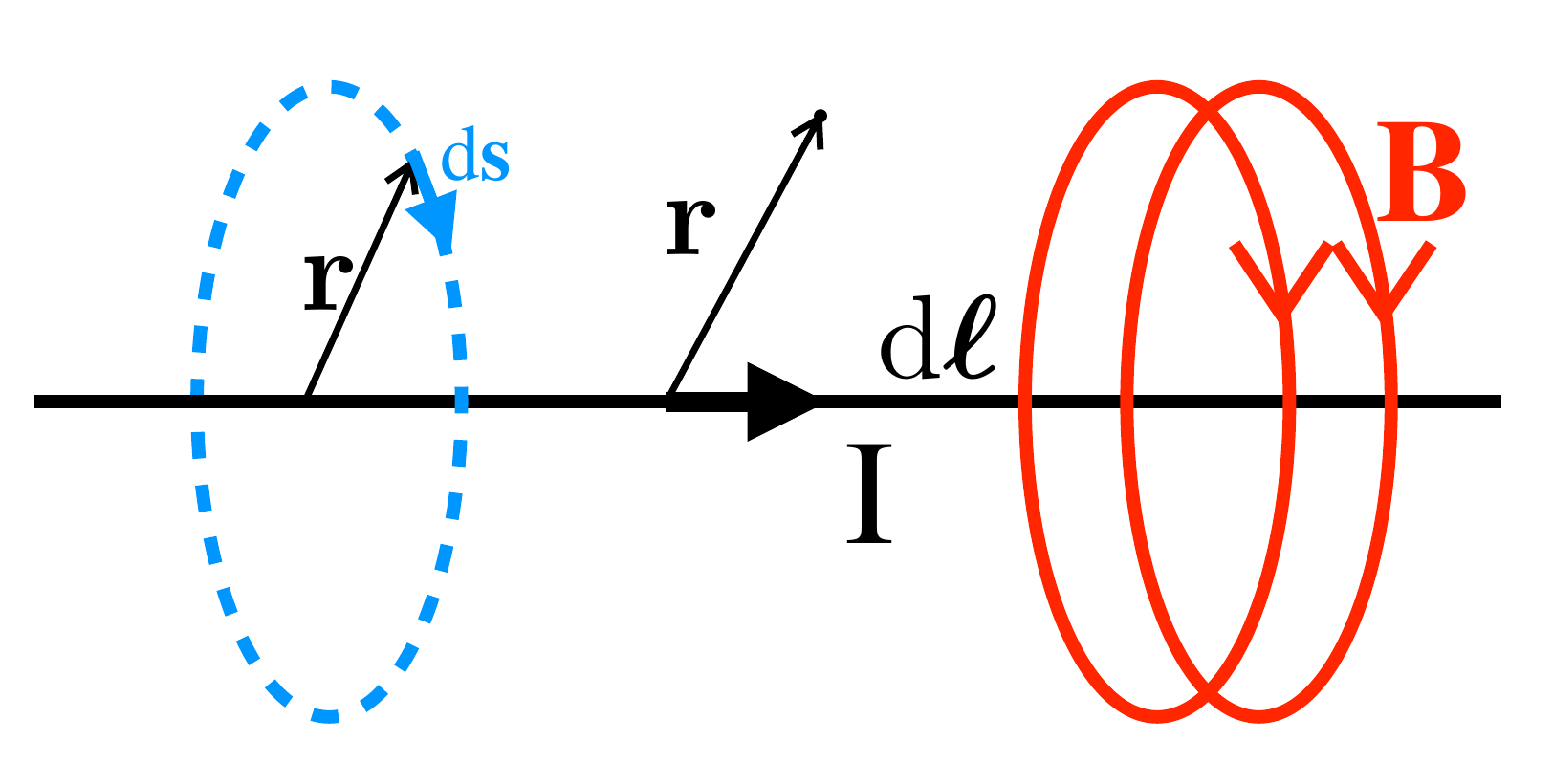}\caption{{\small{}{}The magnetic induction generated by a current I can be
calculated using Biot-Savart's law, see Eq.~\ref{eq:Biot-Savart}.
Ampere's law (see Eq.~\ref{eq:Ampere}) is always valid, but useful
to calculate the $\mathbf{B}$ fields only for cases of particular
symmetry, e.g., an infinite straight wire.}}

\label{fig:BiotSavartAmpere} 
\end{figure}

The magnetic induction at point $\mathbf{r}$ due to a current loop
can be calculated using the Biot-Savart law 
\begin{equation}
{\rm d}\mathbf{B}=\frac{\mu_{0}I}{4\pi r^{2}}{\rm d}\boldsymbol{\ell}\times\frac{\mathbf{r}}{r}\label{eq:Biot-Savart}
\end{equation}
where ${\rm d}\boldsymbol{\ell}$ points in the direction of the current
$I$, see Fig.~\ref{fig:BiotSavartAmpere}. Equivalent to the Biot-Savart
law is Ampere's law, which reads 
\begin{equation}
\oint_{\mathcal{C}}\mathbf{B}\cdot{\rm d}\mathbf{s}=\mu_{0}I\label{eq:Ampere}
\end{equation}
where $I$ is the current enclosed by the closed loop $\mathcal{C}$,
see Fig.~\ref{fig:BiotSavartAmpere}. Ampere's law is general, but
it is useful to \emph{calculate} magnetic fields only in cases of
high symmetry, for example the magnetic field generated by an infinite
straight wire. Using Stoke's theorem, we can put Ampere's law in differential
form 
\begin{equation}
\nabla\times\mathbf{B}=\mu_{0}\mathbf{j}\,.\label{eq:Ampere diff}
\end{equation}
Ampere's law together with the absence of magnetic monopoles condition
\begin{equation}
\nabla\cdot\mathbf{B}=0\label{eq:div B}
\end{equation}
constitute the \emph{Maxwell equations for magnetostatics}. 

These equations give us indeed time independent magnetic fields, and
if we compare the magnetostatic equations with the full microscopic
Maxwell equations 
\begin{eqnarray}
\nabla\cdot\mathbf{B} & = & 0\\
\nabla\cdot\mathbf{E} & = & \frac{\rho}{\varepsilon_{0}}\\
\nabla\times\mathbf{E} & = & -\frac{\partial\mathbf{B}}{\partial t}\\
\nabla\times\mathbf{B} & = & \mu_{0}\left(\mathbf{j}+\varepsilon_{0}\frac{\partial\mathbf{E}}{\partial t}\right)
\end{eqnarray}
(with $\varepsilon_{0}=8.85\times10^{-12}\,{\rm {F}{\rm {m}^{-1}}}$
the vacuum permittivity) we see that we have moreover decoupled the
magnetic and electric fields.

\subsection*{Check points}
\begin{itemize}
\item What is the magnetostatic condition? 
\item Write the magnetostatic Maxwell equations 
\end{itemize}

\section{\label{sec:Magnetic-moment}Magnetic moment}

\begin{figure}[h]
\includegraphics[width=1\textwidth]{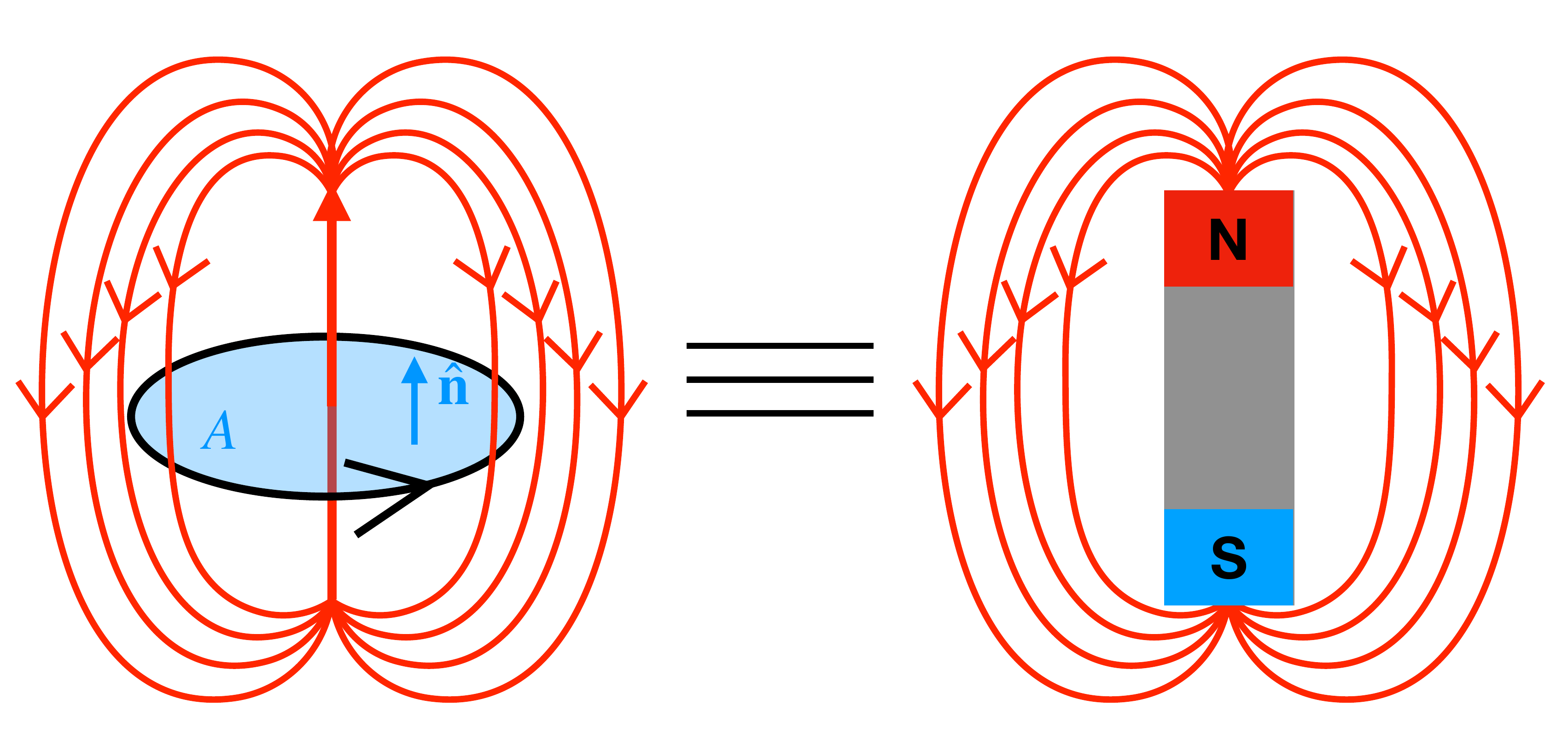}

\caption{{\small{}{}Magnetic dipole: the magnetic field induced by a small
current loop is equivalent to that of a small magnet.}}
\label{fig:current_loop_eq_NS} 
\end{figure}

The magnetic moment of a current loop is defined as 
\begin{equation}
\mathbf{m}=IA\hat{\mathbf{n}}\,,\label{eq:dipole moment}
\end{equation}
where $A$ is the area enclosed by the loop and $\hat{\mathbf{n}}$
is the normal to the surface, with its direction defined from the
circulating current by the right-hand rule, see Fig.~\ref{fig:current_loop_eq_NS}.
$\mathbf{m}$ defines a \emph{magnetic dipole} in the limit of $A\rightarrow0$
but finite moment.

\begin{figure}
\center \includegraphics[width=0.8\textwidth]{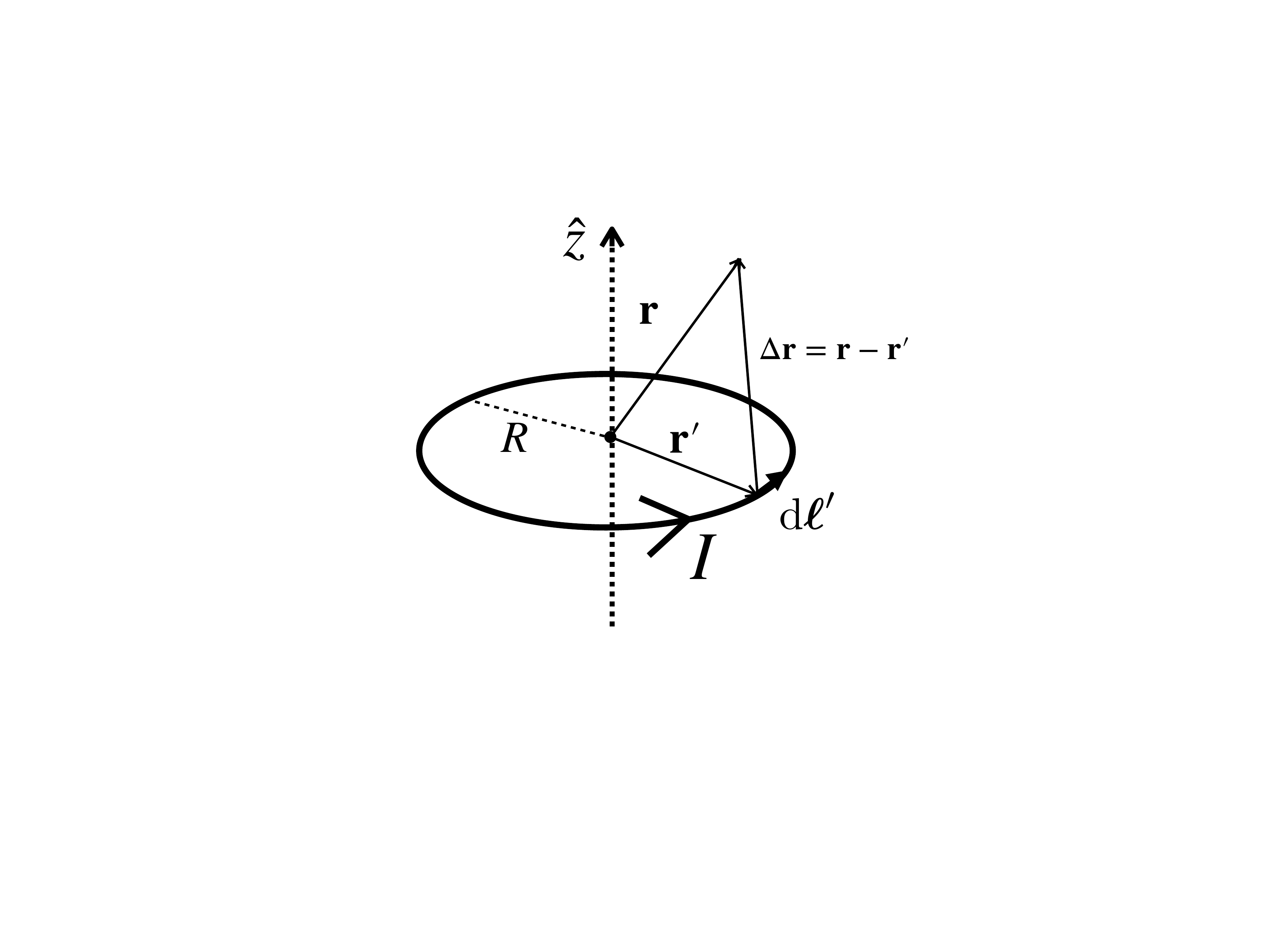}

\caption{{\small{}{}Magnetic induction due to a small circular current loop:
we use Biot-Savart to calculate the $\mathbf{B}$ field. Seen from
" far away" , it is the field of a magnetic
dipole.}}

\label{fig:current_loop_m} 
\end{figure}

Using Eq.~\ref{eq:Biot-Savart} we can calculate the magnetic induction
generated by a small current loop of radius $R$ 
\begin{align}
\mathbf{B}(\mathbf{r}) & =\frac{\mu_{0}I}{4\pi}\int\frac{{\rm d}\boldsymbol{\ell}'\times\Delta\mathbf{r}}{\Delta r^{3}}\label{eq:Biot-Savart-Loop}\\
 & =-\frac{\mu_{0}I}{4\pi}\int{\rm d}\boldsymbol{\ell}'\times\nabla\left(\frac{1}{\Delta\mathbf{r}}\right)\nonumber 
\end{align}
with $\Delta\mathbf{r}=\mathbf{r}-\mathbf{r}'$ (see Fig.~\ref{fig:current_loop_m}).
From Eq.~\ref{eq:div B} we know we can define a vector potential
$\mathbf{A}(\mathbf{r})$ such that $\mathbf{B}(\mathbf{r})=\nabla\times\mathbf{A}(\mathbf{r})$.
By a simple manipulation of Eq.~\ref{eq:Biot-Savart-Loop}, one can
show that in the far field limit ($\Delta\mathbf{r}\gg R$ ), 
\begin{align}
\mathbf{A}(\mathbf{r}) & =\frac{\mu_{0}}{4\pi}\mathbf{m}\times\frac{\hat{\mathbf{r}}}{r^{2}}\label{eq:m dipole field A}\\
\mathbf{B}(\mathbf{r}) & =\frac{\mu_{0}}{4\pi}\frac{3(\mathbf{m}\cdot\mathbf{r})\mathbf{r}-r^{2}\mathbf{m}}{r^{5}}\label{eq:dipole field B}
\end{align}
which is the magnetic induction generated by a magnetic dipole. 

More generally, for an arbitrary current density distribution $\mathbf{j}(\mathbf{r'})$
one can define \cite{john_david_jackson_classical_1998,david_j._griffiths_introduction_2014}
\begin{equation}
\mathbf{m}=\frac{1}{2}\int{\rm d}^{3}\mathbf{r}'\left[\mathbf{r}'\times\mathbf{j}(\mathbf{r}')\right]\label{eq:m gen}
\end{equation}
and Eq.~\ref{eq:m dipole field A} is the lowest non-vanishing term
in a multipole expansion of the vector potential (in the Coulomb gauge,
$\nabla\cdot\mathbf{A}=0$) 
\begin{equation}
\mathbf{A}(\mathbf{r})=\frac{\mu_{0}}{4\pi}\int{\rm d}^{3}\mathbf{r}'\frac{\mathbf{j}(\mathbf{r}')}{|\mathbf{r}-\mathbf{r}'|}\,.\label{eq:A vec gen}
\end{equation}

The energy of a magnetic dipole in a magnetic field is given by 
\begin{equation}
E_{{\rm Z}}=-\mathbf{m}\cdot\mathbf{B}\label{eq:Zeeman energy}
\end{equation}
and therefore is minimized for $\mathbf{m}\parallel\mathbf{B}$. This
is called the \emph{Zeeman Energy}. 
\begin{enumerate}
\item \textbf{\emph{Exercise: derive Eqs.\ref{eq:m dipole field A} and
\ref{eq:dipole field B} (tip: use the " chain rule"
and a multipole expansion).}} 
\item \textbf{\emph{Exercise: show that Eq.~\ref{eq:dipole moment} follows
from \ref{eq:m gen} (tip: 1-D Delta-function distributions have units
of 1/length).}} 
\end{enumerate}

\subsection*{Check points}
\begin{itemize}
\item How do you show the equivalence between the magnetic field of a small
current loop and that of a small magnet? A conceptual explanation
suffices. 
\end{itemize}

\section{\label{sec:Orbital-angular-momentum}Orbital angular momentum}

\begin{figure}
\center \includegraphics[width=0.8\textwidth]{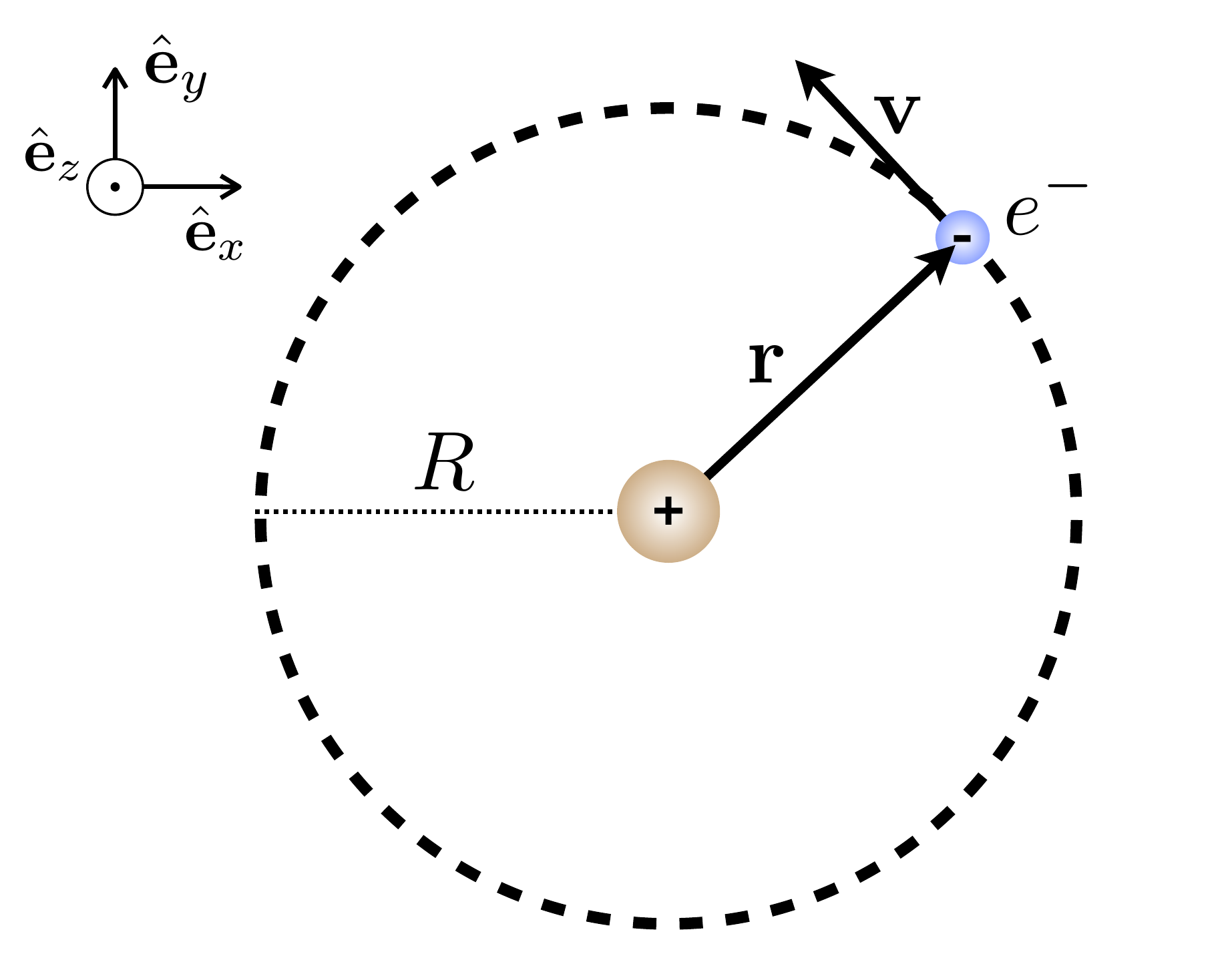}

\caption{{\small{}{}Semiclassical picture used to calculate the orbital angular
momentum of an electron.}}

\label{fig:orbital_L_e} 
\end{figure}

The magnetic moment $\mathbf{m}$ can be related to angular momentum.
In order to do this, we consider the limit of one electron $e$ (with
negative charge $-e$) orbiting around a fixed nucleus, see Fig.~\ref{fig:orbital_L_e}.
Note that here we get the first indication that magnetism is a purely
quantum effect: stable orbits like that are not allowed classically,
and we need quantum mechanics to justify the stability of atoms. Our
argument here is therefore a semiclassical one. 

The average current due to this single electron is 
\begin{equation}
I=-\frac{e}{T}=-\frac{e\omega}{2\pi}\label{eq:I single e}
\end{equation}
where $T$ is one period of revolution. The electron also possesses
orbital angular momentum $\mathbf{L}=m_{e}\mathbf{r}\times\mathbf{v}$.
Measured from the center of the orbit, 
\begin{equation}
\mathbf{L}=m_{e}R^{2}\vec{\omega}\label{eq:L single e}
\end{equation}
and using Eq.~\ref{eq:dipole moment} we obtain 
\begin{equation}
\mathbf{m}=-\frac{e}{2m_{e}}\mathbf{L}\,.\label{eq:m L}
\end{equation}
Therefore we have linked the magnetic moment of a moving charge to
its orbital angular momentum. The coefficient of proportionality is
called the \emph{gyromagnetic ratio 
\begin{equation}
\gamma_{{\rm L}}=-\frac{e}{2m_{e}}\,,\label{eq:gyro ratio L}
\end{equation}
}which is negative due to the negative charge of the electron. Hence
in this case the magnetic moment and angular momentum are antiparallel.
In solids, electrons are the primary source of magnetism due to their
small mass compared to that of the nucleus. Since $m_{p}\approx10^{3}m_{e}$,
the gyromagnetic ratio for the nucleus is strongly suppressed with
respect to the electronic one.

\subsection*{Check points}
\begin{itemize}
\item What is the gyromagnetic ratio? 
\item Why is the gyromagnetic ratio of the nucleus suppressed with respect
to the electronic one? 
\end{itemize}

\section{\label{sec:Spin-angular-momentum}Spin angular momentum}

Although we performed a classical calculation, the result obtained
for the gyromagnetic ratio in Eq.~\ref{eq:gyro ratio L} is consistent
with the quantum mechanical result. We know however that the electron
posses an intrinsic angular momentum, that is, the \emph{spin} $\mathbf{S}$.
The \emph{total} angular momentum of the electron is therefore given
by 
\begin{equation}
\mathbf{J}=\mathbf{L}+\mathbf{S}\,.\label{eq:J tot}
\end{equation}
The spin has no classical analog and the coefficient of proportionality
$\gamma_{{\rm S}}$ between magnetic moment and spin 
\begin{equation}
\mathbf{m}_{{\rm S}}=\gamma_{{\rm S}}\mathbf{S}\label{eq:ms}
\end{equation}
needs to be calculated quantum mechanically \emph{via }the Dirac equation
(see \textit{{e.g.}} Chapter 2 of Ref. \cite{nolting_quantum_2009}).
The results is 
\begin{equation}
\gamma_{{\rm S}}\approx-\frac{e}{m}=2\gamma_{{\rm L}}\,,\label{eq:gyro ratio S}
\end{equation}
where the approximate symbol indicates that there are relativistic
corrections (also contained in the Dirac equation!) to this expression.
The $\gamma_{{\rm S}}$ value agrees with experimental observations.

The total magnetic moment of the electron is therefore given by 
\begin{equation}
\mathbf{m}_{{\rm TOT}}\approx\gamma_{{\rm L}}\left(\mathbf{L}+2\mathbf{S}\right)\label{eq:m tot}
\end{equation}
and hence is not simply proportional to the total angular momentum!
To understand the relation between $\mathbf{m}_{{\rm TOT}}$ and $\mathbf{J}$,
given by the \emph{Landé factor}, we need to resort to quantum mechanics
and the operator representation of angular momentum. We will do that
in the next chapter, where we discuss the atomic origins of magnetism.

\subsection*{Check points}
\begin{itemize}
\item What is the relation between the magnetic moment of an electron and
the angular momentum operators? 
\end{itemize}

\section{\label{sec:Equation-of-motion}Magnetic moment in a magnetic field}

A magnetic moment in a magnetic field experiences a torque 
\begin{equation}
\mathbf{\mathcal{T}}=\mathbf{m}\times\mathbf{B}\,.\label{eq:torque}
\end{equation}
Therefore, the classical equation of motion for the magnetic dipole
(considering for the moment only the orbital angular momentum) is
\begin{equation}
\frac{{\rm d}\mathbf{L}}{{\rm d}t}=\mathbf{m}\times\mathbf{B}=\gamma_{{\rm L}}\mathbf{L}\times\mathbf{B}\,.\label{eq:EOM L}
\end{equation}
\begin{figure}
\center \label{m in B field}\includegraphics[width=0.8\textwidth]{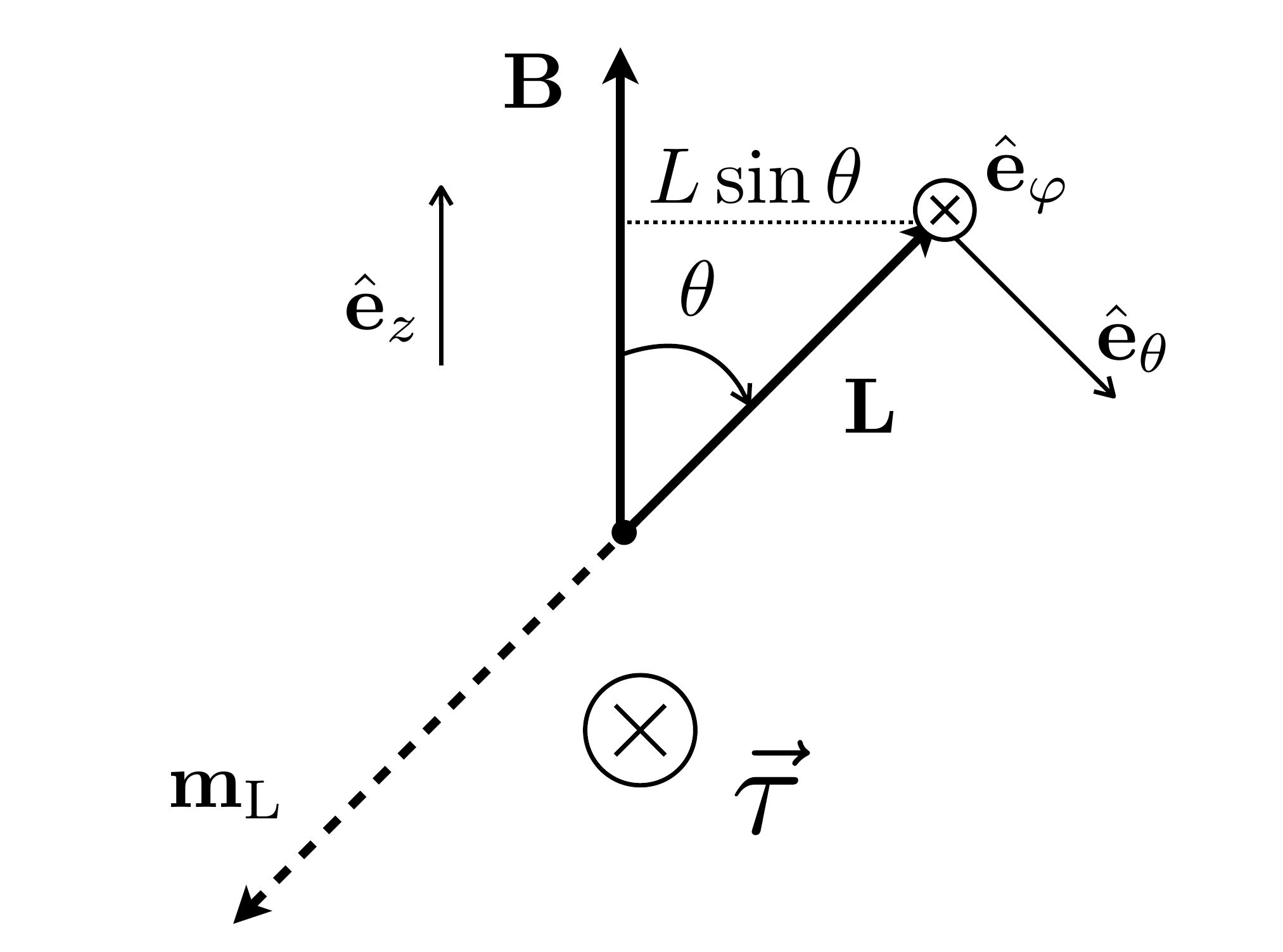}\caption{{\small{}{}A magnetic moment in a $\mathbf{B}$ field experiences
a torque $\mathbf{\mathcal{T}}=\mathbf{m}\times\mathbf{B}$. Remember:
for an electron, $\mathbf{L}$ and $\mathbf{m}$ point in opposite
directions.}}
\end{figure}

\begin{figure}
\center \includegraphics[width=0.8\textwidth]{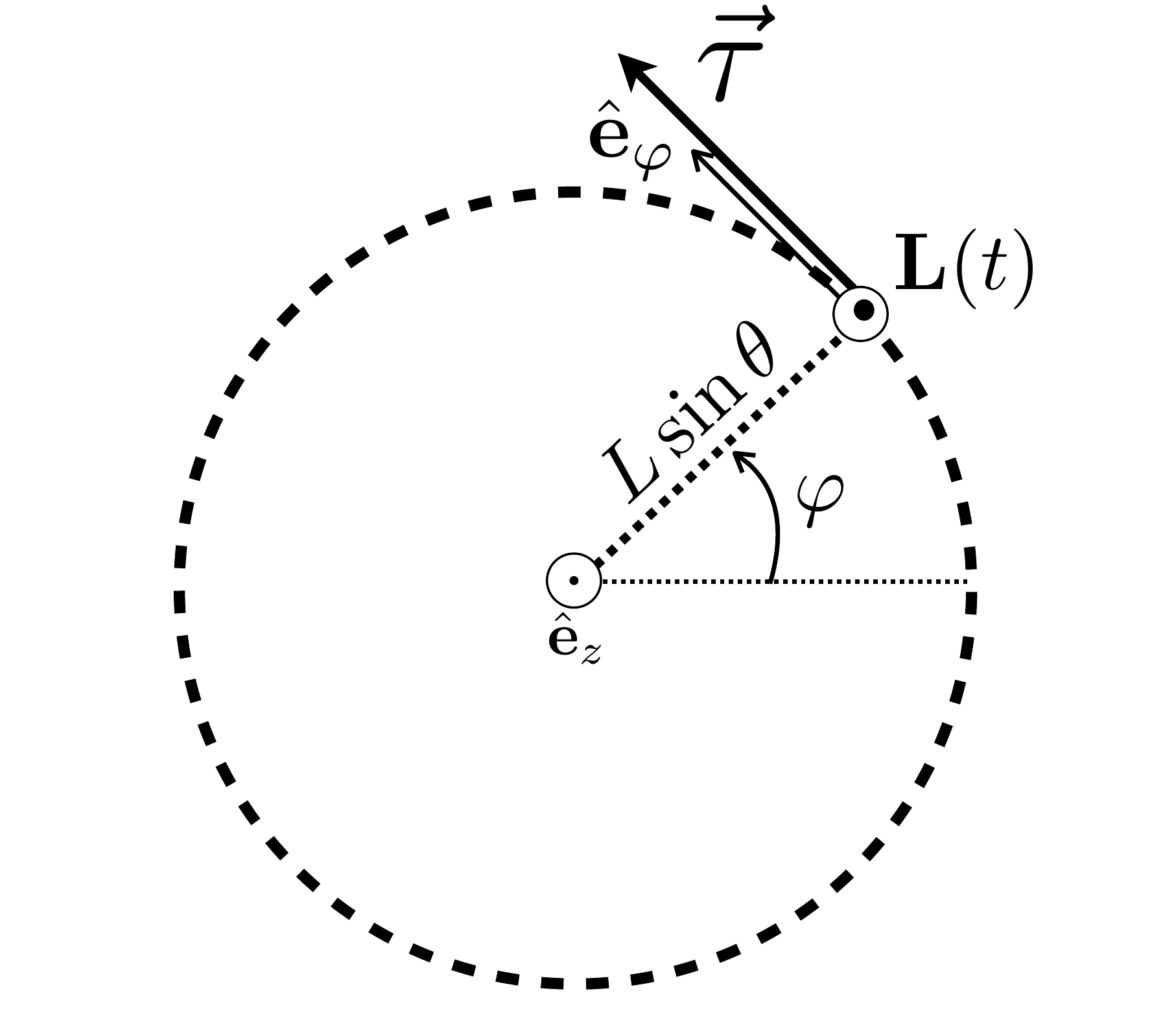}

\caption{{\small{}{}Coordinates used for solving the equation of motion Eq.~\ref{eq:EOM L}viewed
" from above" , in a plane perpendicular
to $\mathbf{B}$. }}

\label{geomEOM} 
\end{figure}

Using the geometry depicted in Fig.~\ref{geomEOM} we obtain 
\begin{align}
\dot{\mathbf{L}} & =L\sin\theta\dot{\varphi}\mathbf{e}_{\phi}\\
\gamma_{{\rm L}}\mathbf{L}\times\mathbf{B} & =|\gamma_{{\rm L}}|LB\sin\theta\mathbf{e}_{\phi}\,.
\end{align}
Hence the magnetic moment precesses around $\mathbf{B}$ at a frequency
\begin{equation}
\omega_{{\rm L}}=\dot{\varphi}=|\gamma_{{\rm L}}\mathbf{B}|\label{eq:Larmor freq}
\end{equation}
which is denominated the \emph{Larmor frequency.} Therefore the angular
momentum will precess around the $\mathbf{B}$ field at a fixed angle
$\theta$ and with constant angular frequency $\omega_{{\rm L}}$.

\begin{figure}
\begin{centering}
\includegraphics[width=0.8\textwidth]{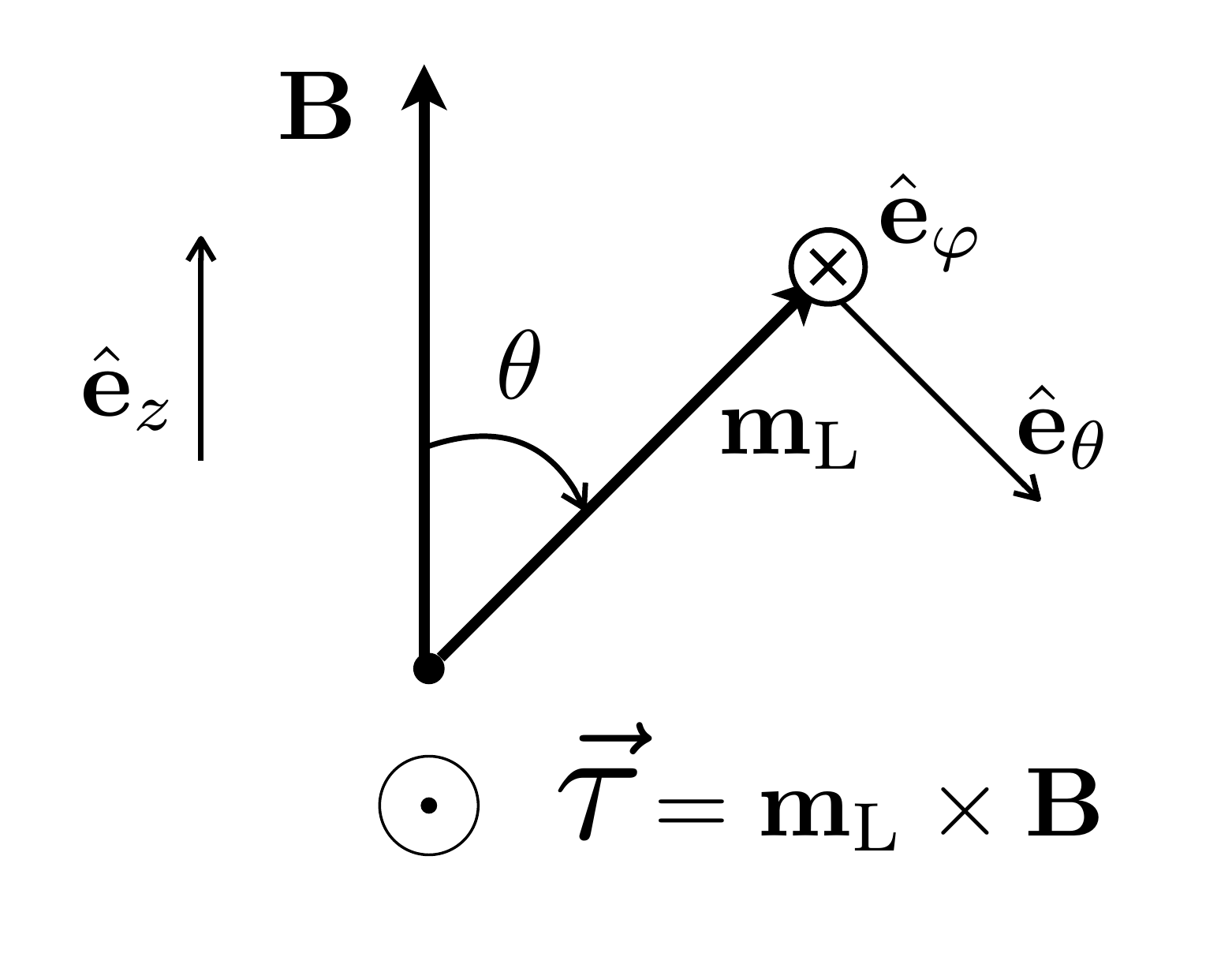} 
\par\end{centering}
\caption{{\small{}{}Coordinate system used to obtain Eq.~\ref{eq:work T}.}}

\label{fig:torque m B} 
\end{figure}

This is consistent with the energy expression defined in Eq.~\ref{eq:Zeeman energy}.
The work per unit time performed by the torque is given by the usual
expression for the power 
\begin{equation}
\frac{{\rm d}W}{{\rm d}t}=\mathcal{T}\cdot\vec{\omega}\label{eq:power T}
\end{equation}
where the angular velocity vector is perpendicular to the plane of
rotation and its direction is given by the right hand rule. We see
therefore that there is no power transfer in the Larmor precession,
since $\omega_{{\rm L}}\hat{\mathbf{e}}_{z}\cdot\mathcal{T}=0$. There
is however an energy cost if we want to change the angle $\theta$
of precession, since the resultant angular velocity $\dot{\theta}\mathbf{e}_{\varphi}$
is collinear with the torque. Using (for simplicity we defined now
$\theta$ as the angle between $\mathbf{m}$ and $\mathbf{B}$, see
Fig.~\ref{fig:torque m B}). 
\begin{align}
\mathcal{T} & =-mB\sin\theta\mathbf{e}_{\varphi}\nonumber \\
\vec{\omega}_{\theta} & =\dot{\theta}\mathbf{e}_{\varphi}\label{eq:torque m b}
\end{align}
we find 
\begin{equation}
\mathcal{T}\dot{\theta}=-mB\sin\theta\frac{{\rm d}\theta}{{\rm d}t}\label{eq:power B}
\end{equation}
and therefore the work exerted to rotate $\mathbf{m}$ up to an angle
$\theta$ is (up to a constant) 
\begin{equation}
W=-\int mB\sin\theta{\rm d}\theta=\mathbf{m}\cdot\mathbf{B}=-E_{{\rm Z}\,.}\label{eq:work T}
\end{equation}
We can therefore take the Zeeman energy $E_{{\rm Z}}$ as the potential
energy associated with the necessary work required to rotate the dipole
$\mathbf{m}$ with respect to an external $\mathbf{B}$ field. 

We will see equations of motion in the form of Eq.~\ref{eq:EOM L}
reappearing throughout this course, even as we treat the angular momenta
as quantum operators. The reason is that, even though the total magnetic
moment $\mathbf{m}_{{\rm TOT}}$ is not proportional to the total
angular momentum $\mathbf{J}$ (see Eq.~\ref{eq:m tot}), their quantum
mechanical expectation values are proportional to each other through
the Landé factor $g$. We will see this more formally when we start
dealing with the quantum mechanical representation of the angular
momenta. For now, we assume that $\mathbf{m}_{{\rm TOT}}$ and $\mathbf{J}$
are related by 
\begin{equation}
\mathbf{m}_{{\rm TOT}}\cdot\mathbf{J}=g\gamma_{{\rm L}}\mathbf{J\cdot\mathbf{J}}\label{eq:Lande g classical}
\end{equation}
from which we can obtain a classical expression for $g$, by replacing
Eqs.\ref{eq:m tot} and \ref{eq:J tot} into \ref{eq:Lande g classical}
and noting that\label{l dot s} 
\[
\mathbf{L}\cdot\mathbf{S}=\frac{1}{2}\left(J^{2}-L^{2}-S^{2}\right)
\]
from $J^{2}=\left(\mathbf{L}+\mathbf{S}\right)^{2}$. One obtains
\begin{equation}
g_{{\rm cl}}=\frac{3}{2}+\frac{S^{2}-L^{2}}{2J^{2}}\,,\label{eq:Lande Classic}
\end{equation}
where the superscript indicates this is a classical approximation
for $g$, which coincides with the quantum mechanical result in the
limit $J^{2}$, $S^{2}$, and $L^{2}$ large \cite{stancil_spin_2009}.

\subsection*{Check points}
\begin{itemize}
\item Write the equation of motion for an angular momentum in the presence
of a magnetic field 
\item What is the dynamics of an angular momentum in the presence of a magnetic
field? 
\item What is the Larmor frequency? 
\end{itemize}

\section{\label{sec:Magnetization}Magnetization}

Inside a material, the magnetic induction $\mathbf{B}$ indicates
the response of the material to the applied magnetic field $\mathbf{H}$.
Both vector fields are related through the \emph{magnetization }in
the sample 
\begin{equation}
\mathbf{B}=\mu_{0}\left(\mathbf{H}+\mathbf{M}\right)\,,\label{eq:B field}
\end{equation}
where the magnetization is defined as the average magnetic moment
per unit volume, 
\begin{equation}
\mathbf{M}(\mathbf{r})=\frac{N\langle\mathbf{m}\rangle_{V}}{V}\,,\label{eq:mag def}
\end{equation}
and where the average indicates that we average over all atomic magnetic
moments in a small volume $V$ around position $\mathbf{r}$ \footnote{We average over a " microscopically large but macroscopically
small" volume $V$.} containing $N$ magnetic ions. In this way, a smooth vectorial function
of position is obtained. From Eq.~\ref{eq:B field}, we see that
the magnetization has the same units as the magnetic field $\mathbf{H}$
(${\rm A}.{\rm m}^{-1}$). In Eq.~\ref{eq:B field}, both $\mathbf{B}$
and $\mathbf{H}$ indicate the fields \emph{inside} the material,
and hence $\mathbf{H}$ contains also the \emph{demagnetizing} fields
(that is, it is not just the external applied field). We will see
more on demagnetization fields in the next section.

The response to the magnetic field of the magnetization and field
induction are characterized by the \emph{magnetic susceptibility}
$\chi$ and the \emph{permeability} $\mu$, respectively 
\begin{align}
\mathbf{M} & =\chi\mathbf{H}\label{eq:susc}\\
\mathbf{B} & =\mu\mathbf{H}\,,\label{eq:perm}
\end{align}
where we have written the simplest expressions for the case in which
all fields are collinear, static (that is, independent of time), and
homogeneous in space ($\mathbf{q}=\omega=0$). In general however
the response functions are tensorial quantities, e.g. $M_{i}=\sum_{j}\chi_{ij}H_{j}$,
and depend on frequency $\omega$ and momentum $\mathbf{q}$. Note
that from Eq.~\ref{eq:B field} we obtain 
\begin{equation}
\mu_{{\rm r}}=\frac{\mu}{\mu_{0}}=1+\chi
\end{equation}
again in the simple collinear case. $\mu_{{\rm r}}$ is the \emph{relative
permeability, }is dimensionless and equals to unity in free space.

The quantities defined in Eqs. \ref{eq:susc} and \ref{eq:perm} are
still allowed to depend on temperature $T$ and magnetic field $\mathbf{H}$.
We will now consider qualitatively the dependence on $\mathbf{H}$.
For linear materials, $\chi$ and $\mu$ are independent of $\mathbf{H}$.
A linear material with negative constant susceptibility is \emph{diamagnetic,
}whereas a positive susceptibility indicates either \emph{paramagnetism}
(no magnetic order) or \emph{antiferromagnetism} (magnetic order with
magnetic moments anti-aligned and zero total magnetization). In these
cases, the magnetization is finite only in the presence of a magnetic
field. On the other hand, if $\chi$ and $\mu$ depend on $\mathbf{H}$,
the relations Eqs.\ref{eq:susc} and \ref{eq:perm} are non linear.
This is the case for magnetically ordered states with net magnetization,
namely \emph{ferromagnets }(magnetic moments aligned and pointing
in the same direction) and \emph{ferrimagnets} (magnetic moments anti-aligned
but of different magnitude, so that there is a net magnetization).
In these materials, the magnetization increases non-linearly with
the applied field and saturates when all the magnetic moments are
aligned. When decreasing the magnetic field, there is a remanent,
finite magnetization at zero field. This process is called \emph{hysteresis}
and it is used to magnetize materials. As we learned in the previous
section, the magnetic moment, and hence the magnetic characteristics
of a material, are related to the total angular momentum of the electrons,
and therefore on the atomic structure. We will learn more about this
in the next chapter.

\subsection*{Check points}
\begin{itemize}
\item What is the relation between magnetic moment and magnetization? 
\end{itemize}

\section{\label{sec:Magnetostatic-Maxwell-equations}Magnetostatic Maxwell
equations in matter}

To calculate the magnetic dipole moment $\mathbf{m}$ from Eq.~\ref{eq:m gen}
we have to know the microscopic current density. In general however
we are not interested in microscopic, fast fluctuations. We already
saw an example in which we considered the average current $I$ generated
by one orbiting electron, to obtain semiclassically the gyromagnetic
ratio $\gamma_{{\rm L}}$ in Sec.\ref{sec:Orbital-angular-momentum}.
We have also defined the magnetization $\mathbf{M}$ as a macroscopic
quantity which entails the average density of the microscopic $\mathbf{m}$.
In a material, in general we have access to the magnetization, which
is due to bound microscopic currents, and to the macroscopic current
density due to free charges, which we will denominate $\mathbf{j}_{{\rm F}}$.
This motivates defining a macroscopic vector potential $\mathbf{A}$
in terms of these two macroscopic quantities, and not the microscopic
currents as in Eq.~\ref{eq:A vec gen} 
\begin{equation}
\mathbf{A}(\mathbf{r})=\frac{\mu_{0}}{4\pi}\int{\rm d}^{3}\mathbf{r}'\left[\frac{\mathbf{j}_{{\rm F}}(\mathbf{r}')}{|\mathbf{r}-\mathbf{r}'|}+\frac{\mathbf{M}(\mathbf{r}')\times(\mathbf{r}-\mathbf{r}')}{|\mathbf{r}-\mathbf{r}'|^{3}}\right]\,.\label{eq:A vec macro}
\end{equation}
Note that this is simply re-writing Eq.~\ref{eq:A vec gen}, separating
the bound- and free- currents contributions. The bound-current contribution,
the second term in Eq.~\ref{eq:A vec macro}, is written in terms
of the magnetization and is equivalent to an averaged Eq.~\ref{eq:m dipole field A}.

Eq.~\ref{eq:A vec macro} allows us to define an effective current
density associated with the magnetization, by noting that \cite{john_david_jackson_classical_1998}
\begin{align}
\int_{\mathcal{V}}{\rm d}^{3}\mathbf{r}'\frac{\mathbf{M}(\mathbf{r}')\times(\mathbf{r}-\mathbf{r}')}{|\mathbf{r}-\mathbf{r}'|^{3}} & =\int_{\mathcal{V}}{\rm d}^{3}\mathbf{r}'\mathbf{M}(\mathbf{r}')\times\nabla'\left(\frac{1}{|\mathbf{r}-\mathbf{r}'|}\right)\label{eq:jm from M 2}\\
 & =\int_{\mathcal{V}}{\rm d}^{3}\mathbf{r}'\nabla'\times\mathbf{M}(\mathbf{r}')\left(\frac{1}{|\mathbf{r}-\mathbf{r}'|}\right)+\oint_{\mathcal{S}}\frac{\mathbf{M}(\mathbf{r}')\times{\rm d}\mathbf{a}'}{|\mathbf{r}-\mathbf{r}'|}\,.\nonumber 
\end{align}
We can therefore define an effective \emph{bound volume current density
\begin{equation}
\mathbf{j}_{{\rm B}}=\nabla\times\mathbf{M}\label{eq:jB}
\end{equation}
}and an effective \emph{bound surface current density 
\begin{equation}
\mathbf{K}_{{\rm B}}=\mathbf{M}\times\hat{\mathbf{n}}\label{eq:KB}
\end{equation}
}where the surface element is defined as ${\rm d\mathbf{a}={\rm d}a}\mathbf{\hat{n}}$.
In the bulk, for a well behaved magnetization function, the surface
integral vanishes and we obtain 
\begin{equation}
\mathbf{A}(\mathbf{r})=\frac{\mu_{0}}{4\pi}\int{\rm d}^{3}\mathbf{r}'\left[\frac{\mathbf{j}_{{\rm F}}(\mathbf{r}')}{|\mathbf{r}-\mathbf{r}'|}+\frac{\mathbf{j}_{{\rm B}}(\mathbf{r}')}{|\mathbf{r}-\mathbf{r}'|}\right]\,.\label{eq:A vec bulk}
\end{equation}
The surface current $\mathbf{K}_{{\rm B}}$ enters usually through
boundary conditions at interfaces.

If we now go back to Ampere's Eq.~\ref{eq:Ampere diff} and separate
the total current density into free and bound contributions $\mathbf{j}=\mathbf{j}_{{\rm F}}+\mathbf{j}_{{\rm B}}$,
we obtain 
\begin{equation}
\nabla\times\mathbf{B}=\mu_{0}\left(\mathbf{j}_{{\rm F}}+\nabla\times\mathbf{M}\right)\label{eq:Ampere B M}
\end{equation}
which defines the magnetic field $\mathbf{H}$ 
\begin{equation}
\mathbf{H}=\frac{1}{\mu_{0}}\mathbf{B}-\mathbf{M}\label{eq:def H}
\end{equation}
such that 
\begin{equation}
\nabla\times\mathbf{H}=\mathbf{j}_{{\rm F}}\,.\label{eq:rot H}
\end{equation}
Therefore the magnetic field $\mathbf{H}$ takes into account in an
average way the bound currents, and has as its only source the free
currents. Eq.~\ref{eq:rot H} is equivalent to Eq.~\ref{eq:Ampere diff},
just re-written in a more convenient form for macroscopic magnetostatics
in matter. Note that $\mathbf{H}$ is, on the contrary to $\mathbf{B}$,
not divergence-free: 
\begin{equation}
\nabla\cdot\mathbf{H}=-\nabla\cdot\mathbf{M}\,.\label{eq:div H}
\end{equation}

The magnetostatic Maxwell equations in matter (also known as " macroscopic" )
therefore read 
\begin{align}
\nabla\cdot\mathbf{B} & =0\nonumber \\
\nabla\times\mathbf{H} & =\mathbf{j}_{{\rm F}}\label{eq:ME in matter}
\end{align}
and have to be complemented by the \emph{constitutive equation} $\mathbf{B}=\mu\mathbf{H}$
in linear media (taking $\mu$ as a constant, independent of $\mathbf{H}$)
or $\mathbf{B}=F(\mathbf{H})$ in non-linear (e.g., ferromagnetic)
media, where $F$ is a characteristic function of the material.

We finish this section by stating the magnetostatic boundary conditions
at an interface between two different media 1 and 2 
\begin{align}
(\mathbf{B}_{2}-\mathbf{B}_{1})\cdot\hat{\mathbf{n}} & =0\label{eq:BC B}\\
\hat{\mathbf{n}}\times(\mathbf{H}_{2}-\mathbf{H}_{1}) & =\mathbf{K}_{{\rm F}}\label{eq:BC H}
\end{align}
where $\mathbf{K}_{{\rm F}}$ is a free surface current density (usually
0). 
\begin{enumerate}
\item \textbf{\emph{Exercise: Prove Eqs \ref{eq:jm from M 2}.}} 
\end{enumerate}

\subsection*{Check points}
\begin{itemize}
\item What is the meaning of the magnetic field $\mathbf{H}$? 
\item What are the magnetostatic Maxwell equations in matter? 
\end{itemize}

\section{\label{sec:Demagnetizing-fields}Demagnetizing fields}

A crucial difference between magnetic and electric fields is the lack
of free magnetic charges, or monopoles.\footnote{Magnetic monopoles, if they exist, have evaded experimental detection
so far. They can however emerge as effective quasiparticles in condensed
matter systems, and have been detected in materials which behave magnetically
as a " spin ice" \cite{castelnovo2008,spinice1,spinice2}.} They are however a useful mathematical construction in some cases,
for example, to calculate the so called demagnetization fields. In
finite systems, we can consider the magnetization as dropping to zero
abruptly at the boundary of the material, giving rise to an accumulated
" magnetic charge density" at the surface
which acts as an extra source of magnetic fields inside of the material.
These fields in general oppose to an externally applied magnetic field
and are therefore dubbed \emph{demagnetizing} fields. A surface magnetic
charge density is energetically costly, and for finite magnetic systems
at the microscale it can determine the spatial dependence of the magnetically
ordered ground state, giving rise to \emph{magnetic textures}.

\begin{figure}
\begin{centering}
\includegraphics[width=0.7\textwidth]{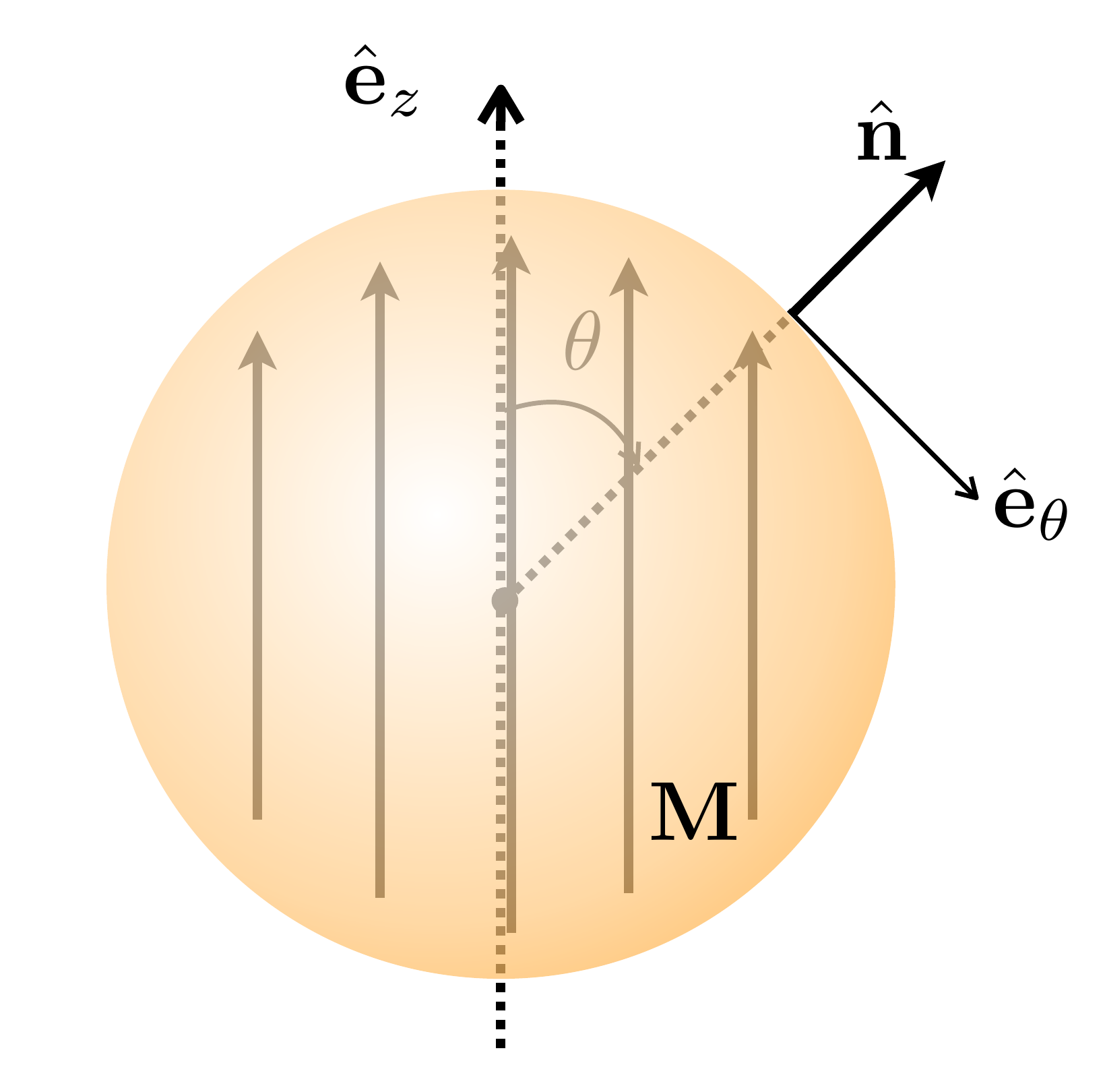} 
\par\end{centering}
\caption{{\small{}{}Coordinate system for the uniformly magnetized sphere
problem.}}

\label{fig:Unif_M_Sphere} 
\end{figure}

If we consider the special case of no free currents, $\mathbf{j}_{{\rm F}}=0$,
Eqs. \ref{eq:ME in matter} imply that we can define a magnetic scalar
potential $\phi_{{\rm M}}$ such that 
\begin{equation}
\mathbf{H}=-\nabla\phi_{{\rm M}}
\end{equation}
and using Eq.~\ref{eq:div H} we obtain a Poisson equation 
\begin{equation}
\nabla^{2}\phi_{{\rm M}}=-\nabla\cdot\mathbf{M}
\end{equation}
with solution 
\begin{equation}
\phi_{{\rm M}}(\mathbf{r})=-\frac{1}{4\pi}\int_{\mathcal{V}}{\rm d}^{3}\mathbf{r}'\frac{\nabla\cdot\mathbf{M}(\mathbf{r}')}{|\mathbf{r}-\mathbf{r}'|}+\frac{1}{4\pi}\oint_{\mathcal{S}}{\rm d}a'\frac{\hat{\mathbf{n}'}\cdot\mathbf{M}(\mathbf{r}')}{|\mathbf{r}-\mathbf{r}'|}\,.\label{eq:Pot phi M}
\end{equation}
Analogous to the case of the vector potential in Eq. \ref{eq:jm from M 2},
this allows us to define an effective \emph{magnetic-charge density}
\begin{equation}
\rho_{{\rm M}}=-\nabla\cdot\mathbf{M}\label{eq:rho M}
\end{equation}
and an effective \emph{magnetic surface-charge density} 
\begin{equation}
\sigma_{{\rm M}}=\mathbf{M}\cdot\mathbf{\hat{n}\,.}\label{eq:sigma m}
\end{equation}
We see that $\rho_{{\rm M}}$ can only be finite for a non-homogeneous
magnetization $\mathbf{M}(\mathbf{r})$, whereas a finite $\sigma_{{\rm M}}$
indicates a discontinuity of $\mathbf{M}$ at the chosen surface $\mathcal{S}$. 
\begin{enumerate}
\item \textbf{\emph{Exercise: Uniformly magnetized sphere}} 
\end{enumerate}
For a ferromagnet at saturation, the magnetization can be considered
as given, so we can in principle calculate the resulting magnetic
field for a given geometry using Eq.~\ref{eq:Pot phi M}. We consider
here as an example the case of a uniformly magnetized sphere as depicted
in Fig.~\ref{fig:Unif_M_Sphere}. 

a) Choosing $\hat{\mathbf{e}}_{z}$ in the direction of $\mathbf{M}$
we can write $\mathbf{M}=M_{0}\hat{\mathbf{e}}_{z}$. Calculate $\rho_{{\rm M}}$
and $\sigma_{{\rm M}}$ and write the Poisson equation for $\phi_{{\rm M}}$
.

b) Show that the scalar potential inside of the sphere is 
\begin{equation}
\phi_{{\rm M}}^{{\rm in}}=\frac{1}{3}M_{0}z\label{eq:phi M in}
\end{equation}
and find the magnetic field $\mathbf{H}^{{\rm in}}$ and magnetic
induction $\mathbf{B}^{{\rm in}}$ inside of the sphere.

c) The magnetic field $\mathbf{H}^{{\rm in}}$ inside of the sphere
opposes the magnetization and it is therefore called a \emph{demagnetizing
field}. The proportionality coefficient between $\mathbf{H}^{{\rm in}}$
and $\mathbf{M}$ is called the \emph{demagnetizing factor} $N$.
What is the value of $N$ in this case? Demagnetization factors are
geometry dependent, and can moreover be defined only in very special
cases with simple geometries \footnote{The demagnetizing fields are always present, but it is only in very
simple geometries that one can describe them with simple numerical
factors.}. Besides the sphere, one can define demagnetization factors for an
infinite plane, an infinite cylinder, and a spheroid.

d) Let's assume that now the sphere is placed in an external magnetic
field $\mathbf{H}_{0}$. Using linearity, write the solution for $\mathbf{H}^{{\rm in}}$
and $\mathbf{B}^{{\rm in}}$ in this case.

e) Let's now consider the case that the sphere is not permanently
magnetized, but we now the material has a permeability $\mu$. From
the constitutive equation 
\begin{equation}
\mathbf{B}^{{\rm in}}=\mu\mathbf{H}^{{\rm in}}\,,\label{eq:constitutive}
\end{equation}
obtain the magnetization as a function of the external magnetic field
$\mathbf{M}_{\mu}(\mathbf{H}_{0})$, where the notation $\mathbf{M}_{\mu}$
implies that in this case we consider the magnetization not as given,
but it depends on the permeability of the material. Show that $\mathbf{M}_{\mu}(0)=0$,
and therefore the obtained expression is not valid for materials with
permanent magnetization.

\subsection*{Check points}
\begin{itemize}
\item Which is the origin of the demagnetization factors? 
\end{itemize}

\chapter{\label{chap:Atomic-Origins-of}Atomic Origins of Magnetism}

In the previous chapter we reviewed the basic concepts of magnetism
and magnetostatics using some semiclassical considerations. In particular,
we attributed the magnetic moment of atoms to " small
current loops" and to the angular momentum of electrons.
In this chapter, we will put these concepts into more solid footing
with the help of quantum mechanics.

\section{\label{sec:Basics-of-Quantum}Basics of quantum mechanics}

We first review some basic concepts of quantum mechanics. In quantum
mechanics, we describe a particle of mass $m$ in a potential $V$
by a wavefunction $\psi(\mathbf{r},t)$ which satisfies the \emph{Schrödinger
equation} 
\begin{equation}
i\hbar\frac{\partial\psi(\mathbf{r},t)}{\partial t}=-\frac{\hbar^{2}}{2m}\nabla^{2}\psi(\mathbf{r},t)+V\,\psi(\mathbf{r},t)\,.\label{eq:schroedinger eq}
\end{equation}
The probability of finding the particle at a time $t$ in a volume
element ${\rm d}^{3}r$ around position $\mathbf{r}$ is given by
$|\psi(\mathbf{r},t)|^{2}{\rm d}^{3}r$ . If the potential $V$ is
independent of time, $\psi(\mathbf{r},t)=\psi(\mathbf{r})f(t)$ and
$\psi(\mathbf{r})$ is an eigenfunction of the \emph{time-independent
Schrödinger equation} 
\begin{equation}
-\frac{\hbar^{2}}{2m}\nabla^{2}\psi(\mathbf{r})+V(\mathbf{r})\,\psi(\mathbf{r})=E\psi(\mathbf{r})\label{eq:time independent schr}
\end{equation}
with energy $E.$ Equivalently, we can write the eigenvalue equation
for the Hamiltonian in Dirac notation 
\begin{equation}
\hat{H}\Ket{\psi}=E\Ket{\psi}\label{eq:H E}
\end{equation}
where the quantum state of the particle is represented by the ket
$\Ket{\psi}$, the respective wavefunction is $\psi(\mathbf{r})=\langle\mathbf{r}|\psi\rangle$,
and the Hamiltonian operator is 
\begin{equation}
\hat{H}=\frac{\hat{p}^{2}}{2m}+V(\hat{\mathbf{r}}).\label{eq:Hamiltonian general}
\end{equation}
In the position representation, $\mathbf{\hat{p}}\rightarrow-i\hbar\nabla$
so that, for example, $\langle\mathbf{r}|\hat{\mathbf{p}}|\psi\rangle=-i\hbar\nabla\psi(\mathbf{r})$
.

In general, for any operator $\hat{A}$ we can write the \emph{eigenvalue
equation $\hat{A}\Ket{\psi_{\alpha}}=\alpha\Ket{\psi_{\alpha}}$ }where
$\Ket{\psi_{\alpha}}$ is an eigenstate with eigenvalue\emph{ $\alpha$.
}For a hermitian operator, $\hat{A}^{\dagger}=\hat{A}$, $\alpha$
is real and the eigenstates form a basis of the Hilbert space where
the operator acts. This is called an \emph{observable}. The expectation
value for $\hat{A}$ if the system is in the eigenstate $\Ket{\psi_{\alpha}}$
then is simply $\Bra{\psi_{\alpha}}\hat{A}\Ket{\psi_{\alpha}}=\alpha$.
If we consider a second operator $\hat{B}$ acting on the same Hilbert
space, it is only possible to find a common basis of eigenstates of
$\hat{A}$ and $\hat{B}$ if and only if the two operators commute:
$[\hat{A},\hat{B}]=\hat{A}\hat{B}-\hat{B}\hat{A}=0$. In this case,
the two operators can be measured simultaneously to (in principle)
arbitrary precision. If the operators do not commute, then we run
into the \emph{Heisenberg uncertainty principle. }The most well known
example is that of the momentum and position operators, which satisfy
$[\hat{x},\hat{p}]=i\hbar$. How precise we measure one of the operators
will determine the precision up to which we can know the value of
the other: $\Delta x\Delta p\ge\hbar/2$. In general 
\begin{equation}
\Delta A\Delta B\ge\frac{1}{2}\left|\langle[\hat{A},\hat{B}]\rangle\right|
\end{equation}
where $\Delta A=\sqrt{\langle\hat{A}^{2}\rangle-\langle\hat{A}\rangle^{2}}$
corresponds to the standard variation of $\hat{A}$ and analogously
for operator $\hat{B}$.

\section{\label{sec:Orbital-Angular-Momentum}Orbital angular momentum in
quantum mechanics}

\begin{figure}
\begin{centering}
\includegraphics[width=0.8\textwidth]{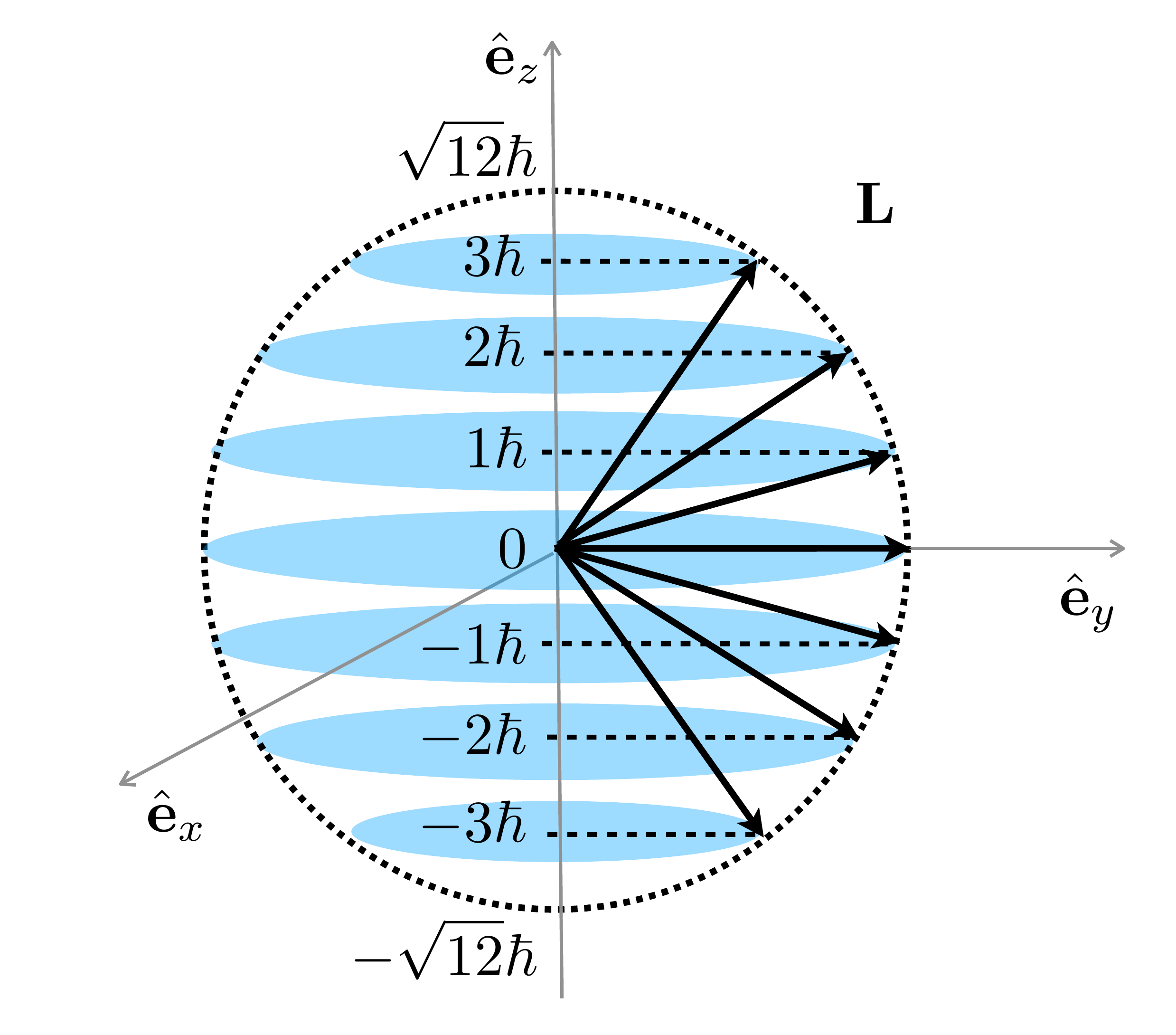} 
\par\end{centering}
\caption{{\small{}{}Pictorial representation of the spatial quantization of
angular momentum. In this example, $l=3$. $\mathbf{L}$ precesses
around the $z$ axis and has a definite projection on it, that can
take one of the allowed values $-3\le m_{l}\le3$. Note that the maximal
quantum mechanically allowed value of the projection ($3\hbar$ in
this case) is smaller than the classically allowed one.}}
\label{Fig:PicL} 
\end{figure}

The orbital angular momentum operator expression in quantum mechanics
is inherited from its classical expression, $\hat{\mathbf{L}}=\hat{\mathbf{r}}\times\hat{\mathbf{p}}$.
In the position representation it is given by 
\begin{equation}
\hat{\mathbf{L}}=-i\hbar\mathbf{r}\times\nabla\,.\label{eq:L op}
\end{equation}
From this expression it is easy to verify that the different components
of $\mathbf{\hat{L}}$ do not commute with each other. Instead, one
obtains 
\begin{equation}
[\hat{L}_{i},\hat{L_{j}}]=i\hbar\epsilon_{ijk}\hat{L}_{k}\,,\label{eq:L commutator}
\end{equation}
where $\epsilon_{ijk}$ is the Levi-Civita tensor and the Einstein
convention for the implicit sum or repeated indices has been used.
Therefore, it is not possible to measure simultaneously with arbitrary
precision all components of the angular momentum. Let's assume we
choose to measure $\hat{L}_{z}$ . In this case, the Heisenberg uncertainity
principle reads 
\begin{equation}
\Delta L_{x}\Delta L_{y}\ge\frac{\hbar}{2}|\langle L_{z}\rangle|\,.\label{eq:heisenberg L}
\end{equation}
It is however possible to find a common basis for $\hat{L}^{2}$ and
one of the angular momentum components, since $[\hat{L}^{2},\hat{L}_{i}]=0$.
Typically, $\hat{L}_{z}$ is taken and the respective eigenvalues
are labelled by $l$ and $m$. These are called the \emph{quantum
numbers}. The eigenstates satisfy 
\begin{align}
\hat{L}^{2}\Ket{\psi_{lm_{l}}} & =\hbar^{2}l(l+1)\Ket{\psi_{lm_{l}}}\nonumber \\
\hat{L}_{z}\Ket{\psi_{lm_{l}}} & =\hbar m\Ket{\psi_{lml}}\,.\label{eq:l and m}
\end{align}
For the orbital angular momentum, $l$ is an integer and $-l\le m_{l}\le l$.
These conditions can be depicted pictorically as in Fig.~\ref{Fig:PicL}.

The angular momentum vector has a magnitude $\hbar\sqrt{l(l+1)}$
and its projection on the \emph{z}-axis is quantized and takes one
of the possible values $\hbar m_{l}$. The maximum value of $L_{z}$
is $\hbar l$, instead of $\hbar\sqrt{l(l+1)}$ as one would expect
classically. We recover the classical expectation in the limit $l\gg1$
. The $L_{x}$ and $L_{y}$ components do not have a definite value
and are represented as a precession of $\mathbf{L}$ around the \emph{z}-axis.

\subsection*{Check points}
\begin{itemize}
\item Explain graphically the properties of an angular momentum operator
and how it differs from a classical angular momentum. 
\end{itemize}

\section{\label{sec:Hydrogen-Atom}Hydrogen atom}

For a problem with rotational symmetry the angular momentum is conserved:
$[\hat{H},\hat{\mathbf{L}}]=0$. Hence we can find a basis of eigenstates
that are also energy eigenstates, such that 
\begin{equation}
\hat{H}\Ket{\psi_{nlm}}=E_{n}\Ket{\psi_{nlm}}\,.\label{eq:E_n}
\end{equation}
The quantum numbers are denominated as \emph{principal}, \emph{azimuthal},
and \emph{magnetic} respectively for $n,$ $l$, and $m$. If we ignore
spin, we have all the tools to solve the energy levels and orbitals
for the hydrogen atom, in which the electron is subject to the Coulomb
potential 
\begin{equation}
V_{C}(r)=-\frac{e^{2}}{4\pi\varepsilon_{0}r}\label{eq:Coulomb Pot}
\end{equation}
due to the nucleus. Due to the spherical symmetry of the problem,
is is convenient to write the wavefunction in spherical coordinates.
From Eq.~\ref{eq:time independent schr} it can be shown that 
\[
\psi_{nlm}(r,\theta,\phi)=R_{nl}(r)Y_{lm}(\theta,\phi)
\]
The principal number $n=1,2,3...$ gives the quantization of energy
$E_{n}\propto-1/n^{2}$. The $R_{nl}(r)$ are \emph{associated Laguerre
functions }and determine the radial profile of the probability distribution
for the electron. $Y_{lm}(\theta,\phi)$ are spherical harmonics,
which can also be written in terms of the \emph{associated Legendre
functions}, $Y_{lm}(\theta,\phi)=P_{l}^{m}(\cos\theta)e^{im\phi}$.
For $m=0$, $Y_{l0}(\theta)=P_{l}(\cos\theta)$ are simply the Legendre
polynomials. The azimuthal number $l=0,1,...,n-1$ labels the usual
s, p, d, ... orbitals. The s orbitals are spherically symmetric, since
$Y_{00}(\theta,\phi)=1/\sqrt{4\pi}$. The higher the azimuthal number,
the higher the probability to find the electron further away from
the nucleus, whereas $n$ gives the number of nodes of the wavefunction
in the radial direction.

That the azimuthal quantum number $l$ is quantized was demonstrated
experimentally in what we now know as the \emph{Zeeman effect}. In
order to see why, we come back to the relation between the orbital
angular momentum and the magnetic moment. We can write 
\begin{align}
m_{{\rm L}} & =\mu_{{\rm B}}\sqrt{l(l+1)}\nonumber \\
\mathbf{m}_{{\rm L}}\cdot\mathbf{z} & =-\mu_{{\rm B}}m_{l}\label{eq:m_L quantized}
\end{align}
where $\mu_{{\rm B}}=\hbar\gamma_{{\rm L}}$ is the \emph{Bohr magneton
}and the expressions correspond to expectation values\emph{. }If the
atom is placed in an external magnetic field, there will be an extra
contribution to the energy due to the Zeeman term, see Eq.~\ref{eq:Zeeman energy}.
We can take the \emph{z}-axis to coincide with the magnetic field,
hence 
\begin{equation}
E_{{\rm Z}}=\mu_{{\rm B}}m_{l}B\label{eq:Zeeman ml}
\end{equation}
and we say that the degeneracy of the $l$ level, originally $2l+1$,
is split for all $l>0$. This splitting can be measured in the absorption
spectrum of atoms with total spin angular momentum equal to zero.
To add the effect of the spin degree of freedom, we have however first
to understand how to combine angular momentum operators in quantum
mechanics.

\subsection*{Check points}
\begin{itemize}
\item What are the quantum numbers for the Hydrogen atom and what do they
tell us? 
\end{itemize}

\section{\label{sec:Addition-of-Angular}Addition of angular momentum and
magnetic moment}

The orbital angular momentum is the generator of rotations in position
space. In general however, we can define an angular momentum simply
by its algebra, determined by the commutation relation \ref{eq:L commutator}.
The spin angular momentum generates rotations in spin space and satisfies
\begin{align*}
[\hat{S}_{i},\hat{S_{j}}] & =i\hbar\epsilon_{ijk}\hat{S}_{k}\\
\hat{S}^{2}\Ket{s} & =\hbar^{2}s(s+1)\Ket{s}\\
\hat{S}_{z}\Ket{s} & =\hbar m_{s}\Ket{s}\,.
\end{align*}
In contrast to the orbital angular momentum, the spin quantum number
$s$ is not constrained to be an integer, and can take also half-integer
values. Fermions (\emph{e.g.}, the electron) have half-integer values
of spin and bosons integer values. For electrons, $s=1/2$ and $m_{s}=\pm1/2$
.

The spin operator commutes with the orbital angular momentum 
\[
[\hat{\mathbf{S}},\hat{\mathbf{L}}]=0
\]
since they act on different state spaces. One would be therefore tempted
to choose $\left\{ \hat{\mathbf{S}}^{2},\hat{\mathbf{L}}^{2},\hat{L}_{z},\hat{S}_{z}\right\} $
as a set of commuting observables. Spin and orbital angular momentum
however interact \emph{via }the\emph{ spin-orbit interaction}, a relativistic
correction to the Hamiltonian \ref{eq:Hamiltonian general} which
reads 
\begin{equation}
\hat{H}_{{\rm SO}}=\frac{\hbar^{2}}{2m_{e}^{2}c^{2}}\frac{1}{r}\frac{\partial V}{\partial r}\hat{\mathbf{S}}\cdot\hat{\mathbf{L}}\,.\label{eq:SO Ham}
\end{equation}
For an atomic system, $V$ is the Coulomb potential. This correction
is usually small, but it increases with the atomic number. Since $\hat{\mathbf{S}}$
and $\hat{\mathbf{L}}$ are coupled by Eq.~\ref{eq:SO Ham}, they
are not conserved and they do not commute separately with the Hamiltonian.
The total angular momentum $\hat{\mathbf{J}}=\hat{\mathbf{S}}+\hat{\mathbf{L}}$
however is conserved. We choose therefore $\left\{ \hat{\mathbf{S}}^{2},\hat{\mathbf{L}}^{2},\hat{\mathbf{J}}^{2},\hat{J}_{z}\right\} $
as a set of commuting observables, and $s$, $l$ , $j$ and $m_{j}$
are our quantum numbers. The change of basis is achieved via the \emph{Clebsch-Gordan
coefficients 
\begin{equation}
\Ket{sljm_{j}}=\sum_{m_{s}m_{l}}\Ket{sm_{s}lm_{l}}\langle sm_{s}lm_{l}|sljm_{j}\rangle\,.\label{eq:Clebsh Gordan}
\end{equation}
}We now have the task of relating the magnetic dipolar moment $\mathbf{m}_{{\rm TOT}}$
to the total angular momentum operator $\mathbf{\hat{J}}$. From Eq.~\ref{eq:m tot}
we know that $\mathbf{m}_{{\rm TOT}}$ is not collinear with $\mathbf{\hat{J}}$,
but it is proportional to $\mathbf{\hat{L}+2\hat{\mathbf{S}}}$. 
\begin{figure}
\begin{centering}
\includegraphics[width=0.8\textwidth]{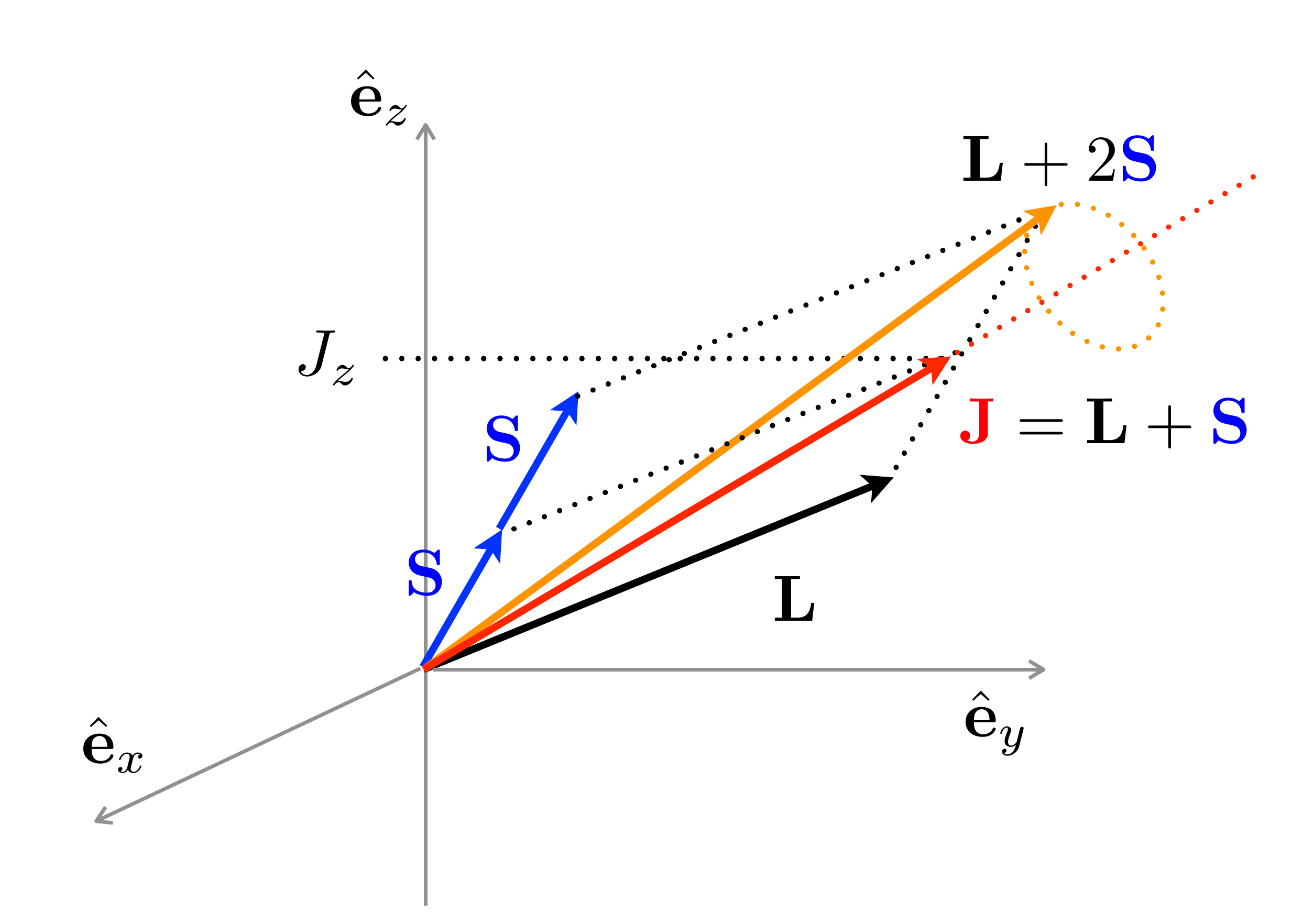} 
\par\end{centering}
\caption{{\small{}{}Pictorial depiction of the precession of $\mathbf{\hat{L}+2\hat{\mathbf{S}}}=\mathbf{\hat{m}}_{{\rm TOT}}/\gamma_{{\rm L}}$
around $\mathbf{\hat{J}}$.} }

\label{fig:Lplus2S} 
\end{figure}

We note however that the projection $\hat{\mathbf{m}}_{{\rm TOT}}\cdot\hat{\mathbf{J}}$
is well defined, since 
\begin{align}
[\hat{\mathbf{L}}\cdot\hat{\mathbf{J}},\hat{\mathbf{J}}^{2}] & =[\hat{\mathbf{L}}\cdot\hat{\mathbf{J}},\hat{\mathbf{L}}^{2}]=[\hat{\mathbf{L}}\cdot\hat{\mathbf{J}},\hat{\mathbf{S}}^{2}]=[\hat{\mathbf{L}}\cdot\hat{\mathbf{J}},\hat{\mathbf{J}_{z}}]=0\nonumber \\{}
[\hat{\mathbf{S}}\cdot\hat{\mathbf{J}},\hat{\mathbf{J}}^{2}] & =[\hat{\mathbf{S}}\cdot\hat{\mathbf{J}},\hat{\mathbf{L}}^{2}]=[\hat{\mathbf{S}}\cdot\hat{\mathbf{J}},\hat{\mathbf{S}}^{2}]=[\hat{\mathbf{S}}\cdot\hat{\mathbf{J}},\hat{\mathbf{J}_{z}}]=0\,.\label{eq:LdotJ SdotJ}
\end{align}
The magnetic moment therefore precesses around $\mathbf{\hat{J}}$,
as depicted in Fig.~\ref{fig:Lplus2S}.

The Wigner-Eckart theorem allows us to relate $\mathbf{\hat{m}}_{{\rm TOT}}$
with $\mathbf{\hat{J}}$ in terms of expectation values by noting
that \cite{ashcroft_solid_1976} 
\[
\gamma_{{\rm L}}\langle sljm_{j}|\hat{\mathbf{L}}+2\hat{\mathbf{S}}|sljm_{j}'\rangle=\gamma_{{\rm L}}g(slj)\langle sljm_{j}|\hat{\mathbf{J}}|sljm_{j}'\rangle\,.
\]
Therefore in the $(2j+1)$ degenerate subspace with fixed quantum
numbers $(s,l,j)$ we can think of $\mathbf{\hat{m}}_{{\rm TOT}}$
as being proportional to $\mathbf{\hat{J}}$. In the literature, this
is sometimes denoted as "$\mathbf{\hat{m}}_{{\rm TOT}}=\gamma_{{\rm L}}g\mathbf{\hat{J}}${}
" in an abuse of notation. The value of $g(slj)$ we
can find by projecting $\hat{\mathbf{m}}_{{\rm TOT}}\cdot\hat{\mathbf{J}}$
\[
\langle sljm_{j}|\left(\hat{\mathbf{L}}+2\hat{\mathbf{S}}\right)\cdot\hat{\mathbf{J}}|sljm_{j}\rangle=g(slj)\langle sljm_{j}|\hat{\mathbf{J}}^{2}|sljm_{j}\rangle
\]
and using Eqs.\ref{eq:LdotJ SdotJ} we obtain the \emph{Landé factor
\begin{equation}
g(s,l,j)=\frac{3}{2}+\frac{s(s+1)-l(l+1)}{2j(j+1)}\,.\label{eq:Lande factor}
\end{equation}
}We see that this expression coincides with the classical one obtained
in Eq.~\ref{eq:Lande Classic} in the limit of $s,l,j\gg1$.

Going back to the Zeeman splitting, we see that in general the Zeeman
correction to the atomic energy in the presence of a magnetic field
is given by 
\begin{equation}
E_{{\rm Z}}=\mu_{{\rm B}}g(s,l,j)m_{j}B\label{eq:Zeeman mj}
\end{equation}
and therefore the Zeeman splitting depends on all the atomic orbital
numbers, and it is not simply $\mu_{{\rm B}}B$. This is denominated
the \emph{anomalous Zeeman effect }since at the time of the experiments
the spin was still not known, and it was not possible to explain the
effect. Note however that the normal Zeeman effect can be observed
only for atoms with zero total spin, and therefore the anomalous one
is much more common. 
\begin{enumerate}
\item \textbf{\emph{Exercise: Prove Eqs.\ref{eq:LdotJ SdotJ}}} 
\item \textbf{\emph{Exercise: Derive Eq.~\ref{eq:Lande factor}.}} 
\end{enumerate}

\subsection*{Check points}
\begin{itemize}
\item How do you relate the magnetic moment of an electron to its total
angular momentum? Why? 
\item What is the Landé factor? 
\end{itemize}

\section{\label{sec:Generalization-to-Many}Generalization to many electrons}

Generalizing the previous concepts beyond the hydrogen atom / single
electron problem is impossible to do in an exact manner, since the
problem turns into a many-body problem: the many electrons interact
not only with the nucleus, but among themselves. We can however make
analytical progress by doing some reasonable approximations. The first
one is called the \emph{Hartree }approximation, in which we consider
that each electron moves in an effective central potential $V_{{\rm eff}}(r)$
generated by the nucleus plus all the other electrons. The other electrons
are said to \emph{screen }the potential of the nucleus, since their
charge is opposite.

The second approximation concerns the spin-orbit coupling Eq.~\ref{eq:SO Ham}.
For all except the heaviest atoms, the spin-orbit interaction is weak
and can be treated within perturbation theory. In this case, we can
first neglect this interaction, an consider that $\hat{\mathbf{L}}_{i}$
and $\hat{\mathbf{S}}_{i}$ for each electron $i$ are independent.
Hence, we can calculate the total orbital and spin angular momentum
simply by summing them separately: $\hat{\mathbf{L}}_{{\rm TOT}}=\sum_{i}\hat{\mathbf{L}}_{i}$and
$\hat{\mathbf{S}}_{{\rm TOT}}=\sum_{i}\hat{\mathbf{S}}_{i}$ , and
then proceed to calculate the total angular momentum $\hat{\mathbf{J}}_{{\rm TOT}}=\hat{\mathbf{L}}_{{\rm TOT}}+\hat{\mathbf{S}}_{{\rm TOT}}$
and the corresponding magnetic moment. $\hat{\mathbf{S}}_{{\rm TOT}}^{2}$
has eigenvalues $\hbar S(S+1)$ with $S=\sum_{i}m_{s,i}$ and, respectively,
$\hat{\mathbf{L}}_{{\rm TOT}}^{2}$ has eigenvalues $\hbar L(L+1)$
with $L=\sum_{i}m_{l,i}$. The allowed values of $\hat{\mathbf{J}}_{{\rm TOT}}$
are given by the angular momentum summation rules: $J:|L-S|,|L-S|+1,...L+S$.
This approximation is denominated the \emph{Russell-Saunders coupling
}and it gives rise to the well known\emph{ Hund's rules}. To calculate
$\hat{\mathbf{J}}_{{\rm TOT}}$ we need first a prescription to obtain
$\hat{\mathbf{S}}_{{\rm TOT}}$ and $\hat{\mathbf{L}}_{{\rm TOT}}$.

A closed shell means that we have occupied all $2(2l+1)$ levels in
it, where the factor of 2 comes for the spin $s=\pm1$. Therefore
both the total orbital and spin angular momentum of the shell are
zero, and hence also the total angular momentum $\hat{\mathbf{J}}$
in the shell. The total angular momentum of the atom, and therefore
its magnetic properties, will be determined by the last, partially
unoccupied shell. If all shells are closed (that is, full), then $\hat{\mathbf{J}}_{{\rm TOT}}=0$
and the atom is diamagnetic. The Hund rules tell us how to distribute
our $n\le2(2l+1)$ " left over" electrons
in the last shell.The total spin $\hat{\mathbf{S}}_{{\rm TOT}}$ we
obtain by applying the Pauli principle: since electrons are fermions,
their wavefunction is antisymmetric and two electrons can not have
the same quantum numbers. Each orbital characterized by $l$ can then
be occupied by only two electrons, one with spin up, and one with
spin down. The\emph{ first Hund rule} tells us to maximize S, since
this will tend to put one electron with in each orbital until half
filling, $(2l+1)$, and then continue with spin down. This minimizes
Coulomb repulsion by putting electrons, in average, as far apart as
possible. The\emph{ second Hund rule }tells us to maximize the orbital
angular momentum, once the spin is maximized. This also minimizes
Coulomb repulsion, by making the electrons orbit as far apart as possible.
The\emph{ third Hund rule }sounds more mysterious: if the shell is
more than half filled ($n\ge2l+1$) then $J=L+S$, and for a less
than half filled shell ($n\le2l+1$), $J=|L-S|$ . This is actually
due to the spin-orbit interaction $\lambda\hat{\mathbf{L}}_{{\rm TOT}}\cdot\hat{\mathbf{S}}_{{\rm TOT}}$,
where it can be shown that $\lambda$'s sign changes between these
two configurations. Therefore to minimize spin-orbit coupling requires
$\hat{\mathbf{S}}_{{\rm TOT}}$ and $\hat{\mathbf{L}}_{{\rm TOT}}$
parallel or antiparallel, depending on the filling.

To finish this section, we point out that for the heavier elements
the Russell-Saunders coupling prescription is not valid anymore, due
to the strong spin-orbit coupling. For these elements, a different
prescription, denominated \emph{jj coupling}, is used. There $\hat{\mathbf{J}_{i}}$
for each electron is first calculated, and then the total $\hat{\mathbf{J}}_{{\rm TOT}}$.

As we mentioned, if all shells are closed the total angular momentum
is zero and the atom is diamagnetic. Diamagnetism can be understood
by the Faraday law: as a magnetic field is turned on, it induces a
change in the orbital motion of the electrons which opposes the change
in magnetic flux. Diamagnetism is usually a weak effect and it is
overshadowed by paramagnetism in atoms with partially unfilled shells.
Once the atoms are ordered in a lattice and form a solid, it can happen
that magnetic order develops as the temperature is lowered. This will
depend on the electronic interactions, as we will see in the next
chapter.

\subsection*{Check points}
\begin{itemize}
\item What is the Russell-Saunders coupling scheme? 
\end{itemize}

\chapter{\label{chap:Magnetism-in-Solids}Magnetism in Solids}

In the previous chapter we showed how to calculate the magnetic moment
of an atom. We saw that the problem is already quite involved even
for a single atom if we go beyond a hydrogen-like one. When atoms
come together to form a solid, to treat the magnetic problem atom
per atom is not only impossible but also not correct, since we have
to take into account the binding between the atoms that form the solid,
and what matters is the collective behavior of the material. In a
solid, the orbitals of the constituent atoms overlap to form bands
instead of discrete energy levels. Depending on the character of the
orbitals involved in the magnetic response of a material, we can divide
the problem into two big subsets: metals and insulators. In metals,
the orbitals are extended and have a good amount of overlap, so the
electrons are \emph{delocalized }and free to move around the solid.
In insulators, the orbitals are narrow and we talk about \emph{localized
magnetic moments}. Of course, this is an oversimplified view: there
are systems in which both localized and delocalized electrons participate
in magnetism, or in which magnetism and electric conduction are due
to different groups of orbitals. An example is that of \emph{heavy
fermion} systems, which can be modeled as a \emph{Kondo Lattice}:
a lattice of magnetic impurities embedded in a sea of free electrons.
Itinerant magnetism (that one due to delocalized electrons) and mixed
systems belong to what one calls \emph{strongly correlated systems},
highly complex many-body problems in which electron-electron interactions
have to be taken into account. In this course we will concentrate
mostly on magnetic insulators. Moreover, to understand \emph{magnetic
ordering} we have to introduce electronic interactions and take into
account the Pauli principle.

\section{\label{sec:The-Curie-Weiss-law}The Curie-Weiss law}

We consider first a system of localized, $N$ identical non interacting
magnetic moments, and calculate their collective paramagnetic response.\footnote{As we saw in the last chapter, all atoms present a diamagnetic behavior,
but since it is very weak compared to the paramagnetic response, only
atoms with closed shells (e.g. noble gases) present an overall diamagnetic
response. An exception is of course superconducting materials, which
can have a perfect diamagnetic response. This is however a collective,
macroscopic response due to the superconducting currents which oppose
the change in magnetic flux.} We can obtain the magnetization from the Helmoltz free energy (see
e.g. Ref. \cite{ashcroft_solid_1976}) 
\begin{align}
F & =-k_{{\rm B}}T\ln Z\label{eq:F free energy}\\
M & =-\frac{N}{V}\frac{\partial F}{\partial B}
\end{align}
where $k_{{\rm B}}$ is the Boltzmann constant, $T$ is the temperature,
and $Z$ the canonical partition function for one magnetic moment
\begin{equation}
Z=\sum_{n}e^{-\frac{E_{n}}{k_{{\rm B}}T}}\,.\label{eq:partition function}
\end{equation}
If we consider a magnetic moment with total momentum $J$, we have
$2J+1$ possible $J_{z}$ values and 
\begin{equation}
Z=\sum_{J_{z}=-J}^{J}e^{-\frac{1}{k_{{\rm B}}T}g\mu_{{\rm B}}J_{z}B}\,.\label{eq:Z Jz}
\end{equation}
The sum can be performed since it is a geometric one. Putting all
together, one finds 
\begin{equation}
M=\frac{N}{V}g\mu_{{\rm B}}J\mathcal{B}_{J}(\frac{\mu_{{\rm B}}gJB}{k_{{\rm B}}T})\label{eq:mag B(x)}
\end{equation}
where 
\[
\mathcal{B}_{J}(x)=\frac{2J+1}{2J}\coth(\frac{2J+1}{2J}x)-\frac{1}{2J}\coth(\frac{1}{2J}x)
\]
is the Brillouin function. For $\mu_{{\rm B}}B\gg k_{{\rm B}}T$ this
function goes to 1 and the magnetization saturates: all momenta are
aligned with the B-field. For large temperatures instead, $\mu_{{\rm B}}B\ll k_{{\rm B}}T$,
one obtains the inverse-temperature dependence of the magnetization
known as the \emph{Curie law},\emph{ }characterized by the susceptibility\emph{
\begin{equation}
\chi_{{\rm c}}=\frac{M}{H}=\frac{N}{V}\frac{\mu_{0}\left(g\mu_{{\rm B}}\right)^{2}}{3}\frac{J(J+1)}{k_{{\rm B}}T}\,.\label{eq:Curie Susc}
\end{equation}
}

Experimentally, it is found that this result is good for describing
insulating crystals containing rare-earth ions (e.g. Yb, Er), whereas
for transition metal ions in an insulating solid agreement is found
only if one takes $J=S$. This is due to an effect denominated \emph{angular
momentum quenching}, where effectively $L=0$. This quenching is due
to \emph{crystal fields}: since now the atoms are located in a crystalline
environment, rotational symmetry is broken and each atom is located
in electric fields due to the other atoms in the crystal. In the case
of rare-earth ions, the magnetic moments come from f-shells, which
are located deep inside the atom and therefore better isolated form
crystal fields. For transition metals instead, the magnetic moments
come from the outermost d-shells, which are exposed to the symmetry
breaking fields. These however are spatial dependent fields which
do not affect directly the spin degree of freedom. As a consequence
of the breaking of spatial rotational invariance, the orbital angular
momentum is not conserved and precesses instead around the crystal
fields, averaging to zero.

The Curie law is followed by all materials with net magnetic moment
for large enough temperature. For low temperatures however, in certain
materials magnetic order develops and there is a deviation from the
Curie law in the susceptibility. Weiss postulated the existence of
\emph{molecular fields}, which are proportional to the magnetization
in the material and are responsible for the magnetic ordering. The
total magnetic field acting on a magnetic moment within this picture
is therefore $\mathbf{H}_{tot}=\mathbf{H}_{ext}+\lambda\mathbf{M}$
and from the Curie susceptibility 
\begin{equation}
\chi_{{\rm c}}=\frac{M}{H_{tot}}=\frac{M}{H_{ext}+\lambda M}=\frac{C}{T}\label{eq:Curie susc}
\end{equation}
with 
\begin{equation}
C=\frac{N}{V}\frac{\mu_{0}\left(g\mu_{{\rm B}}\right)^{2}}{3}\frac{J(J+1)}{k_{{\rm B}}}\label{eq:Curie const}
\end{equation}
we obtain 
\begin{equation}
M=\frac{CH_{ext}}{T-\lambda C}\label{eq:M Weiss}
\end{equation}
and therefore the susceptibility in the external field follows the
\emph{Curie-Weiss law} 
\begin{equation}
\chi_{{\rm W}}=\frac{C}{T-\lambda C}\,.\label{eq:Weiss susc}
\end{equation}
We see that for large temperatures this susceptibility follows the
Curie inverse law, but as the temperature approaches a critical temperature
$T_{C}=\lambda C$, $\chi_{{\rm W}}$ diverges indicating a phase
transition to the magnetically ordered phase.

\begin{figure}
\begin{centering}
\includegraphics[width=0.8\textwidth]{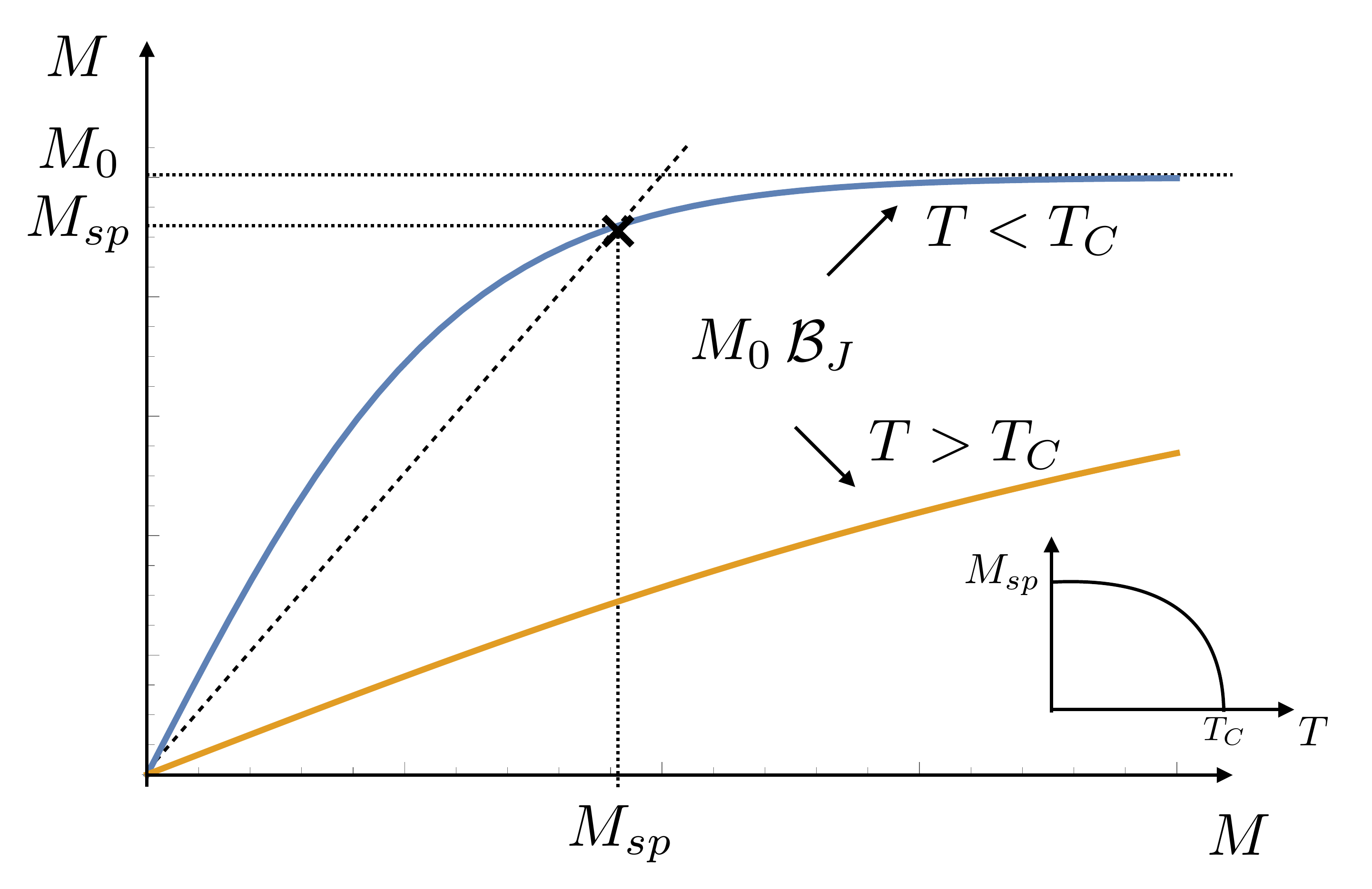} 
\par\end{centering}
\caption{Possibility of spontaneous magnetic order according to the condition
Eq.~\ref{eq:slope condition}. For temperatures lower than the critical
temperature $T_{C}$, there is a non-trivial solution of Eq. \ref{eq:impl Eq M},
indicating magnetic order with a spontaneous magnetization $M_{sp}$.
The inset depicts schematically the behavior of the spontaneous magnetization
$M_{sp}$ with temperature.}

\label{Fig:M order} 
\end{figure}

A magnetically ordered phase implies $M\neq0$ for an external field
$B_{ext}=0$. From Eq.~\ref{eq:mag B(x)} we see that, if we do not
consider the molecular fields postulated by Weiss, $M(0)=0$. Let's
now consider $B_{ext}=0$ but the existence of a molecular field $B_{W}=\mu_{0}\lambda M$.
Inserting this field in Eq.~\ref{eq:mag B(x)} we obtain an implicit
equation for $M$: 
\begin{equation}
M=M_{0}\mathcal{B}_{J}(\frac{\mu_{{\rm B}}\mu_{0}gJ\lambda M}{k_{{\rm B}}T})\,,\label{eq:impl Eq M}
\end{equation}
where we have defined the saturation magnetization 
\begin{equation}
M_{0}=\frac{N}{V}g\mu_{{\rm B}}J\label{eq:sat mag}
\end{equation}
since $\mathcal{B}_{J}(x\rightarrow\infty)=1$. Eq.~\ref{eq:impl Eq M}
has still a solution $M=0$. However, if 
\begin{equation}
\left[\frac{{\rm d}(M_{0}\mathcal{B}_{J})}{{\rm d}M}\right]_{M=0}\ge1\label{eq:slope condition}
\end{equation}
we see that a second solution to the transcendental equation is possible.
From 
\[
\mathcal{B}_{J}(x\rightarrow0)\approx\frac{J+1}{3J}x
\]
we obtain 
\[
M_{sp}=\lambda C\frac{M_{sp}}{T_{C}}
\]
and therefore $T_{C}=\lambda C$, in agreement with Eq.~\ref{eq:Weiss susc}.
The subscript $sp$ indicates this is a \emph{spontaneous magnetization},
not induced by an external magnetic field. For $T>T_{C}$, Eq.~\ref{eq:slope condition}
is not fulfilled and $M=0$ is the only solution. This can be seen
graphically in Fig.~\ref{Fig:M order}. 

At the time, the origin of these postulated molecular fields was not
known. Naively, one could expect the dipole-dipole interaction between
the magnetic moments to be the origin of the magnetic ordering. This
energy scale is however too small to explain magnetism at room temperature.
The potential energy of one magnetic dipole $\mathbf{m}_{2}$ in the
magnetic field created by another dipole $\mathbf{m}_{1}$, $V_{pot}=-\mathbf{m}_{2}\cdot\mathbf{B}_{1}(\mathbf{r})$
is 
\begin{equation}
V_{pot}=-\mathbf{m}_{2}\cdot\frac{\mu_{0}}{4\pi}\frac{3(\mathbf{m}_{1}\cdot\mathbf{r})\mathbf{r}-r^{2}\mathbf{m}_{1}}{r^{5}}\,.\label{eq:dip pot}
\end{equation}
A quick estimate corresponds to taking the distance between the dipoles
as the interatomic distance, and the dipolar moments simply as Bohr
magnetons. Equating this energy to $k_{{\rm B}}T_{{\rm C}}$ results
in a critical temperature for magnetic ordering of $T_{{\rm C}}\approx1K$.
Therefore dipolar interactions could be responsible magnetic ordering
only below this temperature, which is very low. We know however that
magnetic ordering at room temperature is possible. Instead, a rough
estimate of the repulsive Coulomb energy between two electrons gives
\[
\frac{U_{c}}{k_{{\rm B}}}=\frac{e^{2}}{k_{{\rm B}}4\pi\varepsilon_{0}a}\approx10^{5}K
\]
which is very large! This could provide us with the necessary energy
scale. In the following section we will see that magnetic ordering
is due to a combination of the electrostatic energy and a very quantum
effect: the Pauli principle.
\begin{enumerate}
\item \textbf{\emph{Exercise: Prove Eq.~\ref{eq:mag B(x)}.}} 
\item \textbf{\emph{Exercise: estimate $T_{{\rm C}}$ from the dipole-dipole
interaction.}} 
\end{enumerate}

\subsection*{Check points}
\begin{itemize}
\item What is the Curie law and when is it valid? 
\item What is the Curie-Weiss law? 
\item Why the dipolar-dipolar interaction cannot in general explain magnetic
ordering? 
\end{itemize}

\section{\label{sec:Exchange-interaction}Exchange interaction}

Let's consider first the case of two electrons subject to a Hamiltonian
\begin{equation}
H=\frac{p_{1}^{2}}{2m_{e}}+\frac{p_{2}^{2}}{2m_{e}}+V(\mathbf{r}_{1},\mathbf{r}_{2})\,.\label{eq:two e Hamiltonian}
\end{equation}
This Hamiltonian is independent of spin, however the Pauli exclusion
principle imposes a spatial symmetry on the wavefunction $\psi(\mathbf{r}_{1},\mathbf{r}_{2})$
solution of $H\psi(\mathbf{r}_{1},\mathbf{r}_{2})=E\psi(\mathbf{r}_{1},\mathbf{r}_{2})$.
Since two electrons with the same quantum numbers are not allowed
at the same place, the total wavefunction $\psi(\mathbf{r}_{1},\mathbf{r}_{2};s_{1},s_{2})$
must be antisymmetric with respect to exchange of both spin and space:
\[
\psi(\mathbf{r}_{1},\mathbf{r}_{2};s_{1},s_{2})=-\psi(\mathbf{r}_{2},\mathbf{r}_{1};s_{2},s_{1})\,.
\]
If there is no spin-orbit coupling in our Hamiltonian, we can separate
the wavefunction 
\[
\psi(\mathbf{r}_{1},\mathbf{r}_{2};s_{1},s_{2})=\psi(\mathbf{r}_{1},\mathbf{r}_{2})\chi(s_{1},s_{2})
\]
and therefore we see that the antisymmetry implies a correlation between
the spin and orbital parts of the wavefunction. This is a constraint
of the problem and solutions that do not fulfill this constraint are
infinite in energy. For two electrons, we have the total $S=0,1$
and correspondently $S_{z}=0,-1,0,1$. We can write the corresponding
states in explicit symmetric and antisymmetric linear combinations,
the antisymmetric singlet state 
\begin{equation}
|0;0\rangle\frac{1}{\sqrt{2}}\left(|\uparrow\downarrow\rangle-|\downarrow\uparrow\rangle\right)\label{eq:singlet}
\end{equation}
and the symmetric triplet state with $S=1$, $S_{z}=-1,0,1$ :

\begin{align}
|1;1\rangle & =|\uparrow\uparrow\rangle\label{eq:triplet}\\
|1;0\rangle & =\frac{1}{\sqrt{2}}\left(|\uparrow\downarrow\rangle+|\downarrow\uparrow\rangle\right)\nonumber \\
|1;-1\rangle & =|\downarrow\downarrow\rangle\,\nonumber 
\end{align}
where the states are labeled as $|S;S_{z}\rangle$. Since the Hamiltonian
does not depend on spin, we can work simply with the spatial component
of the wavefunction, but imposing the right symmetry. The spatial
part of the wavefunction corresponding to the singlet configuration,
$\psi_{singlet}(\mathbf{r}_{1},\mathbf{r}_{2})$ will therefore be
symmetric in space coordinates, whereas $\psi_{triplet}(\mathbf{r}_{1},\mathbf{r}_{2})$
has to be antisymmetric 
\begin{align}
\psi_{singlet}(\mathbf{r}_{1},\mathbf{r}_{2}) & =\psi_{singlet}(\mathbf{r}_{2},\mathbf{r}_{1})\nonumber \\
\psi_{triplet}(\mathbf{r}_{2},\mathbf{r}_{1}) & =-\psi_{triplet}(\mathbf{r}_{2},\mathbf{r}_{1})\,,\label{eq:s t WF}
\end{align}
and we can write the full state as

\begin{align}
|\Psi_{singlet}\rangle & =|\psi_{singlet}\rangle|0;0\rangle\nonumber \\
|\Psi_{triplet}(S_{z})\rangle & =|\psi_{triplet}\rangle|1;S_{z}\rangle\label{eq:triplet siglet full WV}
\end{align}
The Schrödinger equation therefore reads 
\begin{align*}
H|\psi_{singlet}\rangle & =E_{s}|\psi_{singlet}\rangle\\
H|\psi_{triplet}(S_{z})\rangle & =E_{t}|\psi_{triplet}\rangle
\end{align*}
and if $E_{s}\neq E_{t}$, the ground state is spin-dependent even
though H is not. If that is the case, since spatial and spin sectors
are correlated, we search for a Hamiltonian operating in spin space
$\tilde{H}$, coupling $\hat{\mathbf{s}}_{1}$ and $\hat{\mathbf{s}}_{2}$
and such that it is equivalent to $H$:

\begin{align}
\tilde{H}|0;0\rangle & =E_{s}|0;0\rangle\label{eq:equiv ham}\\
\tilde{H}|1;S_{z}\rangle & =E_{t}|1;S_{z}\rangle\,.\nonumber 
\end{align}
The Hamiltonian that does the trick is \cite{nolting_quantum_2009}
\begin{equation}
\tilde{H}=\frac{1}{4}\left(E_{s}+3E_{t}\right)-\frac{1}{\hbar^{2}}\left(E_{s}-E_{t}\right)(\hat{\mathbf{s}}_{1}\cdot\hat{\mathbf{s}}_{2})\label{eq:H tilde}
\end{equation}
since 
\begin{equation}
\frac{\hat{\mathbf{s}}_{1}\cdot\hat{\mathbf{s}}_{2}}{\hbar^{2}}=\frac{1}{2}S(S+1)-\frac{3}{4}=\begin{cases}
-\frac{3}{4} & S=0\\
\frac{1}{4} & S=1
\end{cases}\,.\label{eq:S1.S2}
\end{equation}
We have therefore constructed a Hamiltonian acting on spin space which
is in principle equivalent to the interacting Hamiltonian of Eq.~\ref{eq:two e Hamiltonian},
which gives an effective interaction between the spins. This is called
the \emph{molecular Heisenberg model} and can be written as\label{H tilde J12}
\[
\tilde{H}=J_{0}-J_{12}\hat{\mathbf{s}}_{1}\cdot\hat{\mathbf{s}}_{2}
\]
with 
\begin{equation}
J_{12}=\frac{1}{\hbar^{2}}\left(E_{s}-E_{t}\right)\,.\label{eq:J12}
\end{equation}
If $J_{12}>0$, this interaction favors a ferromagnetic alignment
of the spins, consistent with the fact that the singlet energy is
higher than the triplet one. We now show an example in which $E_{s}\neq E_{t}$
and calculate explicitly $J_{12}$. We will see that part of $J_{12}$
has no classical analog and comes from interchanging particles 1 and
2, since in quantum mechanics they are indistinguishable. 
\begin{enumerate}
\item \textbf{\emph{Exercise: Prove Eq.~\ref{eq:S1.S2} and show that $\tilde{H}$
given in Eq.~\ref{eq:H tilde} satisfies \ref{eq:equiv ham}.}} 
\end{enumerate}

\subsection*{Check points}
\begin{itemize}
\item Why can you write a Hamiltonian in spin space which is equivalent
to the Hamiltonian defined in position space? 
\item How do you impose the equivalence for the two-electrons system? 
\end{itemize}

\section{\label{sec:Hydrogen-molecule}Hydrogen molecule}

We consider now two hydrogen atoms that are brought close together
to form a hydrogen molecule. We consider the nuclei, $a$ and $b$,
as fixed at positions $\mathbf{R}_{a}$ and $\mathbf{R}_{b}$, whereas
the electrons, at positions $\mathbf{r}_{1}$ and $\mathbf{r}_{2}$
are subject to the nuclei Coulomb potential plus the repulsive Coulomb
interaction between them . This corresponds to taking the nuclei mass
$m_{a},m_{b}\rightarrow\infty$ and it is an example of the denominated
\emph{Born Oppenheimer approximation}. The Hamiltonian for each, separate
hydrogen atom $a$ and $b$ are given by 
\begin{align*}
H_{a} & =-\frac{\hbar^2\nabla_{1}^{2}}{2m_{e}}-\frac{e^{2}}{4\pi\varepsilon_{0}}\frac{1}{|\mathbf{R}_{a}-\mathbf{r}_{1}|}\\
H_{b} & =-\frac{\hbar^2\nabla_{2}^{2}}{2m_{e}}-\frac{e^{2}}{4\pi\varepsilon_{0}}\frac{1}{|\mathbf{R}_{b}-\mathbf{r}_{2}|}
\end{align*}
and we know the respective eigenfunctions, $\phi_{a,b}$ with eigenenergies
$E_{a,b}$. If the distance between the two atoms $|\mathbf{R}_{a}-\mathbf{R}_{b}|\rightarrow\infty$,
these are the exact solutions. If however the atoms are brought close
together to form a molecule, there will be an interaction term 
\[
H_{I}=\frac{e^{2}}{4\pi\varepsilon_{0}}\frac{1}{|\mathbf{R}_{a}-\mathbf{R}_{b}|}-\frac{e^{2}}{4\pi\varepsilon_{0}}\frac{1}{|\mathbf{R}_{b}-\mathbf{r}_{1}|}-\frac{e^{2}}{4\pi\varepsilon_{0}}\frac{1}{|\mathbf{R}_{a}-\mathbf{r}_{2}|}+\frac{e^{2}}{4\pi\varepsilon_{0}}\frac{1}{|\mathbf{r}_{2}-\mathbf{r}_{1}|}
\]
and the total Hamiltonian is given by 
\[
H_{tot}=H_{a}+H_{b}+H_{I}\,.
\]
Even though this is a quite simple system, this problem cannot be
solved exactly and we have to resort to approximations. We treat $H_{I}$
as a perturbation and use the atomic wavefunctions $\phi_{a,b}$ as
a basis for a variational solution of the full wavefunction. This
implies we are assuming the electrons are quite localized at their
respective atom. This approach is denominated the \emph{Heitler-London
}method.

If we consider the unperturbed Hamiltonian 
\[
H_{0}=H_{a}+H_{b}
\]
the eigenfunctions will be simply linear combinations of the product
of the original orbitals, $\phi_{a}\phi_{b}$, with eigenenergy $E_{a}+E_{b}$
since the two systems do not interact with each other. To preserve
the indistinguishability of the particles, we cannot simply write
a solution as $\phi_{a}(\mathbf{r}_{1})\phi_{b}(\mathbf{r}_{2})$,
since the probability density $|\phi_{a}(\mathbf{r}_{1})|^{2}|\phi_{b}(\mathbf{r}_{2})|^{2}$
is not invariant under exchanging $\mathbf{r}_{1}\longleftrightarrow\mathbf{r}_{2}$.
The symmetric and antisymmetric combinations fulfill however the indistinguishability
condition 
\begin{align}
\psi_{s} & (\mathbf{r}_{1},\mathbf{r}_{2})=\frac{1}{\sqrt{2}}\left(\phi_{a}(\mathbf{r}_{1})\phi_{b}(\mathbf{r}_{2})+\phi_{b}(\mathbf{r}_{1})\phi_{a}(\mathbf{r}_{2})\right)\nonumber \\
\psi_{t} & (\mathbf{r}_{1},\mathbf{r}_{2})=\frac{1}{\sqrt{2}}\left(\phi_{a}(\mathbf{r}_{1})\phi_{b}(\mathbf{r}_{2})-\phi_{b}(\mathbf{r}_{1})\phi_{a}(\mathbf{r}_{2})\right)\label{eq:variational WF}
\end{align}
where with the notation $\psi_{s}$, $\psi_{t}$ we have anticipated
that the symmetric (antisymmetric) solution in space corresponds to
the singlet (triplet) solution in spin space, with a full wavefunction
as given in Eq.~\ref{eq:triplet siglet full WV}. Note that both
$\psi_{s}$, $\psi_{t}$ are eigenfunctions of $H_{0}$ with eigenvalue
$E_{a}+E_{b}$, and therefore without the interaction $H_{I}$, $E_{t}=E_{s}=E_{a}+E_{b}$
and the soulutions are 4-fold degenerate.

We now calculate $E_{t}$ and $E_{s}$ perturbatively in the presence
of the interaction $H_{I}$ and using $\psi_{s}$, $\psi_{t}$ as
variational wavefunctions 
\begin{equation}
E_{s/t}=\frac{\langle\psi_{s/t}|H_{tot}|\psi_{s/t}\rangle}{\langle\psi_{s/t}|\psi_{s/t}\rangle}\,.\label{eq:variational E s t}
\end{equation}
We are interested in the ground state solutions, so that $\phi_{a,b}$
are solutions of each hydrogen atom with $E_{a,b}=E_{0}$. The variational
principle tells us that the energies calculated by Eq.~\ref{eq:variational E s t}
are always greater or equal that the true groundstate.\footnote{Note that in general the variational procedure would be to write $\psi_{s/t}(\mathbf{r}_{1},\mathbf{r}_{2})=c_{1}\phi_{a}(\mathbf{r}_{1})\phi_{b}(\mathbf{r}_{2})\pm c_{2}\phi_{b}(\mathbf{r}_{1})\phi_{a}(\mathbf{r}_{2})$
and find the coefficients $c_{1,2}$ by minimizing Eq.~\ref{eq:variational E s t}.
One then finds $c_{1}=c_{2}=1/\sqrt{2}$. In Eq.~\ref{eq:variational WF}
we used our knowledge of the symmetry of the problem plus the normalization
of the wavefunctions $\phi_{a,b}$ to write the result immediately.}

Let's first analyze the simple overlap 
\begin{equation}
\langle\psi_{s/t}|\psi_{s/t}\rangle=\int{\rm d}^{3}r_{1}{\rm d}^{3}r_{2}\psi_{s/t}^{*}(\mathbf{r}_{1},\mathbf{r}_{2})\psi_{s/t}(\mathbf{r}_{1},\mathbf{r}_{2})\,.\label{eq:overlap s t}
\end{equation}
If the two atoms are infinitely apart $|\mathbf{R}_{a}-\mathbf{R}_{b}|\rightarrow\infty$,
then the orbitals corresponding to different atoms have zero overlap,
$\langle\phi_{a}|\phi_{b}\rangle=0$, and 
\begin{equation}
\langle\psi_{s/t}|\psi_{s/t}\rangle_{0}=\int{\rm d}^{3}r_{1}{\rm d}^{3}r_{2}|\phi_{a}(\mathbf{r}_{1})|^{2}|\phi_{b}(\mathbf{r}_{2})|^{2}=1\,.\label{eq:overlap a b}
\end{equation}
When the atoms are brought close together, their orbitals will overlap:
$\langle\phi_{a}|\phi_{b}\rangle\neq0$. We do not calculate this
overlap explicitly, but simply note that it will be finite and denote
it by $O$ 
\begin{equation}
O^{2}=\int{\rm d}^{3}r_{1}{\rm d}^{3}r_{2}\phi_{a}^{*}(\mathbf{r}_{1})\phi_{b}(\mathbf{r}_{1})\phi_{a}(\mathbf{r}_{2})\phi_{b}^{*}(\mathbf{r}_{2})\label{eq:O 2}
\end{equation}
and hence 
\begin{equation}
\langle\psi_{s/t}|\psi_{s/t}\rangle=1\pm O^{2}\,.\label{eq:ove s t O}
\end{equation}
We now turn to the numerator in Eq.~\ref{eq:variational E s t}.
We already know that the non-interacting contribution is 
\begin{equation}
\frac{\langle\psi_{s/t}|H_{0}|\psi_{s/t}\rangle}{\langle\psi_{s/t}|\psi_{s/t}\rangle}=2E_{0}\,.\label{eq:H0 exp}
\end{equation}
The correction to this non-interacting energy is given by the term
containing $\langle\psi_{s/t}|H_{I}|\psi_{s/t}\rangle$, which we
see contains two kind of terms. One of them is simply the Coulomb
electrostatic interaction between the two atoms, assuming that they
are close enough to interact but electron 1(2) still "
belongs" to atom $a$($b$) 
\begin{equation}
K=\int{\rm d}^{3}r_{1}{\rm d}^{3}r_{2}\phi_{a}^{*}(\mathbf{r}_{1})\phi_{b}(\mathbf{r}_{2})H_{I}\phi_{a}(\mathbf{r}_{1})\phi_{b}^{*}(\mathbf{r}_{2})\,.\label{eq:K coulomb}
\end{equation}
The remaining term has no classical analog, and it measures the Coulomb
energy cost upon exchanging the two electrons 
\begin{equation}
X=\int{\rm d}^{3}r_{1}{\rm d}^{3}r_{2}\phi_{a}^{*}(\mathbf{r}_{1})\phi_{b}(\mathbf{r}_{1})H_{I}\phi_{a}(\mathbf{r}_{2})\phi_{b}^{*}(\mathbf{r}_{2})\,.\label{eq:X Coulomb}
\end{equation}
and it is called the \emph{exchange integral}, or \emph{exchange interaction}.
Putting all together we obtain 
\begin{equation}
E_{s/t}=2E_{0}+\frac{K\pm X}{1\pm O^{2}}\label{eq:E s t}
\end{equation}
where the $+$ ($-$) corresponds to the singlet (triplet) solution
$\psi_{s}$ ($\psi_{t}$ ). In general $O\ll1$ and we can replace
the denominator by 1. We have therefore shown that $E_{s}-E_{t}\neq0$
and hence for this problem 
\begin{equation}
J_{12}\approx\frac{2}{\hbar^{2}}X\,,\label{eq:J 12}
\end{equation}
which justifies the name \emph{exchange parameter }for $J_{12}$.

\subsection*{Check points}
\begin{itemize}
\item What is the Heitler-London model? 
\item What is the exchange interaction and why does it not have a classical
analog? 
\end{itemize}

\section{\label{sec:Heisenberg,-Ising,-and}Heisenberg, Ising, and XY models}

In our variational solution for the Hydrogen molecule from the last
section, double occupation is forbidden, so two electrons cannot be
in the same atom at the same time. This indicates our treatment is
valid for insulators, where electrons are quite localized, but in
turn leads necessarily to small values of the exchange parameter,
since $J_{12}$ relies on the overlap of the single-atom orbitals.
The above example therefore must be taken as a toy model which reveals
the character of the ferromagnetic interaction. In general, the exchange
constant is generated by more complex interactions, \emph{e.g. superexchange
}where the ferromagnetic exchange interaction between two spins is
mediated by an exchange interaction with an atom in between those
with a net angular momentum.

We further postulate that our model can be generalized to $N$ multielectron
atoms 
\begin{equation}
H=-\frac{1}{2}\sum_{ij}J_{ij}\hat{\mathbf{S}}_{i}\cdot\hat{\mathbf{S}}_{j}\label{eq:Heisenberg Ham}
\end{equation}
where the exchange coefficient $J_{ij}$ is taken as a parameter of
the model, that has to be calculated for each particular material.
The Hamiltonian in Eq.~\ref{eq:Heisenberg Ham} is the \emph{Heisenberg
Hamiltonian}. The factor of $1/2$ accounts for the double-counting
in the sum. We write $\hat{\mathbf{S}}$ by convention, and we refer
to the magnetic moment as " spins" in
an abuse of language: in reality, unless the orbital angular momentum
is quenched, the total angular momentum of the ions is meant. The
spin operators follow the angular momentum algebra when located at
the same site, and commute with operators at different sites: 
\begin{equation}
[\hat{S}_{i}^{\alpha},\hat{S_{j}}^{\beta}]=i\hbar\delta_{ij}\epsilon_{\alpha\beta\gamma}\hat{S}_{i}^{\gamma}\label{eq:comm diff sites}
\end{equation}
where $\alpha$, $\beta$, $\gamma$ indicate the spatial components
of the angular momentum $x$, $y$, $z$. These commutation relations
make the quantum Heisenberg model, despite its simple appearance,
quite a rich model and exactly solvable only for a few simple cases.
The input of the model is the lattice connectivity and dimensionality,
and the exchange parameter $J_{ij}$ .

Besides insulators, the Heisenberg Hamiltonian is a valid model for
localized magnetic moments embedded in a metal. In that case, the
exchange interaction is mediated by the conduction electrons, which
gives rise to the \emph{RKKY interaction} (Ruderman--Kittel--Kasuya--Yosida).
The calculated exchange function $J_{ij}$ is an oscillating function
of position, alternating between positive and negative values, and
is longer ranged than in the insulating case.

If there are no local magnetic moments but the system still presents
magnetic order, the conduction electrons are also responsible for
the magnetic order. In this case the magnetism is denominated \emph{itinerant
}and it is described by a different Hamiltonian: the \emph{Hubbard
Hamiltonian}. This model takes into account the kinetic energy of
the electrons, who can " jump" from lattice
site to lattice site, and penalizes double occupation with a local
Coulomb repulsion term. The Hubbard model takes an effective Heisenberg
form in the particular case of a half-filled band and strong Coulomb
interaction.

The Heisenberg Hamiltonian is the " father"
Hamiltonian of other well known models in magnetism. In a crystal,
crystal fields can give rise to anisotropies in the exchange parameter
$J_{ij}$. If the anisotropy is along only one direction one can write
\begin{equation}
H=-\sum_{ij}\tilde{J}_{ij}\left(\hat{S}_{i}^{x}\hat{S}_{j}^{x}+\hat{S}_{i}^{y}\hat{S}_{j}^{y}+\Delta\,\hat{S}_{i}^{z}\hat{S}_{j}^{z}\right)\,.\label{eq:Anis z ham}
\end{equation}
If $\Delta>1$, magnetic ordering occurs along the $z$ axis, which
is denominated the \emph{easy axis}. For $\Delta\gg1$ the Hamiltonian
turns effectively into the \emph{Ising Model 
\begin{equation}
H_{{\rm Ising}}=-\sum_{ij}J_{ij}\,S_{i}^{z}S_{j}^{z}\,.\label{eq: Ising ham}
\end{equation}
} Note that in this particular case, all operators in the Hamiltonian
commute, and therefore the model is in this sense classical. If $\Delta<1$
then we have an \emph{easy plane} ordering. For $\Delta\ll1$ the
system is effectively two-dimensional and isotropic, which is termed
the \emph{XY }model

\begin{equation}
H_{{\rm XY}}=-\sum_{ij}J_{ij}\,\left(\hat{S}_{i}^{x}\hat{S}_{j}^{x}+\hat{S}_{i}^{y}\hat{S}_{j}^{y}\right)\,.\label{eq:XY ham}
\end{equation}

If the system is placed in an external magnetic field, a Zeeman term
is added to the Hamiltonian 
\begin{align}
H & =-\frac{1}{2}\sum_{ij}J_{ij}\hat{\mathbf{S}}_{i}\cdot\hat{\mathbf{S}}_{j}-g\mu_{{\rm B}}\mathbf{B}\cdot\sum_{i}\hat{\mathbf{S}}_{i}\nonumber \\
 & =-\frac{1}{2}\sum_{ij}J_{ij}\hat{\mathbf{S}}_{i}\cdot\hat{\mathbf{S}}_{j}-g\mu_{{\rm B}}B\sum_{i}\hat{S}_{i}^{z}\,.\label{eq:Heisenberg in B field}
\end{align}
In Eq. (\ref{eq:Heisenberg in B field}) the spin operators are dimensionless
and $\hbar$ has been absorbed in $\mu_{{\rm B}}$ (correspondingly,
in the commutation relation Eq. (\ref{eq:comm diff sites}) $\hbar$
should be set to 1 if this convention is used). Note that the Zeeman
energy is $-\mathbf{m}\cdot\mathbf{B}$ (see Eq. (\ref{eq:Zeeman energy}))
and tends to align the magnetic moment with the magnetic field, and
anti-align the angular momentum. Sometimes by convention the extra
minus sign is not used, since from now onwards one always works with
the angular momentum operators, and one takes $g\mu_{{\rm B}}>0$.
This corresponds simply to transform $\mathbf{B}\rightarrow-\mathbf{B}$
if we want to translate into the magnetization or magnetic moments,
and does not affect the results.

\subsection*{Check points}
\begin{itemize}
\item Write the Heisenberg Hamiltonian in a magnetic field 
\end{itemize}

\section{\label{sec:Mean-field-theory}Mean field theory }

Once we have a Hamiltonian that models our system, we want in principle
to (i) find the ground state, that is, the lowest energy eigenstate
of the system, which is the only state populated at zero temperature,
(ii) find the excitations on top of this ground state, which will
determine the behavior of the system at $T\neq0$, and (iii) study
phase transitions, either at $T=0$ (denominated a \emph{quantum phase
transition}) where by changing some other external parameter like
the magnetic field, the ground state of the system changes abruptly,
or at finite temperature, where an order parameter of the system (given
by a quantum-statistical average of some relevant quantity to describe
the system), goes to zero as a function of temperature or other external
parameters. An example of the latter is the magnetization $M(B,T)$.
We go back now to the issue of magnetic ordering armed with the Heisenberg
Hamiltonian. As we pointed out above, this is a very rich model, and
there are very few general statements that can be made about the three
points mentioned above. We turn therefore first to a well know approximation
denominated \emph{mean field theory}. This approximation is in general
good only for long-range interactions and high dimensions, it is however
widely used to get a first idea of, for example, what kind of phases
and phase transitions our model can present. For the Heisenberg model,
we will see that mean field theory will give us a microscopic justification
of the Weiss molecular fields we introduced at the beginning of this
chapter.

We start by assuming that $\langle\hat{\mathbf{S}}_{i}\rangle$ is
finite. For example, for a ferromagnetic ground state, $\langle\hat{\mathbf{S}}_{i}\rangle$
is uniform and such that 
\begin{equation}
\mathbf{M}=\frac{N}{V}g\mu_{{\rm B}}\langle\hat{\mathbf{S}}_{i}\rangle\,,\label{eq:mag Si}
\end{equation}
where $N$ is the number of lattice sites and $V$ the total volume.
We write now $\hat{\mathbf{S}}_{i}$ in the suggestive form 
\begin{equation}
\hat{\mathbf{S}}_{i}=\langle\hat{\mathbf{S}}_{i}\rangle+\left(\hat{\mathbf{S}}_{i}-\langle\hat{\mathbf{S}}_{i}\rangle\right)\,,\label{eq:Si MF}
\end{equation}
which corresponds to splitting the operator into its quantum statistical
average value and the fluctuations with respect to this average. In
mean field theory, these fluctuations are assumed to be small. The
\emph{mean field Hamiltonian} is obtained from the Heisenberg Hamiltonian
by keeping only terms up to first order in the fluctuation, 
\begin{equation}
H_{{\rm MF}}=\frac{1}{2}\sum_{ij}J_{ij}\langle\hat{\mathbf{S}}_{i}\rangle\cdot\langle\hat{\mathbf{S}}_{j}\rangle-\sum_{ij}J_{ij}\langle\hat{\mathbf{S}}_{i}\rangle\cdot\hat{\mathbf{S}}_{j}-g\mu_{{\rm B}}\mathbf{B}\cdot\sum_{i}\hat{\mathbf{S}}_{i}\,,\label{eq:mean field ham}
\end{equation}
where we already included an external magnetic field $\mathbf{B}$.
The first term in Eq. (\ref{eq:mean field ham}) is simply a constant
shift in the energy. The second term corresponds to a spin a site
$j$ in the presence of a magnetic field generated by all other spins.
We can therefore define an effective magnetic field 
\begin{equation}
\mathbf{B}_{{\rm eff}}=\mathbf{B}+\frac{1}{g\mu_{{\rm B}}}\sum_{i}J_{ij}\langle\hat{\mathbf{S}}_{i}\rangle
\end{equation}
and our problem is reduced from an interacting problem (where spins
interact with each other via the exchange interaction), to that of
non-interacting spins in the presence of a magnetic field. If we go
back to Eq. (\ref{eq:Curie susc}), we see that now we can give a
microscopic explanation to the Weiss molecular fields, which were
assumed to be proportional to to the magnetization. In particular
we find that we can write 
\begin{equation}
\mathbf{B}_{{\rm eff}}=\mathbf{B}+\mu_{0}\lambda\text{\textbf{M}}
\end{equation}
with 
\begin{equation}
\lambda=\frac{1}{\mu_{0}}\frac{V}{N}\frac{1}{\left(g\mu_{{\rm B}}\right)^{2}}J_{0}
\end{equation}
where we have used translational symmetry and defined $J_{0}=\sum_{i}J_{ij}$
independent of $j$. This system therefore presents a transition to
an ordered state as a function of temperature and magnetic field as
discussed for the Curie-Weiss law, but we now have a microscopic explanation
for the phenomenological model.

\subsection*{Check points}
\begin{itemize}
\item What is the meaning of the mean field theory? 
\item How is it related to the Curie-Weiss law? 
\end{itemize}

\section{\label{sec:Ground-state-of}Ground state of the ferromagnetic Heisenberg
Hamiltonian}

For the particular case in which $J_{ij}\ge0\,\forall\,i,\,j$ it
is possible to find the ground state of the Heisenberg Hamiltonian
without further specifications. This is called the \emph{ferromagnetic
Heisenberg model} since the ground state is ferromagnetically ordered,
as we show in the following.

We consider hence the Hamiltonian 
\begin{equation}
\hat{H}=-\frac{1}{2}\sum_{ij}J_{ij}\hat{\mathbf{S}}_{i}\cdot\hat{\mathbf{S}}_{j}\,\,,\,{\rm with}\,\,J_{ij}=J_{ji}\ge0\,.\label{eq:Heis FM}
\end{equation}
f the spins were classical vectors, the state of lowest energy would
be that one with all $N$ spins are aligned. Hence a natural candidate
for the ground state of $\hat{H}$ is 
\begin{equation}
|0\rangle=|S,S\rangle_{1}|S,S\rangle_{2}...|S,S\rangle_{N}\label{eq:ground state}
\end{equation}
where $\hat{\mathbf{S}}_{i}^{2}|S,S\rangle_{i}=S(S+1)|S,S\rangle_{i}$
and $\hat{S}_{i}^{z}|S,S\rangle_{i}=S|S,S\rangle_{i}$, that is all
spins take their maximum projection of $\hat{S}^{z}$ (we consider
all spins identical). The individual spin operators however do not
commute with the Hamiltonian \ref{eq:Heis FM}, 
\[
\left[\hat{\mathbf{S}}_{i},\hat{H}\right]\neq0
\]
and therefore a product states of the form 
\begin{equation}
|\psi\rangle=|S,m_{1}\rangle|S,m_{2}\rangle...|S,m_{N}\rangle\label{eq:product state}
\end{equation}
(with $\hat{S}_{i}^{z}|S,m_{i}\rangle=m_{i}|S,m_{i}\rangle$) span
a basis of the Hilbert space, but are not necessarily eigenstates
of $\hat{H}$. The total spin operator $\hat{\mathbf{S}}_{{\rm TOT}}$
however does commute with $\hat{H}$ 
\[
\left[\hat{\mathbf{S}}_{{\rm TOT}},\hat{H}\right]=0
\]
and we can construct an eigenbasis for $\hat{H}$, $\hat{\mathbf{S}}_{{\rm TOT}}^{2}$,
and $\hat{\mathbf{S}}_{{\rm TOT}}^{z}$. The state \ref{eq:ground state}
is a state with maximum $\hat{\mathbf{S}}_{{\rm TOT}}$. We will now
prove that (i) \ref{eq:ground state} is an eigenstate of \ref{eq:Heis FM},
and (ii) that there is no eigenstate with higher energy \cite{ashcroft_solid_1976}.

To follow with the proof it is convenient to recast $\hat{H}$ in
term of \emph{ladder operators} 
\begin{align}
\hat{S}_{i}^{\pm} & =\hat{S}_{i}^{x}\pm i\hat{S}_{i}^{y}\label{eq:ladder op}\\
\hat{S}_{i}^{\pm}|S_{i},m_{i}\rangle & =\sqrt{\left(S_{i}\mp m_{i}\right)\left(S_{i}+1\pm m_{i}\right)}|S_{i},m_{i}\pm1\rangle\,,\nonumber 
\end{align}
from which it is clear why they are call ladder operators: $\hat{S}^{+}$
($\hat{S}^{-}$) increases (decreases) the projection of $\hat{S}^{z}$
by one unit. In terms of \ref{eq:ladder op}, the Hamiltonian \ref{eq:Heis FM}
reads 
\begin{equation}
\hat{H}=-\frac{1}{2}\sum_{ij}J_{ij}\left[\frac{1}{2}\left(\hat{S}_{i}^{+}\hat{S}_{j}^{-}+\hat{S}_{i}^{-}\hat{S}_{j}^{+}\right)+\hat{S}_{i}^{z}\hat{S}_{j}^{z}\right]\,.\label{eq:Heis ladder}
\end{equation}
With this expression it is now straightforward to show that $|0\rangle$
is an eigenstate of $\hat{H}$ 
\[
\hat{H}|0\rangle=-\frac{1}{2}\sum_{ij}J_{ij}\hat{S}_{i}^{z}\hat{S}_{j}^{z}|0\rangle=-\frac{S}{2}^{2}\sum_{ij}J_{ij}|0\rangle
\]
since $\hat{S}_{i}^{+}|0\rangle=0\,\forall\,i$ (remember that spin
operators at different sites commute, and $J_{ii}=0$). Therefore
\[
\hat{H}|0\rangle=E_{0}|0\rangle
\]
with 
\[
E_{0}=-\frac{S}{2}^{2}\sum_{ij}J_{ij}\,.
\]
We now need to prove that $E_{0}$ is the minimum possible energy.
For that we consider the expectation value of $\hat{H}$ with any
other arbitrary eigenstate $|\psi\rangle$
\[
E_{0}'=\langle\psi|\hat{H}|\psi\rangle\ge-\frac{1}{2}\sum_{ij}J_{ij}\max\langle\psi|\hat{\mathbf{S}}_{i}\cdot\hat{\mathbf{S}}_{j}|\psi\rangle\ge-\frac{S^{2}}{2}\sum_{ij}J_{ij}
\]
where we have used that $J_{ij}\ge0$, $J_{ii}=0$, and that $\langle\psi|\hat{\mathbf{S}}_{i}\cdot\hat{\mathbf{S}}_{j}|\psi\rangle\le S^{2}$
for $i\neq j$\cite{ashcroft_solid_1976}. Hence 
\[
E_{0}'\ge E_{0}\,,
\]
which proves that the fully polarized state $|0\rangle$ given in
Eq.~\ref{eq:ground state} is indeed the ground state. This ground
state is however not unique, but it is $\left(2S_{{\rm TOT}}+1\right)$
degenerate in spin space, with $S_{{\rm TOT}}=NS$. This can be easily
visualized in the 2-spins case, where for the triplet state, $S_{{\rm TOT}}=1$
and the state has 3 possible projections of $\hat{S}_{z}$, see Eq.
\ref{eq:triplet}. Note that each of these states is moreover infinitely
degenerate in position space, since we are able to choose the quantization
axis freely. This is an example of what is called \emph{spontaneous
symmetry breaking}, which occurs when the ground state has a lower
symmetry than the Hamiltonian \cite{auerbach_interacting_1994}. The
ground state of $\hat{H}$ is a specific realization of the $\left(2S_{{\rm TOT}}+1\right)$
possible ground states, and therefore has " picked"
a preferred direction in spin space, which is not determined by the
symmetry of $\hat{H}$. We mention in passing that our results are
valid in all its generality strictly for $T=0$. For $T>0$, the \emph{Mermin-Wagner
theorem} tells us that, in 1 and 2 dimensions and for short range
interactions continuous symmetries cannot be spontaneously broken.

If we add a magnetic field to $\hat{H}$, our total Hamiltonian is
the one in Eq.~\ref{eq:Heisenberg in B field}. In this case, it
is favorable for the system also to maximize the projection of $\hat{S}_{z}$,
and in this case $|0\rangle$ given in Eq.~\ref{eq:ground state}
is the \emph{only} ground state, with energy 
\begin{equation}
E_{0}(B)=-\frac{S}{2}^{2}\sum_{ij}J_{ij}\,-g\mu_{{\rm B}}BNS\,.\label{eq:E0 in B}
\end{equation}
In this case the symmetry is not spontaneously broken, since is is
already broken at the Hamiltonian level by the applied magnetic field.

\subsection*{Check points}
\begin{itemize}
\item What is the ground state of the ferromagnetic Heisenberg Hamiltonian? 
\item How do you prove it is the ground state? 
\item What happens in the presence of an external magnetic field to the
degeneracy of the state? 
\end{itemize}

\section{\label{sec:Ground-state-of-1}Ground state of the antiferromagnetic
Heisenberg Hamiltonian}

Aside from the ferromagnetic case treated in the previous section,
finding the ground state of the Heisenberg Hamiltonian is in general
difficult and has to be studied case by case. The ground state will
depend on the nature of the interactions (short or long range, sign,
anisotropy), from the lattice structure, and from the dimensionality
of the system. To illustrate this difficulty, we discuss here briefly
the antiferromagnetic Heisenberg model on a bipartite lattice \cite{auerbach_interacting_1994}.
In this case $J_{ij}\le0\,\forall\,i,\,j$ and the lattice can be
sub-divided in two sublattices $A$ and $B$, such that $J_{ij}$
is finite only when $i$ and $j$ belong to two different sublattices.
The simplest example is that of a square lattice with nearest neighbor
interactions, where all nearest neighbors of $A$ belong to the $B$
sublattice, and vice-versa, see Fig.

\begin{figure}
\centering{}\includegraphics[width=0.8\textwidth]{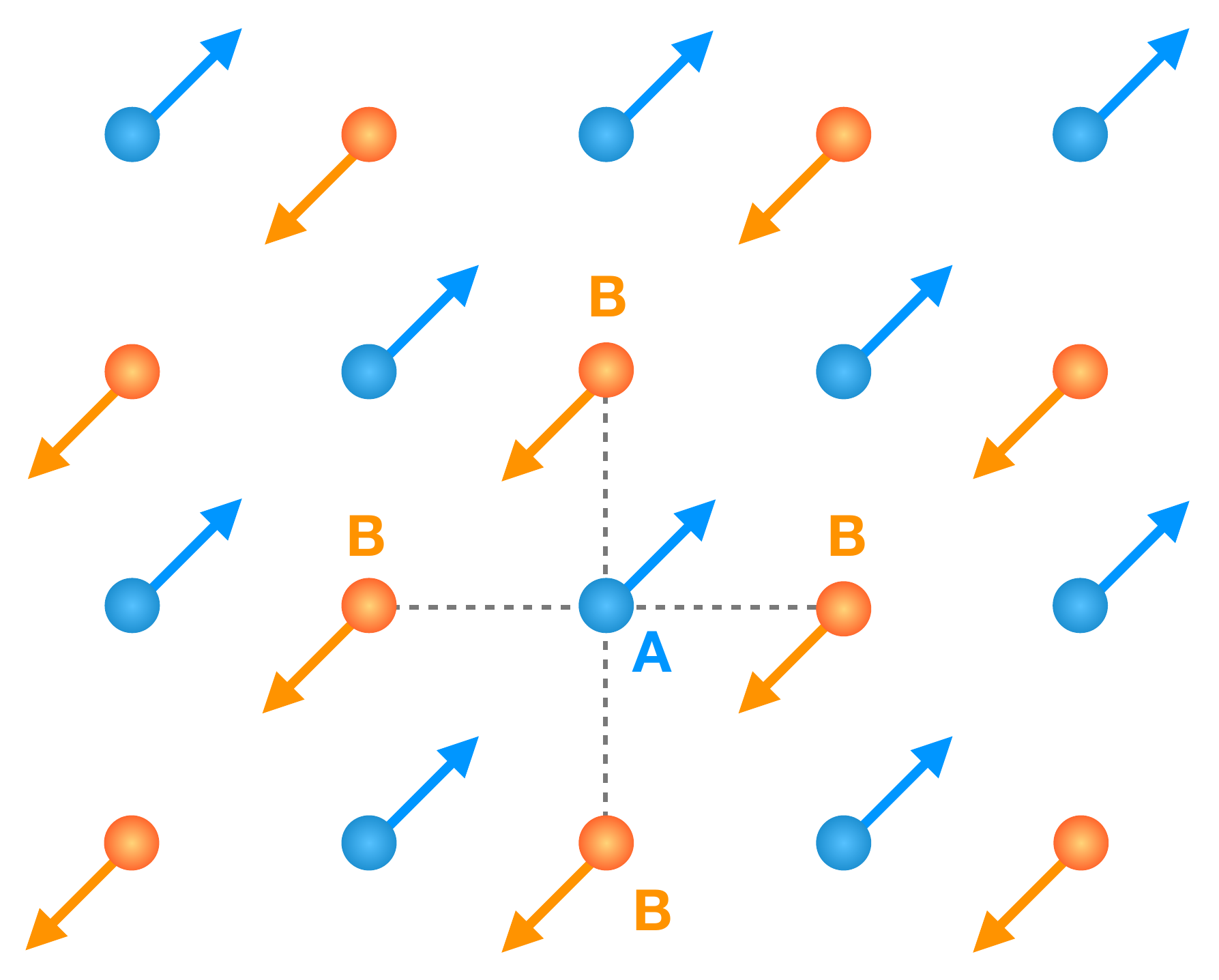}\caption{Antiferromagnetic Heisenberg model on a bipartite square lattice,
the dotted lines indicate nearest-neighbor interactions. Example of
a Néel state.}
\end{figure}

A guess of the ground state based on the classical model is the
so called \emph{Néel state}: the two sublattices are fully polarized,
but in opposite directions 
\[
|0?\rangle_{{\rm AF}}=\prod_{i\in A}|S,S\rangle_{i}\prod_{j\in B}|S,-S\rangle_{j}\,.
\]
It is easy to see however that this state is not an eigenstate of
the Heisenberg Hamiltonian. Using the representation of $\hat{H}$
in terms of ladder operators, see Eq.~\ref{eq:Heis ladder}, we see
that 
\[
\hat{H}|0?\rangle_{{\rm AF}}=-\frac{1}{2}\sum_{i\in A}\sum_{j\in B}J_{ij}\left[\frac{1}{2}\left(\hat{S}_{i}^{+}\hat{S}_{j}^{-}|0?\rangle_{{\rm AF}}+\hat{S}_{i}^{-}\hat{S}_{j}^{+}|0?\rangle_{{\rm AF}}\right)+\hat{S}_{i}^{z}\hat{S}_{j}^{z}|0?\rangle_{{\rm AF}}\right]\,.
\]
The first term is simply zero, since the sublattices $A$ and $B$
are fully polarized in " the right way"
with respect to the operators. The last term is simply proportional
to $|0?\rangle_{{\rm AF}}$. However for the second term

\[
\hat{S}_{i}^{-}\hat{S}_{j}^{+}|0?\rangle_{{\rm AF}}\propto|S,\rangle...|S,-S+1\rangle_{j}...|S,S-1\rangle_{i}...|S,m_{N}\rangle
\]
which shows that $|0?\rangle_{{\rm AF}}$ cannot be an eigenstate.
In general, one can prove that the true ground state $|0\rangle_{{\rm AF}}$
is non-degenerate and a singlet of total spin: $\hat{S}_{{\rm TOT}}|0\rangle_{{\rm AF}}=0$.
This is called the \emph{Marshall's Theorem}. Which kind of singlet
is the actual ground state is however not determined.

\subsection*{Check points}
\begin{itemize}
\item What is a Néel state? 
\item Why it is not the ground state of the antiferromagnetic Heisenberg
Hamiltonian? 
\end{itemize}

\section{\label{sec:Ground-state-of-2}Ground state of the classical Heisenberg
model}

The classical Heisenberg model is obtained by replacing the spin operators
$\hat{\mathbf{S}}_{i}$ by simple vectors $\mathbf{S}_{i}$ with fixed
length $|\mathbf{S}_{i}|=S$. As we saw in previous sections, this
kind of approximation is valid in the limit of large spin $S$. For
the classical model it is straightforward to find the ground state
for translational invariant systems, 
\begin{equation}
J_{ij}=J_{ji}=J(\mathbf{R}_{i}-\mathbf{R}_{j})\label{eq:jij Jji}
\end{equation}
where $\mathbf{R}_{i}$ indicates the points on a Bravais lattice.
In Fourier space 
\begin{align}
\mathbf{S}_{i} & =\frac{1}{\sqrt{N}}\sum_{\mathbf{k}}\mathbf{S}_{\mathbf{k}}e^{i\mathbf{k}\cdot\mathbf{R}_{i}}\label{eq:FT S s}\\
\mathbf{S}_{\mathbf{k}} & =\frac{1}{\sqrt{N}}\sum_{i}\mathbf{S}_{i}e^{-i\mathbf{k}\cdot\mathbf{R}_{i}}\nonumber 
\end{align}
and hence 
\[
H_{{\rm cl}}=-\frac{1}{2N}\sum_{\mathbf{R}_{i}\mathbf{R}_{j}}J(\mathbf{R}_{i}-\mathbf{R}_{j})\sum_{\mathbf{k}\mathbf{k}'}\left(e^{i\mathbf{k}\cdot\mathbf{R}_{i}}\mathbf{S_{\mathbf{k}}}\cdot\mathbf{S}_{\mathbf{k}'}e^{i\mathbf{k'}\cdot\mathbf{R}_{j}}\right)
\]
Defining $\Delta\mathbf{R}=\mathbf{R}_{i}-\mathbf{R}_{j}$ and using
\begin{equation}
\sum_{\mathbf{R}_{i}}e^{i\mathbf{k}\cdot\mathbf{R}_{i}}=N\delta_{\mathbf{k},0}\label{eq:Kronecken sum}
\end{equation}
being $\delta_{\mathbf{k},\mathbf{k'}}$ the Kronecker Delta, we obtain
\begin{align*}
H_{{\rm cl}} & =-\frac{1}{2N}\sum_{\mathbf{R}_{i}\Delta\mathbf{R}}J(\Delta\mathbf{R})\sum_{\mathbf{k}\mathbf{k}'}\left(e^{i\mathbf{k}\cdot\mathbf{R}_{i}}\mathbf{S_{\mathbf{k}}}\cdot\mathbf{S}_{\mathbf{k}'}e^{i\mathbf{k'}\cdot\mathbf{R}_{i}}e^{-i\mathbf{k'}\cdot\Delta\mathbf{R}}\right)\\
 & =-\frac{1}{2N}\sum_{\Delta\mathbf{R}}J(\Delta\mathbf{R})\sum_{\mathbf{k}\mathbf{k}'}N\delta_{\mathbf{k}+\mathbf{k}',0}\mathbf{S_{\mathbf{k}}}\cdot\mathbf{S}_{\mathbf{k}'}e^{-i\mathbf{k'}\cdot\Delta\mathbf{R}}\\
 & =-\frac{1}{2}\sum_{\Delta\mathbf{R}}J(\Delta\mathbf{R})\sum_{\mathbf{k}}\mathbf{S_{\mathbf{k}}}\cdot\mathbf{S}_{-\mathbf{k}}e^{i\mathbf{k}\cdot\Delta\mathbf{R}}
\end{align*}
and defining 
\begin{align}
J(\mathbf{k}) & =\sum_{i}J(\Delta\mathbf{R})e^{i\mathbf{k}\cdot\Delta\mathbf{R}}\,\label{eq:FT J}
\end{align}
we obtain 
\begin{equation}
H_{{\rm cl}}=-\frac{1}{2}\sum_{\mathbf{k}}J(\mathbf{k})\mathbf{S}_{\mathbf{k}}\cdot\mathbf{S}_{-\mathbf{k}}\label{eq:heis class}
\end{equation}
Imposing the constraint $\mathbf{S}^{2}=S^{2}$, one can show that
the minimum of the energy given by Eq.~\ref{eq:heis class} is given
by setting $\mathbf{k}=\mathbf{Q}$, where $\mathbf{Q}$ determines
the global maximum of $J(\mathbf{Q})$ \cite{lyons_method_1960}.
One obtains the equilibrium configuration 
\[
\mathbf{S}_{i}=S\left(\cos(\mathbf{Q}.\mathbf{R}_{i}),\sin(\mathbf{Q}.\mathbf{R}_{i}),0\right)
\]
which is in general a planar helical state. Since $\mathbf{Q}$ is
determined by the maximum of $J(\mathbf{Q})$, the order is not necessarily
commensurate with the lattice, unless this maximum coincides with
a high symmetry point in the Brillouin zone. Special cases are $\mathbf{Q}=0$,
where we recover ferromagnetic order, or $\mathbf{Q}$ taking a value
at the edge of the Brillouin zone, in which case the order is antiferromagnetic.
Applying a magnetic field in the $z$ direction tilts the magnetic
order out of plane.

\subsection*{Check points}
\begin{itemize}
\item What is the classical Heisenberg model and how does one in general
find its ground state? 
\end{itemize}

\section{\label{sec:Dipole-dipole-interactions}Dipole-dipole interactions}

In the Heisenberg Hamiltonian we have not included the dipole-dipole
interaction term 
\begin{equation}
\hat{H}_{d-d}=-\frac{\mu_{0}}{4\pi}\frac{\left(\mu_{{\rm B}}g\right)^{2}}{2}\sum_{i\neq j}\frac{1}{|\mathbf{r}_{i}-\mathbf{r}_{j}|^{3}}\left[\frac{3\hat{\mathbf{S}}_{i}\cdot\left(\mathbf{r}_{i}-\mathbf{r}_{j}\right)\hat{\mathbf{S}}_{j}\cdot\left(\mathbf{r}_{i}-\mathbf{r}_{j}\right)}{|\mathbf{r}_{i}-\mathbf{r}_{j}|^{2}}-\hat{\mathbf{S}}_{i}\cdot\hat{\mathbf{S}}_{j}\right]\,.\label{eq:dipole dipole}
\end{equation}
We argued this term is too small to justify magnetic ordering at the
observed temperatures, and we found out that the exchange interaction
is orders of magnitude larger. However the exchange interaction depends
on the overlap of orbitals, and as such is generally short ranged:
it decays exponentially with distance, at least for insulators. From
Eq.~\ref{eq:dipole dipole} we see that the dipole-dipole interaction
decays instead algebraically, as $1/r^{3}$, and that means that it
will be dominant at large distances. This gives rise, in large samples,
to the formation of domains. In intermediate-size samples, usually
micrometer sized, the competition between dipole-dipole interactions
and exchange interactions leads to textured ground states. Being long-ranged,
the dipolar interactions are sensitive to the boundaries of the sample,
where the spins try to align with the boundary, so as to minimize
the " magnetic monopoles" at the surface
as introduced in Chapter \ref{chap:Electromagnetism}. Dipole-dipole
interaction terms can be included in the Heisenberg Hamiltonian in
the long-wavelength limit as demagnetization fields.

\chapter{\label{chap:Spin-waves-and}Spin Waves and Magnons}

In the last chapter we dealt with the ground state of the Heisenberg
model for a few solvable examples. The ground state is the only eigenstate
that is occupied at strictly zero temperature. The excitations on
top of this ground state will determine the behavior of the system
at low temperatures. These are collective excitations of the magnetic
system, and their quanta are denominated magnons \cite{kittel_introduction_2004}.
One can draw an analogy to another physical phenomenon which is perhaps
more intuitive: that of mechanical vibrations. There, all atoms participate
in the collective mechanical vibration, and we call phonons the respective
quanta. In the next sections we will study the magnetic elementary
excitations in more detail. We will focus on the simplest case of
the Heisenberg ferromagnet, but the concepts are general.

\section{\label{sec:Excitations-of-the}Excitations of the Heisenberg ferromagnet}

For the Heisenberg ferromagnet at $T=0$ we found that all spins are
aligned with and the magnetization is given by the saturation magnetization
\[
M_{{\rm s}}=g\mu_{{\rm B}}\frac{N}{V}S\,.
\]
At $T\neq0$ some spins will " flip" or,
more generally, decrease by one unit, which in turn will decrease
the magnetization from it saturation value. An obvious candidate for
an excited state is therefore 
\begin{equation}
|i\rangle=|S,S\rangle_{1}...|S,S-1\rangle_{i}...|S,S\rangle_{N}\label{eq:excited S i}
\end{equation}
which can be obtained from the ground state Eq.~\ref{eq:ground state}
as 
\begin{equation}
|i\rangle=\frac{1}{\sqrt{2S}}\hat{S}_{i}^{-}|0\rangle\,.\label{eq:i st 0 st}
\end{equation}
One can easily see that $|i\rangle$ is an eigenstate of the $\hat{S}^{z}$
operator 
\begin{align*}
\hat{S}_{i}^{z}|i\rangle & =(S-1)|i\rangle\\
\hat{S}_{j}^{z}|i\rangle & =S|i\rangle\quad\forall\quad j\neq i\,.
\end{align*}
However $\hat{S}_{i}^{+}|i\rangle\neq0$, instead one obtains 
\[
\hat{S}_{j}^{-}\hat{S}_{i}^{+}|i\rangle=2S|j\rangle\,.
\]
Therefore the Heisenberg Hamiltonian \ref{eq:Heis ladder} shifts
the site of the " flipped" spin $i\rightarrow j$,
and $|i\rangle$ as given in Eq. \ref{eq:excited S i} is not an eigenstate
of \ref{eq:Heis ladder}. It is also not a good approximation to an
excited eigenstate, since flipping a spin in such a manner has a very
high energy cost, of the order of the exchange interaction. For example,
for nearest neighbor interaction with exchange constant $J$, flipping
a spin has an energy cost of $\Delta E\sim zJS$, where $z$ is the
\emph{coordination number} (that is, $z$ is the number of nearest
neighbors, e.g. for a square lattice $z=4$). We have already seen
that the exchange constant is of the order of the critical temperature,
quite a high energy for regular ferromagnets.

One can prove that an actual eigenstate of \ref{eq:Heis ladder} is
given by 
\begin{equation}
|\mathbf{k}\rangle=\frac{1}{\sqrt{N}}\sum_{\mathbf{R}_{i}}e^{-i\mathbf{R}_{i}\cdot\mathbf{k}}|i\rangle\,,\label{eq:k state}
\end{equation}
where $\mathbf{R}_{i}$ are the lattice sites. We see therefore that
an eigenstate is formed by distributing the " flipped"
spin over all sites, and therefore it is a \emph{collective excitation}
of the system. This collective excitation has well defined momentum
$\hbar\mathbf{k}$ (up to a reciprocal lattice vector) and energy
$\hbar\omega(\mathbf{k})$ and we call it a \emph{quasiparticle}.
In particular for magnetic systems these quasiparticles are denominated
\emph{magnons.} In the following we will show that $|\mathbf{k}\rangle$
is an eigenstate of \ref{eq:Heis ladder} and will calculate its \emph{dispersion
relation} $\hbar\omega(\mathbf{k})$.

In order to do this, we first perform a Fourier transform of the Hamiltonian
\ref{eq:Heis ladder} by defining the spin operators in momentum space
(in analogy to the classical case, see Eqs. \ref{eq:FT J}) 
\begin{align}
\hat{S}_{\mathbf{k}}^{\alpha} & =\frac{1}{\sqrt{N}}\sum_{\mathbf{R}_{i}}e^{-i\mathbf{k}\cdot\mathbf{R}_{i}}\hat{S}_{i}^{\alpha}\label{eq:FT ops S}\\
\hat{S}_{i}^{\alpha} & =\frac{1}{\sqrt{N}}\sum_{\mathbf{k}}\hat{S}_{\mathbf{k}}^{\alpha}e^{i\mathbf{k}\cdot\mathbf{R}_{i}}\nonumber 
\end{align}
where $\alpha=x,\,y,\,z,\,\pm$. Note that 
\begin{align*}
\left(\hat{S}_{\mathbf{k}}^{+}\right)^{\dagger} & =\hat{S}_{-\mathbf{k}}^{-}\\
\left(\hat{S}_{\mathbf{k}}^{z}\right)^{\dagger} & =\hat{S}_{-\mathbf{k}}^{z}\,
\end{align*}
and one can verify that the commutators now read 
\begin{align}
\left[\hat{S}_{\mathbf{k}_{1}}^{+},\hat{S}_{\mathbf{k}_{2}}^{-}\right] & =\frac{2}{\sqrt{N}}\hat{S}_{\mathbf{k}_{1}+\mathbf{k}_{2}}^{z}\label{eq:comms S k}\\
\left[\hat{S}_{\mathbf{k}_{1}}^{z},\hat{S}_{\mathbf{k}_{2}}^{\pm}\right] & =\pm\frac{1}{\sqrt{N}}\hat{S}_{\mathbf{k}_{1}+\mathbf{k}_{2}}^{\pm}\,.\nonumber 
\end{align}
Using that $J_{ij}=J_{ji}$ and that spin operators at different sites
commute, we re-write slightly the Heisenberg Hamiltonian \ref{eq:Heis ladder}
as 
\begin{equation}
\hat{H}=-\frac{1}{2}\sum_{ij}J_{ij}\left[\hat{S}_{i}^{+}\hat{S}_{j}^{-}+\hat{S}_{i}^{z}\hat{S}_{j}^{z}\right]\,.\label{eq:Heis ladder 2}
\end{equation}
From Eqs. \ref{eq:FT ops S} and using again Eq.~\ref{eq:jij Jji}
we obtain 
\begin{align*}
\hat{H} & =-\frac{1}{2N}\sum_{\mathbf{R}_{i}\mathbf{R}_{j}}J(\mathbf{R}_{i}-\mathbf{R}_{j})\sum_{\mathbf{k}\mathbf{k}'}\left(\hat{S}_{\mathbf{k}}^{+}e^{i\mathbf{k}\cdot\mathbf{R}_{i}}\hat{S}_{\mathbf{k'}}^{-}e^{i\mathbf{k'}\cdot\mathbf{R}_{j}}+\hat{S}_{\mathbf{k}}^{z}e^{i\mathbf{k}\cdot\mathbf{R}_{i}}\hat{S}_{\mathbf{k'}}^{z}e^{i\mathbf{k'}\cdot\mathbf{R}_{j}}\right)\\
 & -\frac{g\mu_{{\rm B}}}{\sqrt{N}}B_{0}\sum_{\mathbf{R}_{i}}\sum_{\mathbf{k}}\hat{S}_{\mathbf{k}}^{z}e^{i\mathbf{k}\cdot\mathbf{R}_{i}}
\end{align*}
where we added an external magnetic field. We use $\Delta\mathbf{R}=\mathbf{R}_{i}-\mathbf{R}_{j}$
and Eq.~\ref{eq:Kronecken sum} to write 
\begin{align*}
\hat{H}= & -\frac{1}{2N}\sum_{\Delta\mathbf{R}}J(\Delta\mathbf{R})\sum_{\mathbf{R}_{i}}\sum_{\mathbf{k}\mathbf{k}'}e^{i\mathbf{\left(k+\mathbf{k}'\right)}\cdot\mathbf{R}_{i}}\left(\hat{S}_{\mathbf{k}}^{+}\hat{S}_{\mathbf{k'}}^{-}+\hat{S}_{\mathbf{k}}^{z}\hat{S}_{\mathbf{k'}}^{z}\right)e^{-i\mathbf{k'}\cdot\Delta\mathbf{R}}\\
 & -g\mu_{{\rm B}}\sqrt{N}B_{0}\sum_{\mathbf{k}}\hat{S}_{\mathbf{k}}^{z}\delta_{\mathbf{k},0}\\
= & -\frac{1}{2N}\sum_{\Delta\mathbf{R}}J(\Delta\mathbf{R})N\sum_{\mathbf{k}\mathbf{k}'}\delta_{\mathbf{k}+\mathbf{k'},0}\left(\hat{S}_{\mathbf{k}}^{+}\hat{S}_{\mathbf{k'}}^{-}+\hat{S}_{\mathbf{k}}^{z}\hat{S}_{\mathbf{k'}}^{z}\right)e^{-i\mathbf{k'}\cdot\Delta\mathbf{R}}-g\mu_{{\rm B}}\sqrt{N}B_{0}\hat{S}_{0}^{z}\\
= & -\frac{1}{2}\sum_{\Delta\mathbf{R}}J(\Delta\mathbf{R})\sum_{\mathbf{k}}\left(\hat{S}_{\mathbf{k}}^{+}\hat{S}_{\mathbf{-k}}^{-}+\hat{S}_{\mathbf{k}}^{z}\hat{S}_{\mathbf{-k}}^{z}\right)e^{i\mathbf{k}\cdot\Delta\mathbf{R}}-g\mu_{{\rm B}}\sqrt{N}B_{0}\hat{S}_{0}^{z}\,.
\end{align*}
Using the definition for $J(\mathbf{k})$ given in Eq.~\ref{eq:FT J}
we finally obtain an expression for the Heisenberg Hamiltonian in
momentum space 
\begin{equation}
\hat{H}=-\frac{1}{2}\sum_{\mathbf{k}}J(\mathbf{k})\left(\hat{S}_{\mathbf{k}}^{+}\hat{S}_{\mathbf{-k}}^{-}+\hat{S}_{\mathbf{k}}^{z}\hat{S}_{\mathbf{-k}}^{z}\right)-g\mu_{{\rm B}}\sqrt{N}B_{0}\hat{S}_{0}^{z}\,.\label{eq:heis Ham k space}
\end{equation}

First we check the action of $\hat{H}$ given in Eq.~\ref{eq:heis Ham k space}
on the ground state $|0\rangle$. For that we note 
\begin{align*}
\hat{S}_{\mathbf{k}}^{z}|0\rangle=\frac{1}{\sqrt{N}}\sum_{\mathbf{R}_{i}}e^{-i\mathbf{k}\cdot\mathbf{R}_{i}}\hat{S}_{i}^{z}|0\rangle & =\frac{S}{\sqrt{N}}\sum_{\mathbf{R}_{i}}e^{-i\mathbf{k}\cdot\mathbf{R}_{i}}|0\rangle=S\sqrt{N}\delta_{\mathbf{k},0}|0\rangle\\
\hat{S}_{i}^{+}|0\rangle & =0\quad\Rightarrow\quad\hat{S}_{\mathbf{k}}^{+}|0\rangle=0
\end{align*}
and examine the action of each term in Eq.~\ref{eq:heis Ham k space}
on $|0\rangle$. Using the commutators in Eq.~\ref{eq:comms S k},
for the first term in the sum we obtain 
\begin{align*}
-\frac{1}{2}\sum_{\mathbf{k}}J(\mathbf{k})\hat{S}_{\mathbf{k}}^{+}\hat{S}_{\mathbf{-k}}^{-}|0\rangle & =-\frac{1}{2}\sum_{\mathbf{k}}J(\mathbf{k})\left(\hat{S}_{-\mathbf{k}}^{-}\hat{S}_{\mathbf{k}}^{+}+\frac{2}{\sqrt{N}}\hat{S}_{0}^{z}\right)|0\rangle\\
 & =-\frac{1}{2\sqrt{N}}\sum_{\mathbf{k}}J(\mathbf{k})\left(2\hat{S}_{0}^{z}\right)|0\rangle=-S\sum_{\mathbf{k}}J(\mathbf{k})|0\rangle=0
\end{align*}
where the last equality stems from 
\[
\sum_{\mathbf{k}}J(\mathbf{k})=\sum_{\mathbf{k}}\sum_{\Delta\mathbf{R}}J(\Delta\mathbf{R})e^{i\mathbf{k}\cdot\Delta\mathbf{R}}=\sum_{\Delta\mathbf{R}}J(\Delta\mathbf{R})N\delta_{\Delta\mathbf{R},0}=NJ(\Delta\mathbf{R}=0)=0\,.
\]
For the second term in Eq.~\ref{eq:heis Ham k space} 
\begin{align*}
-\frac{1}{2}\sum_{\mathbf{k}}J(\mathbf{k})\hat{S}_{\mathbf{k}}^{z}\hat{S}_{\mathbf{-k}}^{z}|0\rangle & =-\frac{1}{2}\sqrt{N}S\sum_{\mathbf{k}}J(\mathbf{k})\hat{S}_{\mathbf{k}}^{z}\delta_{\mathbf{-k},0}|0\rangle\\
 & =-\frac{1}{2}\sqrt{N}SJ(\mathbf{k}=0)\hat{S}_{0}^{z}|0\rangle\\
 & =-\frac{1}{2}NS^{2}J(\mathbf{k}=0)|0\rangle\,.
\end{align*}
We now note that 
\[
NJ(\mathbf{k}=0)=N\sum_{\Delta\mathbf{R}}J(\Delta\mathbf{R})=\sum_{\mathbf{R}_{i}\Delta\mathbf{R}}J(\Delta\mathbf{R})=\sum_{\mathbf{R}_{i}\mathbf{R}_{j}}J(\mathbf{R}_{i}-\mathbf{R}_{j})=\sum_{ij}J_{ij}
\]
and therefore 
\[
-\frac{1}{2}\sum_{\mathbf{k}}J(\mathbf{k})\hat{S}_{\mathbf{k}}^{z}\hat{S}_{\mathbf{-k}}^{z}|0\rangle=-\frac{S^{2}}{2}\sum_{ij}J{}_{ij}|0\rangle\,.
\]
For the third term in Eq.~\ref{eq:heis Ham k space} we obtain simply
\[
-g\mu_{{\rm B}}\sqrt{N}B_{0}\hat{S}_{0}^{z}|0\rangle=-g\mu_{{\rm B}}NB_{0}S|0\rangle\,.
\]
Putting all together we obtain $\hat{H}|0\rangle=E_{0}|0\rangle$
with 
\[
E_{0}\left(B_{0}\right)=-\frac{S}{2}^{2}\sum_{ij}J_{ij}\,-g\mu_{{\rm B}}B_{0}NS
\]
which coincides with our result in Eq.~\ref{eq:E0 in B} as expected.

We now want to show that $|\mathbf{k}\rangle$ given in Eq.~\ref{eq:k state}
is a eigenstate of \ref{eq:heis Ham k space}. Using \ref{eq:i st 0 st}
we can write $|\mathbf{k}\rangle$ in terms of the ladder operators
in momentum space, 
\begin{equation}
|\mathbf{k}\rangle=\frac{1}{\sqrt{2S}}\hat{S}_{\mathbf{k}}^{-}|0\rangle\,,\label{eq:k ladder}
\end{equation}
hence it is enough to show that $\hat{S}_{\mathbf{k}}^{-}|0\rangle$
is an eigenstate \cite{nolting_quantum_2009}. Writing 
\begin{align*}
\hat{H}\hat{S}_{\mathbf{k}}^{-}|0\rangle & =\left(\hat{S}_{\mathbf{k}}^{-}\hat{H}+\left[\hat{H},\hat{S}_{\mathbf{k}}^{-}\right]\right)|0\rangle=\left(\hat{S}_{\mathbf{k}}^{-}E_{0}+\left[\hat{H},\hat{S}_{\mathbf{k}}^{-}\right]\right)|0\rangle\\
 & =E_{0}\hat{S}_{\mathbf{k}}^{-}|0\rangle+\left[\hat{H},\hat{S}_{\mathbf{k}}^{-}\right]|0\rangle\,,
\end{align*}
we see that that amounts to showing that $\left[\hat{H},\hat{S}_{\mathbf{k}}^{-}\right]|0\rangle\propto\hat{S}_{\mathbf{k}}^{-}|0\rangle$.
Using similar manipulations as above, and taking into account that
$J(\mathbf{k})=J(-\mathbf{k}),$ one obtains 
\begin{equation}
\left[\hat{H},\hat{S}_{\mathbf{k}}^{-}\right]|0\rangle=\left[g\mu_{{\rm B}}B_{0}-S\left(J(\mathbf{k})-J(\mathbf{k}=0)\right)\right]\hat{S}_{\mathbf{k}}^{-}|0\rangle\label{eq:Comm H Sk}
\end{equation}
and therefore $|\mathbf{k}\rangle$ is an eigenstate of \ref{eq:heis Ham k space}
with eigenvalue 
\begin{equation}
E(\mathbf{k})=E_{0}(B_{0})+g\mu_{{\rm B}}B_{0}-S\left(J(\mathbf{k})-J(\mathbf{k}=0)\right)\,.\label{eq:E(k)}
\end{equation}

The energy on top of the ground state is the \emph{excitation energy}
\begin{equation}
\hbar\omega(\mathbf{k})=g\mu_{{\rm B}}B_{0}-S\left[J(\mathbf{k})-J(\mathbf{k}=0)\right]\label{eq:w(k)}
\end{equation}
which is simply the\emph{ energy of one magnon} --- the quasiparticle
energy is always defined with respect to the ground state energy,
in field theory it is common to call the ground state of a system
the " vacuum" . Eq.~\ref{eq:w(k)} is
also called the \emph{dispersion relation}, since it gives the dependence
of the energy with the wave vector. Note that for $B_{0}=0$, $\hbar\omega(0)=0$
and therefore any infinitesimal temperature will cause excitations
with $\mathbf{k}=0$. This " zero mode"
is an example of a \emph{Goldstone mode}, which is always present
when there is spontaneous symmetry breaking in the system. In general,
Goldstone modes are massless bosons quasiparticles which appear in
systems where there is a spontaneously broken continuous symmetry.

If we compare with the ground state energy $E_{0}(B_{0})$, in particular
the Zeeman term, we see that the magnetic moment of the system has
been modified by one unit. We can therefore conclude that a \emph{magnon
has spin 1}, and therefore it is a \emph{bosonic} quasiparticle. The
expectation value of a local spin operator $\hat{S}_{i}^{z}$ with
respect to the one-magnon state $|\mathbf{k}\rangle$ can be shown
to be 
\begin{equation}
\langle\mathbf{k}|\hat{S}_{i}^{z}|\mathbf{k}\rangle=S-\frac{1}{N}=\langle0|\hat{S}_{i}^{z}|0\rangle-\frac{1}{N}\quad\forall\quad i,\,\mathbf{k}\label{eq:distributed spin}
\end{equation}
which shows that the spin reduction is indeed of one unit and it is
distributed uniformly over all sites $\mathbf{R}_{i}$. Semiclassically,
one can picture each spin in the lattice precessing around the $z$
axis with a projection of $\hbar(S-1/N)$. However, except for $\mathbf{k}=0$,
the spins do not precess in phase but instead they have a phase difference
of $e^{-i\mathbf{k}\cdot(\mathbf{R}_{i}-\mathbf{R}_{j})}$, forming
a \emph{spin wave}. This semiclassical picture can be better understood
if we look at the equations of motion for the spins, which we do in
the following section.

We finishing this section by stating that our intuition fails again
if we want to construct two-magnon states. The obvious choice 
\begin{equation}
|\mathbf{k},\mathbf{k}'\rangle\propto\hat{S}_{\mathbf{k}}^{-}\hat{S}_{\mathbf{k'}}^{-}|0\rangle\label{eq:2 magnon state}
\end{equation}
is actually not an eigenstate of the Heisenberg Hamiltonian. This
is due to magnon-magnon interactions present in the Hamiltonian, which
are not taken into account in a simple product state as \ref{eq:2 magnon state}.
The same of course holds for multiple-magnon states. Therefore, when
having multiple magnons excited in a system, they can interact and
magnon states will decay due to magnon-magnon interactions. These
excited states therefore have a certain \emph{lifetime}. Besides magnon-magnon
interactions, scattering with impurities or phonons in a material
will determine the lifetime of magnon states.

\subsection*{Check points}
\begin{itemize}
\item What is a magnon? 
\item How do you obtain the energy dispersion of a magnon? 
\end{itemize}

\section{\label{sec:Equation-of-motion-1}Equation of motion approach}

In the Heisenberg picture, the time dependence is included in the
operators. Instead of the Schrödinger equation for the state vectors,
we write the \emph{Heisenberg equation of motion} for the spin operators
\begin{equation}
\hbar\frac{d\hat{\mathbf{S}}_{i}}{dt}=i[\hat{H},\hat{\mathbf{S}}_{i}]\,.\label{eq:heis EOM}
\end{equation}
Using the Heisenberg Hamiltonian Eq.~\ref{eq:Heisenberg in B field}
and the commutation relations for the spin operators, it is straightforward
to show 
\begin{equation}
\hbar\frac{d\hat{\mathbf{S}}_{i}}{dt}=-\left(\sum_{j}J_{ij}\hat{\mathbf{S}}_{j}+g\mu_{{\rm B}}\mathbf{B_{0}}\right)\times\hat{\mathbf{S}}_{i}\,.\label{eq:EOM Si}
\end{equation}
This equation is exact. We can compare it however with the classical
equation of motion for the angular momentum given in Eq.~\ref{eq:EOM L},
and we see that, by recovering the units of the spin and absorbing
the Planck constant into the definition of the exchange constant,
Eq.~\ref{eq:EOM Si} can be directly translated into a classical
equation of motion by taking the expectation values of the spin operators,
and in particular 
\[
\mathbf{B}_{eff}=\mathbf{B_{0}}+\frac{1}{g\mu_{{\rm B}}}\langle\sum_{j}J_{ij}\hat{\mathbf{S}}_{j}\rangle
\]
is the effective magnetic field including the Weiss molecular fields.

We turn now to a different approximation, in which we retain for now
the operator character of $\hat{\mathbf{S}}_{i}$. We are, as in the
previous section, interested in the low energy excitations of the
system on top of the ground state. Since the ground state is fully
polarized, we expect the projection $\hat{S}_{i}^{z}$ to remain almost
constant and close to $S$. From Eq.~\ref{eq:distributed spin} we
see this is valid as long as $NS\gg1$, that is, the total spin number
is much larger than the number of excitations in the system. From
Eq.~\ref{eq:EOM Si} we obtain 
\begin{align}
\hbar\frac{d\hat{S}_{i}^{x}}{dt} & \approx-S\sum_{j}J_{ij}\left(\hat{S}_{j}^{y}-\hat{S}_{i}^{y}\right)+g\mu_{{\rm B}}B_{0}\hat{S}_{i}^{y}\label{eq:approx EOM}\\
\hbar\frac{d\hat{S}_{i}^{y}}{dt} & \approx-S\sum_{j}J_{ij}\left(\hat{S}_{i}^{x}-\hat{S}_{j}^{x}\right)-g\mu_{{\rm B}}B_{0}\hat{S}_{i}^{x}\nonumber \\
\hbar\frac{d\hat{S}_{i}^{z}}{dt} & \approx0\,.\nonumber 
\end{align}
These equations are decoupled for the ladder operators 
\begin{equation}
\hbar\frac{d\hat{S}_{i}^{\pm}}{dt}=\mp i\left[S\sum_{j}J_{ij}\left(\hat{S}_{i}^{\pm}-\hat{S}_{j}^{\pm}\right)+g\mu_{{\rm B}}B_{0}\hat{S}_{i}^{\pm}\right]\,,\label{eq:EOM ladder}
\end{equation}
where we write the equal sign in the understanding that the equation
is valid in the limit established for Eqs. \ref{eq:approx EOM}.

Eqs. \ref{eq:EOM ladder} still couple spin operators at different
sites. To decouple them, we go once more to the Fourier representation.
Using Eq.~\ref{eq:FT ops S}, for the lowering operator one obtains
\begin{equation}
\hbar\frac{d\hat{S}_{\mathbf{k}}^{-}}{dt}=iS\left[J(\mathbf{k=0})-J(\mathbf{k})\right]\hat{S}_{\mathbf{k}}^{-}+ig\mu_{{\rm B}}B_{0}\hat{S}_{\mathbf{k}}^{-}\,.\label{eq:Lower EOM}
\end{equation}
Hence we see that the equation of motion for each wave vector $\mathbf{k}$
is decoupled from the rest, in the spirit of a normal modes decomposition.
Eq.~\ref{eq:Lower EOM} is easily solved by 
\begin{equation}
\hat{S}_{\mathbf{k}}^{-}=\hat{M}_{\mathbf{k}}e^{i\omega(\mathbf{k})t+i\alpha_{\mathbf{k}}}\label{eq:solution Sminus}
\end{equation}
with the time dependence given entirely by the exponential term, and
$\omega(\mathbf{k})$ is given by Eq.~\ref{eq:w(k)} ($\alpha_{\mathbf{k}}$
is an arbitrary phase). We have therefore re-derived the dispersion
relation obtained in the previous section for the one-magnon state.

In the semiclassical picture and considering only one excited mode,
for the real-space components of the spin we obtain 
\begin{align*}
S_{i}^{x} & =\frac{M_{\mathbf{k}}}{\sqrt{N}}\cos\left(\mathbf{k}\cdot\mathbf{R}_{i}+\omega(\mathbf{k})t\right)\\
S_{i}^{y} & =\frac{M_{\mathbf{k}}}{\sqrt{N}}\sin\left(\mathbf{k}\cdot\mathbf{R}_{i}+\omega(\mathbf{k})t\right)\\
S_{i}^{z} & =S
\end{align*}
which are the components of a plane wave with frequency $\omega(\mathbf{k})$.

Coming back to the Heisenberg equation of motion for $\hat{S}_{\mathbf{k}}^{-}$
\[
\hbar\frac{d\hat{S}_{\mathbf{k}}^{-}}{dt}=i[\hat{H},\hat{\mathbf{S}}_{\mathbf{k}}^{-}]
\]
from Eq.~\ref{eq:solution Sminus} we can write 
\[
\hbar\omega(\mathbf{k})\hat{S}_{\mathbf{k}}^{-}|0\rangle=[\hat{H},\hat{\mathbf{S}}_{\mathbf{k}}^{-}]|0\rangle=\hat{H}\hat{S}_{\mathbf{k}}^{-}|0\rangle-\hat{S}_{\mathbf{k}}^{-}E_{0}|0\rangle
\]
and therefore we recover 
\[
\hat{H}\hat{S}_{\mathbf{k}}^{-}|0\rangle=\left[E_{0}-\hbar\omega(\mathbf{k})\right]\hat{S}_{\mathbf{k}}^{-}|0\rangle
\]
as in the exact result.

As an example, we give the dispersion relation for spins on a cubic
lattice with nearest-neighbor interactions. For this case, 
\[
J(\mathbf{k})=2J\left(\cos(k_{x}a)+\cos(k_{y}a)+\cos(k_{z}a)\right)
\]
and therefore 
\[
\hbar\omega(\mathbf{k})=2JS\left(3-\cos(k_{x}a)+\cos(k_{y}a)+\cos(k_{z}a)\right)+g\mu_{{\rm B}}B_{0\,.}
\]
For small $\mathbf{k}$ we obtain a quadratic dispersion 
\begin{equation}
\hbar\omega(\mathbf{k})\approx JSk^{2}a^{2}+g\mu_{{\rm B}}B_{0\,,}\label{eq:quad dispersion}
\end{equation}
which is gapped as long as $B_{0}\neq0$. Note that this dispersion
is different from the usual acoustic phonon dispersion in solids,
which is linear in $k$. The dispersion \ref{eq:quad dispersion}
is actually similar to that encountered for \emph{flexural} phonon
modes in materials with reduced dimensionality: for example, the out-of-plane
phonon modes of a graphene membrane. The $k^{2}$ dispersion is a
signature of the rotational symmetry of the problem.

\subsection*{Check points}
\begin{itemize}
\item Derive the Heisenberg equation of motion for the spins from the Heisenberg
Hamiltonian 
\item Relate the concept of magnon from the previous section, with the semiclassical
picture given by the equations of motion. 
\end{itemize}

\section{Holstein-Primakoff transformation\label{sec:Holstein-Primakoff-transformatio}}

Due to their algebra (that is, their commutation relations), angular
momentum operators are difficult to treat in an interacting theory.
There are however transformations which write the angular momentum
operators in terms of second-quantization creation and annihilation
operators, either fermionic or bosonic. The idea of these transformations
is to simplify the commutation rules, so that one can use well known
methods of second quantization. The price to pay is that the transformations
are non linear. In this section we go over one of these transformations
which is widely used, the Holstein-Primakoff transformation. Within
this transformation, the angular momentum operators are written as
non-linear functions of bosonic creation and annihilation operators,
that is, a collection of harmonic oscillators.

Before writing the transformation explicitly we remind briefly the
properties of \emph{creation} ($\hat{a}_{i}^{\dagger}$) and \emph{annihilation}
($\hat{a}_{i}$) harmonic-oscillator operators. The subscript $i$
indicates the lattice site, that is, we have a harmonic oscillator
at every site on the lattice. The commutation relations for these
bosonic operators are 
\begin{align}
\left[\hat{a}_{i},\,\hat{a}_{j}^{\dagger}\right] & =\delta_{ij}\label{eq:comm a a}\\
\left[\hat{a}_{i},\,\hat{a}_{j}\right] & =\left[\hat{a}_{i}^{\dagger},\,\hat{a}_{j}^{\dagger}\right]=0\nonumber 
\end{align}
The Hamiltonian of a single oscillator reads 
\begin{equation}
H_{{\rm osc}}=\hbar\omega_{i}\left(\hat{a}_{i}^{\dagger}\hat{a}_{i}+\frac{1}{2}\right)\,,\label{eq:single osc}
\end{equation}
where $\omega_{i}$ is the frequency of oscillator $i$. If the oscillators
are independent, the total Hamiltonian is simply the sum over the
respective Hamiltonians. Usually however the original operators are
not independent, but if the Hamiltonian is \emph{quadratic} in these,
one can find a linear transformation which diagonalizes the Hamiltonian,
so that in the new basis the Hamiltonian is a sum of harmonic oscillators
Eq.~\ref{eq:single osc}. A state $|n_{i}\rangle$ with $n_{i}$
" particles" at site $i$ can be constructed
from the vacuum $|0_{i}\rangle$ by applying $\hat{a}_{i}^{\dagger}$
(below we will identify $n_{i}$ with the number of flipped spins
at site $i$). In general, we have 
\begin{align}
\hat{a}_{i}|0_{i}\rangle & =0\label{eq:a ops}\\
\hat{a}_{i}^{\dagger}|n_{i}\rangle & =\sqrt{n_{i}+1}|n_{i}+1\rangle\nonumber \\
\hat{a}_{i}|n_{i}\rangle & =\sqrt{n_{i}}|n_{i}-1\rangle\nonumber \\
\hat{a}_{i}^{\dagger}\hat{a}_{i}|n_{i}\rangle & =n_{i}|n_{i}\rangle=\hat{n}_{i}|n_{i}\rangle\,,\nonumber 
\end{align}
where $\hat{n}_{i}=\hat{a}_{i}^{\dagger}\hat{a}_{i}$ is the \emph{number
operator} at site $i.$ Therefore $|n_{i}\rangle$ is an eigenstate
of $H_{{\rm osc}}$ with energy $E(n_{i})=\hbar\omega_{i}\left(n_{i}+1/2\right)$,
with $n_{i}=0,\,1,\,2,\,...$.

Comparing Eqs. \ref{eq:a ops} with Eq.~\ref{eq:ladder op} and $\hat{S}_{i}^{z}$
\begin{align*}
\hat{S}_{i}^{\pm}|S,m_{i}\rangle & =\sqrt{\left(S\mp m_{i}\right)\left(S+1\pm m_{i}\right)}|S,m_{i}\pm1\rangle\\
\hat{S}_{i}^{z}|S,m_{i}\rangle & =m_{i}|S,m_{i}\rangle
\end{align*}
we see that the creation and annihilation bosonic operators act in
a similar way to the spin ladder operators, whereas the number operator
is diagonal in this basis just as $\hat{S}_{i}^{z}$. However we cannot
replace simply $\hat{S}_{i}^{\pm}$ by $\hat{a}_{i}^{\dagger}$, $\hat{a}_{i}$
since this would not satisfy the commutation relations for the spin.
This is accounted for by using a non-linear transformation, the \emph{Holstein-Primakoff
transformation} 
\begin{align}
\hat{S}_{i}^{+} & =\sqrt{2S}\sqrt{1-\frac{\hat{a}_{i}^{\dagger}\hat{a}_{i}}{2S}}\hat{a}_{i}\nonumber \\
\hat{S}_{i}^{-} & =\sqrt{2S}\hat{a}_{i}^{\dagger}\sqrt{1-\frac{\hat{a}_{i}^{\dagger}\hat{a}_{i}}{2S}}\nonumber \\
\hat{S}_{i}^{z} & =\left(S-\hat{a}_{i}^{\dagger}\hat{a}_{i}\right)\,.\label{eq:Holstein Primakoff}
\end{align}
As we anticipated, in this case the number operator counts the number
of flipped spins, which we can see from the last equality in Eqs.
\ref{eq:Holstein Primakoff} 
\[
\hat{S}_{i}^{z}|n_{i}\rangle=\left(S-\hat{n}_{i}\right)|n_{i}\rangle=\left(S-n_{i}\right)|n_{i}\rangle\equiv m_{i}|n_{i}\rangle\,.
\]
For $n_{i}=0$, the spin is fully polarized and $m_{i}=S$, hence
the Holstein-Primakoff vacuum $|n_{i}=0\rangle$ corresponds to the
fully polarized spin state. We see however that the spectrum of $\hat{n}_{i}$
is constrained, due to the square root in Eqs. \ref{eq:Holstein Primakoff}
we must impose $n_{i}=0,\,1,\,...2S$. The maximum value of $n_{i}$
corresponds to the spin " fully flipped"
, $m_{i}=-S$. We observe that 
\begin{align*}
\hat{S}_{i}^{-}|m_{i}=-S\rangle & =\sqrt{2S}\hat{a}_{i}^{\dagger}\sqrt{1-\frac{\hat{n}_{i}}{2S}}|n_{i}=2S\rangle=\sqrt{2S}\hat{a}_{i}^{\dagger}\sqrt{1-\frac{2S}{2S}}|n_{i}=2S\rangle=0\\
\hat{S}_{i}^{+}|m_{i}=-S-1\rangle & =\sqrt{2S}\sqrt{1-\frac{\hat{n}_{i}}{2S}}\hat{a}_{i}|n_{i}=2S+1\rangle\\
 & =\sqrt{2S}\sqrt{2S+1}\sqrt{1-\frac{\hat{n}_{i}}{2S}}|n_{i}=2S\rangle=0
\end{align*}
and hence the ladder operators do not connect the physical subspace
$n_{i}=0,\,1,\,...2S$ with the unphysical one $n_{i}>2S$.

Let's now define 
\begin{equation}
\phi(\hat{n}_{i})=\sqrt{2S}\sqrt{1-\frac{\hat{n}_{i}}{2S}}\label{eq:function phi}
\end{equation}
and write the Heisenberg Hamiltonian Eq.~\ref{eq:Heisenberg Ham}
in terms of the Holstein-Primakoff operators, Eqs. \ref{eq:Holstein Primakoff}.
We find 
\begin{equation}
\hat{H}=-\frac{NS^{2}J_{0}}{2}+SJ_{0}\sum_{i}\hat{n}_{i}-S\sum_{ij}J_{ij}\phi(\hat{n}_{i})\hat{a}_{i}\hat{a}_{j}^{\dagger}\phi(\hat{n}_{j})-\frac{1}{2}\sum_{ij}J_{ij}\hat{n}_{i}\hat{n}_{j}\label{eq:Heis Ham phi}
\end{equation}
with $J_{0}=\sum_{i}J_{ij}$. We see that, due to Eq.~\ref{eq:function phi},
this Hamiltonian is not quadratic in the $\hat{a}_{i}^{\dagger}$,
$\hat{a}_{i}$ operators and therefore we cannot write it as a sum
of independent harmonic oscillators. We have hence transformed our
original Heisenberg Hamiltonian of interacting spins into a Hamiltonian
of interacting bosons.

\subsection*{Check points}
\begin{itemize}
\item Write the Holstein-Primakoff transformation 
\item What is special about it? 
\end{itemize}

\section{Spin-wave approximation\label{sec:Spin-wave-approximation}}

We will now proceed to reformulate our Hamiltonian into a non-interacting
term (namely, non-interacting magnons), plus interaction terms. If
we expand Eq.~\ref{eq:function phi} as a series 
\begin{equation}
\frac{\phi(\hat{n}_{i})}{\sqrt{2S}}=1-\frac{\hat{n}_{i}}{4S}-\frac{\hat{n}_{i}^{2}}{32S^{2}}-\,...\label{eq:phi exp}
\end{equation}
we can write Eq.~\ref{eq:Heis Ham phi} also as a series 
\begin{equation}
\hat{H}=-\frac{NS^{2}J_{0}}{2}+\sum_{n=1}^{\infty}:\hat{H}_{2n}:\label{eq:heis series}
\end{equation}
with $:\hat{H}_{2n}:$ containing $n$ creation and $n$ annihilation
operators in normal order (all $\hat{a}_{i}^{\dagger}$ to the left,
all $\hat{a}_{i}$ to the right, e.g. for $n=2$, $\hat{a}_{i}^{\dagger}\hat{a}_{j}^{\dagger}\hat{a}_{m}\hat{a}_{l}$).
The terms with $n>1$, that is, beyond quadratic, give rise to magnon-magnon
interactions as we will see below. We will however first study the
\emph{spin-wave approximation}, where only the quadratic, non-interacting
terms are kept in the expansion Eq.~\ref{eq:heis series} 
\begin{equation}
:\hat{H}_{2n}:=SJ_{0}\sum_{i}\hat{n}_{i}-S\sum_{ij}J_{ij}\hat{a}_{i}^{\dagger}\hat{a}_{j}\,.\label{eq:H2}
\end{equation}
This truncation of the series is justified at low temperatures, where
the number of excitations (total number of flipped spins) is small
compared with the total number of spins $NS$. For that to hold, the
average number of flipped spins per site has to be small $n_{i}\ll S$
and therefore we can approximate the square root in Eq.~\ref{eq:function phi}
to 1 and hence $\phi(\hat{n}_{i})\approx\sqrt{2S}$. Within this approximation,
the spin ladder operators are indeed approximated by simple harmonic
oscillators, while the $z$ component is kept at saturation 
\begin{align}
\hat{S}_{i}^{+} & \approx\sqrt{2S}\hat{a}_{i}\label{eq:ladder harmonic}\\
\hat{S}_{i}^{-} & \approx\sqrt{2S}\hat{a}_{i}^{\dagger}\nonumber \\
\hat{S}_{i}^{z} & \approx S\,,\nonumber 
\end{align}
and it can be directly seen that the Heisenberg Hamiltonian Eq.~\ref{eq:Heisenberg Ham}
is quadratic in the $\hat{a}_{i}^{\dagger}$, $\hat{a}_{i}$ operators.
This approximation is completely analogous to the one we performed
when working with the equation of motion, Eq.~\ref{eq:approx EOM}.
Note that, although the Hamiltonian is quadratic, it is not diagonal
in $i,\,j$ (see Eq.~\ref{eq:H2}). Just as we did for the equations
of motion, we need to go to Fourier space to obtain a diagonal Hamiltonian
and therefore decoupled harmonic oscillators. In this case, we transform
simply the bosonic operators 
\begin{align*}
\hat{a}_{\mathbf{k}} & =\frac{1}{\sqrt{N}}\sum_{\mathbf{R}_{i}}e^{-i\mathbf{k}\cdot\mathbf{R}_{i}}\hat{a}_{i}\\
\hat{a}_{\mathbf{k}}^{\dagger} & =\frac{1}{\sqrt{N}}\sum_{\mathbf{R}_{i}}e^{i\mathbf{k}\cdot\mathbf{R}_{i}}\hat{a}_{i}^{\dagger}
\end{align*}
and, within this approximation, we can show that the Heisenberg Hamiltonian
reduces to 
\begin{equation}
\hat{H}_{{\rm sw}}=E(B_{0})+\sum_{\mathbf{k}}\hbar\omega(\mathbf{k})\hat{a}_{\mathbf{k}}^{\dagger}\hat{a}_{\mathbf{k}}\label{eq:SW Ham}
\end{equation}
where we have added an external magnetic field for completeness, and
$E(B_{0})$ and $\omega(\mathbf{k})$ are given by Eqs. \ref{eq:E(k)}
and \ref{eq:w(k)} respectively.

The Hamiltonian \ref{eq:SW Ham} describes a system of uncoupled harmonic
oscillators. Its eigenstates are simply products of one-magnon states,
that can be obtained from the vacuum by applying repeatedly $\hat{a}_{\mathbf{k}}^{\dagger}$
\begin{equation}
|\psi_{{\rm SW}}\rangle=\prod_{\mathbf{k}}\left(\hat{a}_{\mathbf{k}}^{\dagger}\right)^{n_{\mathbf{k}}}|0\rangle\label{eq:psi SW}
\end{equation}
where $n_{\mathbf{k}}$ is the number of magnons with wavevector \textbf{$\mathbf{k}$},
eigenvalue of the number operator in Fourier representation $\hat{n}_{\mathbf{k}}=\hat{a}_{\mathbf{k}}^{\dagger}\hat{a}_{\mathbf{k}}$.
We see that we can write the one-magnon state defined in Eq.~\ref{eq:k state}
as $|\mathbf{k}\rangle=\hat{a}_{\mathbf{k}}^{\dagger}|0\rangle$.
This state is an eigenstate of both the full Heisenberg Hamiltonian
and of the non-interacting Hamiltonian \ref{eq:SW Ham}. States with
$n_{\mathbf{k}}>1$ are however only eigenstates of \ref{eq:SW Ham}.

\subsection*{Check points}
\begin{itemize}
\item What is the meaning of the spin-wave approximation in terms of the
Holstein-Primakoff transformation? 
\item What is a magnon in this language? 
\item What is the meaning of the Heisenberg Hamiltonian in terms of the
bosonic operators? 
\end{itemize}

\section{Magnon-magnon interactions\label{sec:Magnon-magnon-interactions}}

We now proceed to investigate the higher order terms ($n>1$) in Eq.
\ref{eq:heis series}. We consider for simplicity a Heisenberg Hamiltonian
with nearest neighbors interactions 
\begin{equation}
\hat{H}_{{\rm n.n.}}=-\frac{J}{2}\sum_{\langle ij\rangle}\left(\frac{\hat{S}_{i}^{+}\hat{S}_{j}^{-}+\hat{S}_{i}^{-}\hat{S}_{j}^{+}}{2}+\hat{S}_{i}^{z}\hat{S}_{j}^{z}\right)\,.\label{eq:H n n}
\end{equation}
Inserting Eqs. \ref{eq:Holstein Primakoff} generally we obtain 
\begin{align*}
\hat{H}_{{\rm n.n.}} & =-\frac{J}{2}\sum_{\langle ij\rangle}\left[S\sqrt{1-\frac{\hat{a}_{i}^{\dagger}\hat{a}_{i}}{2S}}\hat{a}_{i}\hat{a}_{j}^{\dagger}\sqrt{1-\frac{\hat{a}_{j}^{\dagger}\hat{a}_{j}}{2S}}+S\hat{a}_{i}^{\dagger}\sqrt{1-\frac{\hat{a}_{i}^{\dagger}\hat{a}_{i}}{2S}}\sqrt{1-\frac{\hat{a}_{j}^{\dagger}\hat{a}_{j}}{2S}}\hat{a}_{j}\right]\\
 & -\frac{J}{2}\sum_{\langle ij\rangle}\left(S-\hat{a}_{i}^{\dagger}\hat{a}_{i}\right)\left(S-\hat{a}_{j}^{\dagger}\hat{a}_{j}\right)\,.
\end{align*}
We now keep the first two terms in the expansion of $\phi(\hat{n}_{i})$,
see Eq.~\ref{eq:phi exp}. Therefore, simply inserting into Eq.~\ref{eq:H n n}
\begin{align}
\hat{H}_{{\rm n.n.}} & \approx-\frac{J}{2}\sum_{\langle ij\rangle}S\left(1-\frac{\hat{a}_{i}^{\dagger}\hat{a}_{i}}{4S}\right)\hat{a}_{i}\hat{a}_{j}^{\dagger}\left(1-\frac{\hat{a}_{j}^{\dagger}\hat{a}_{j}}{4S}\right)\label{eq:H n n approx 1}\\
 & -\frac{J}{2}\sum_{\langle ij\rangle}S\hat{a}_{i}^{\dagger}\left(1-\frac{\hat{a}_{i}^{\dagger}\hat{a}_{i}}{4S}\right)\left(1-\frac{\hat{a}_{j}^{\dagger}\hat{a}_{j}}{2S}\right)\hat{a}_{j}\\
 & -\frac{J}{2}\sum_{\langle ij\rangle}\left(S-\hat{a}_{i}^{\dagger}\hat{a}_{i}\right)\left(S-\hat{a}_{j}^{\dagger}\hat{a}_{j}\right)\,.
\end{align}
To be consistent with the approximation we keep terms with up to four
creation/annihilation operators in Eq.~\ref{eq:H n n approx 1}.
One obtains 
\begin{align}
\hat{H}_{{\rm n.n.}} & \approx-Nz\frac{JS}{2}^{2}+JS\sum_{\langle ij\rangle}\left(\hat{a}_{i}^{\dagger}\hat{a}_{i}+\hat{a}_{j}^{\dagger}\hat{a}_{j}-\hat{a}_{i}^{\dagger}\hat{a}_{j}-\hat{a}_{j}^{\dagger}\hat{a}_{i}\right)\label{eq:H n. n. approx 2}\\
 & -J\sum_{\langle ij\rangle}\left[\hat{a}_{i}^{\dagger}\hat{a}_{i}\hat{a}_{j}^{\dagger}\hat{a}_{j}-\frac{1}{4}\left(\hat{a}_{i}^{\dagger}\hat{a}_{i}^{\dagger}\hat{a}_{i}\hat{a}_{j}+\hat{a}_{i}^{\dagger}\hat{a}_{j}^{\dagger}\hat{a}_{j}\hat{a}_{j}+\hat{a}_{j}^{\dagger}\hat{a}_{i}^{\dagger}\hat{a}_{i}\hat{a}_{i}+\hat{a}_{j}^{\dagger}\hat{a}_{j}^{\dagger}\hat{a}_{j}\hat{a}_{i}\right)\right]
\end{align}
with the following terms in the expansion being of order $1/S$ or
higher.

We already saw that the quadratic terms in Eq.~\ref{eq:H n. n. approx 2}
can be diagonalized by going to the Fourier representation of the
operators $\hat{a}_{i}$, after which one obtains the Hamiltonian
\ref{eq:SW Ham} --- in this case with $B_{0}=0$ and $\omega(\mathbf{k})$
the corresponding one for nearest neighbors interaction. Here we pay
attention to the new terms, for simplicity we look at one of them,
e.g. the term containing $\hat{a}_{j}^{\dagger}\hat{a}_{i}^{\dagger}\hat{a}_{i}\hat{a}_{i}$.
We denote with $\mathbf{\Delta}$ the nearest-neighbor vector. Hence
\begin{align*}
\frac{J}{4}\sum_{\langle ij\rangle}\hat{a}_{j}^{\dagger}\hat{a}_{i}^{\dagger}\hat{a}_{i}\hat{a}_{i} & =\frac{J}{4}\sum_{\mathbf{R}_{j},\mathbf{\Delta}}\hat{a}_{j}^{\dagger}\hat{a}_{j+\Delta}^{\dagger}\hat{a}_{j+\Delta}\hat{a}_{j+\Delta}\\
 & =\frac{J}{4N^{2}}\sum_{\mathbf{R}_{j},\mathbf{\Delta}}\sum_{\mathbf{k_{1}},\mathbf{k}_{\mathbf{2}},\mathbf{k_{3}},\mathbf{k_{4}}}\hat{a}_{\mathbf{k_{1}}}^{\dagger}e^{-i\mathbf{k_{1}}\cdot\mathbf{R}_{j}}\hat{a}_{\mathbf{k_{2}}}^{\dagger}e^{-i\mathbf{k_{2}}\cdot\left(\mathbf{R}_{j}+\mathbf{\Delta}\right)}\times\\
 & \times\hat{a}_{\mathbf{k_{3}}}e^{i\mathbf{k_{3}}\cdot\left(\mathbf{R}_{j}+\mathbf{\Delta}\right)}\hat{a}_{\mathbf{k_{4}}}e^{i\mathbf{k_{4}}\cdot\left(\mathbf{R}_{j}+\mathbf{\Delta}\right)}\\
 & =\frac{J}{4N^{2}}\sum_{\mathbf{R}_{j},\mathbf{\Delta}}\sum_{\mathbf{k_{1}},\mathbf{k}_{\mathbf{2}},\mathbf{k_{3}},\mathbf{k_{4}}}e^{-i\left(\mathbf{k_{1}}+\mathbf{k}_{\mathbf{2}}-\mathbf{k_{3}}-\mathbf{k_{4}}\right)\cdot\mathbf{R}_{j}}\times\\
 & \times\hat{a}_{\mathbf{k_{1}}}^{\dagger}\hat{a}_{\mathbf{k_{2}}}^{\dagger}\hat{a}_{\mathbf{k_{3}}}\hat{a}_{\mathbf{k_{4}}}e^{-i\mathbf{\left(\mathbf{k}_{\mathbf{2}}-\mathbf{k_{3}}-\mathbf{k_{4}}\right)}\cdot\mathbf{\Delta}}\\
 & =\frac{J}{4N}\sum_{\mathbf{k_{1}},\mathbf{k}_{\mathbf{2}},\mathbf{k_{3}},\mathbf{k_{4}}}\delta_{\mathbf{k_{1}}+\mathbf{k}_{\mathbf{2}},\mathbf{k_{3}}+\mathbf{k_{4}}}\hat{a}_{\mathbf{k_{1}}}^{\dagger}\hat{a}_{\mathbf{k_{2}}}^{\dagger}\hat{a}_{\mathbf{k_{3}}}\hat{a}_{\mathbf{k_{4}}}\times\\
 & \times\sum_{\mathbf{\Delta}}e^{-i\mathbf{\left(\mathbf{k}_{\mathbf{2}}-\mathbf{k_{3}}-\mathbf{k_{4}}\right)}\cdot\mathbf{\Delta}}\\
 & =\frac{J}{4N}\sum_{\mathbf{k_{1}},\mathbf{k}_{\mathbf{2}},\mathbf{k_{3}},\mathbf{k_{4}}}\delta_{\mathbf{k_{1}}+\mathbf{k}_{\mathbf{2}},\mathbf{k_{3}}+\mathbf{k_{4}}}\hat{a}_{\mathbf{k_{1}}}^{\dagger}\hat{a}_{\mathbf{k_{2}}}^{\dagger}\hat{a}_{\mathbf{k_{3}}}\hat{a}_{\mathbf{k_{4}}}\sum_{\mathbf{\Delta}}e^{i\mathbf{\mathbf{k}_{\mathbf{1}}}\cdot\mathbf{\Delta}}\,.
\end{align*}
The last sum is simply a function of $\mathbf{k}_{1}$, which can
be given explicitly once the lattice is known. For example for a cubic
lattice of lattice constant $a$ 
\begin{align*}
\gamma(\mathbf{\mathbf{k}_{\mathbf{1}}})=\sum_{\mathbf{\Delta}}e^{i\mathbf{\mathbf{k}_{\mathbf{1}}}\cdot\mathbf{\Delta}} & =e^{ia\mathbf{k}_{\mathbf{1}x}}+e^{-ia\mathbf{k}_{\mathbf{1}x}}+e^{ia\mathbf{k}_{\mathbf{1}y}}+e^{-ia\mathbf{k}_{\mathbf{1}y}}+e^{ia\mathbf{k}_{\mathbf{1}z}}+e^{-ia\mathbf{k}_{\mathbf{1}z}}\\
 & =2\left[\cos\left(a\mathbf{k}_{\mathbf{1}x}\right)+\cos\left(a\mathbf{k}_{\mathbf{1}y}\right)+\cos\left(a\mathbf{k}_{\mathbf{1}y}\right)\right]\,.
\end{align*}
The term $\hat{a}_{\mathbf{k_{1}}}^{\dagger}\hat{a}_{\mathbf{k_{2}}}^{\dagger}\hat{a}_{\mathbf{k_{3}}}\hat{a}_{\mathbf{k_{4}}}$
corresponds to two magnons with momentum $\mathbf{k}_{3}$ and $\mathbf{k}_{4}$
respectively being annihilated, and two magnons with momentum $\mathbf{k}_{1}$
and $\mathbf{k}_{2}$ being created in the interaction process. The
Kronecker delta $\delta_{\mathbf{k_{1}}+\mathbf{k}_{\mathbf{2}},\mathbf{k_{3}}+\mathbf{k_{4}}}$
ensures conservation of momentum, $\mathbf{k_{1}}+\mathbf{k}_{\mathbf{2}}=\mathbf{k_{3}}+\mathbf{k_{4}}$.
The other 4-magnon interaction terms in \ref{eq:H n. n. approx 2}
can be treated analogously. One can easily see that for long wavelength
magnons (i.e. $k$ small), the scattering cross section of such processes
go as $\left(ka\right)^{4}$ and is therefore small. Further interaction
terms (6-magnon, etc) are suppressed by factors of increasing order
in $1/S$ \cite{stancil_spin_2009}.

Including dipole-dipole interactions has two main effects: (i) it
modifies the dispersion relation $\omega(\mathbf{k})$, which in that
case depends on the angle between the wavevector $\mathbf{k}$ and
the equilibrium direction of the saturated spins, since the dipole-dipole
interaction is anisotropic. This gives rise to a \emph{spin-wave manifold}.
(ii) New 3-magnon momentum-conserving interaction terms, e.g. $\hat{a}_{\mathbf{k_{1}}}^{\dagger}\hat{a}_{\mathbf{k_{2}}}^{\dagger}\hat{a}_{\mathbf{k_{3}}}$
or $\hat{a}_{\mathbf{k_{1}}}\hat{a}_{\mathbf{k_{2}}}\hat{a}_{\mathbf{k_{3}}}^{\dagger}$
are allowed: one magnon can split into two, and vice-versa.

\subsection*{Check points}
\begin{itemize}
\item Why do we get magnon-magnon interactions? 
\item What conservation rules do they fulfill and where do they come from? 
\end{itemize}

\chapter{Magneto-Optical Effects\label{chap:Magneto-Optical-effects}}

In this chapter we will explore the interaction between light and
magnetism in magnetic insulators. The coupling mechanism is the Faraday
effect, in which the plane of polarization of the light rotates as
it goes through a magnetized material. In turn, the light exerts a
very tiny effective " magnetic field"
on the spins: this is called the \emph{inverse Faraday effect} and
it is an example of \emph{backaction}. In what follows we will go
back to the classical realm to obtain the coupling term. This will
allows us to, by proper quantization of the classical coupling energy
term, obtain a coupling Hamiltonian between magnons and the quanta
of light, photons.

\section{Electromagnetic energy and zero-loss condition \label{sec:Electromagnetic-energy-and}}

We go back now to the full Maxwell equations in matter, in contrast
to the magnetostatic approximation we used throughout Chapter \ref{chap:Electromagnetism}
\begin{align}
\nabla\times\mathbf{H} & =\frac{\partial\mathbf{D}}{\partial t}+\mathbf{j}_{{\rm F}}\label{eq:curl H}\\
\nabla\times\mathbf{E} & =-\frac{\partial\mathbf{B}}{\partial t}\label{eq:curl E}\\
\nabla\cdot\mathbf{D} & =\rho\label{eq:div D}\\
\nabla\cdot\mathbf{B} & =0\,.\label{eq:div B-1}
\end{align}
These equations describe completely an electromagnetic system once
we give the constitutive equations 
\begin{align}
D_{i} & =\varepsilon_{ij}E_{j}\label{eq:constitutive-2}\\
B_{i} & =\mu_{ij}H_{j}\nonumber 
\end{align}
where we used the Einstein convention of summation over repeated indices.
These constitutive equations assume an instantaneous response of the
system: the response does not depend on time and the system has no
memory. This is referred to as a \emph{dispersionless}. For an isotropic
system, the permittivity and permeability tensors are diagonal and
proportional to the identity, $\varepsilon_{ij}=\varepsilon_{0}\varepsilon_{r}\delta_{ij}$,
$\mu_{ij}=\mu_{0}\mu_{r}\delta_{ij}$ and Eqs. \ref{eq:constitutive-2}
reduce to the scalar versions, $\mathbf{D}=\varepsilon_{0}\varepsilon_{r}\mathbf{E}$
and $\mathbf{B}=\mu_{0}\mu_{r}\mathbf{H}$.

We will argue that we can represent the coupling of light and magnetization
just by using the permittivity tensor. Our aim now is to obtain symmetry
conditions on the permittivity tensor $\varepsilon_{ij}$ in the presence
of a static magnetization in the material where the light propagates.
For that we will use conservation of electromagnetic energy, given
by a continuity equation involving the energy flux density. The instantaneous
electromagnetic power per unit area is given by the Poynting vector
\begin{equation}
\mathbf{P}=\mathbf{E}\times\mathbf{H}\,.\label{eq:Poynting}
\end{equation}
If we consider volume $V$ bounded by a surface $\mathcal{S}$, the
energy per unit time entering the volume is given by 
\[
-\oint_{\mathcal{S}}\mathbf{P}\cdot{\rm d}\mathbf{s}
\]
where ${\rm d}\mathbf{s}$ is an area element with vector pointing
outwards. This power can be stored in the volume in the form of an
energy density $W$, or dissipated - we denominate the dissipated
power $P_{d}$. We can hence write 
\[
-\int_{V}\nabla\cdot\mathbf{P}{\rm d}^{3}r=\int_{V}\frac{\partial W}{\partial t}{\rm d}^{3}r+\int_{V}P_{d}{\rm d}^{3}r
\]
where on the LHS we have used the divergence theorem. Since the volume is
arbitrary, we have 
\[
\nabla\cdot\mathbf{P}+\frac{\partial W}{\partial t}+P_{d}=0
\]
which has the form of a continuity equation. It remains to identify
the terms $W$ and $P_{d}$ in terms of the electromagnetic fields.
For that we look at the Maxwell equations and we see that by taking
the scalar product of \ref{eq:curl H} with $\mathbf{E}$, of \ref{eq:curl E}
with $\mathbf{H}$, and subtracting \ref{eq:curl E} from \ref{eq:curl H},
we can obtain an equation for the Poynting vector $\mathbf{P}$ by
using the vector identity 
\begin{equation}
\nabla\cdot\left(\mathbf{A}\times\mathbf{C}\right)=\mathbf{C}\cdot\left(\nabla\times\mathbf{A}\right)-\mathbf{A}\cdot\left(\nabla\times\mathbf{C}\right)\,.\label{eq:vector identity}
\end{equation}
Putting all together we obtain the \emph{Poynting theorem} 
\begin{align}
\nabla\cdot\left(\mathbf{E}\times\mathbf{H}\right)+\mathbf{H}\cdot\frac{\partial\mathbf{B}}{\partial t}+\mathbf{E}\cdot\frac{\partial\mathbf{D}}{\partial t}+\mathbf{E}\cdot\mathbf{j}_{{\rm F}} & =0\,\label{eq:Poynting Theo}\\
\nabla\cdot\left(\mathbf{E}\times\mathbf{H}\right)+H_{i}\frac{\partial\left(\mu_{ij}H_{j}\right)}{\partial t}+E_{i}\frac{\partial\left(\varepsilon_{ij}D_{j}\right)}{\partial t}+\mathbf{E}\cdot\mathbf{j}_{{\rm F}} & =0\,,
\end{align}
where in the last line we have used Eqs. \ref{eq:constitutive-2}.
For dispersionless, isotropic media, we obtain 
\[
\nabla\cdot\left(\mathbf{E}\times\mathbf{H}\right)+\frac{\mu_{0}\mu_{r}}{2}\frac{\partial H^{2}}{\partial t}+\frac{\varepsilon_{0}\varepsilon_{r}}{2}\frac{\partial E^{2}}{\partial t}+\mathbf{E}\cdot\mathbf{j}_{{\rm F}}=0\,,
\]
from where we can identify 
\begin{align*}
P_{d} & =\mathbf{E}\cdot\mathbf{j}_{{\rm F}}\\
W & =\frac{\mu_{0}\mu_{r}}{2}H^{2}+\frac{\varepsilon_{0}\varepsilon_{r}}{2}E^{2}
\end{align*}
as the dissipated power density and instantaneous energy density stored
in the magnetic and electric fields, respectively.

We are interested however in time-averaged quantities. To proceed
further we consider for simplicity monochromatic fields in complex
notation 
\begin{align*}
\mathbf{E}(t) & ={\rm Re}\left\{ \mathbf{E}(\omega)e^{-i\omega t}\right\} \\
\mathbf{H}(t) & ={\rm Re}\left\{ \mathbf{H}(\omega)e^{-i\omega t}\right\} \,.
\end{align*}
It can be easily shown that the time average of the product of two
oscillating fields $A(t)=A_{0}\cos(\omega t)$, $B(t)=B_{0}\cos(\omega t+\phi)$
over one period $T=2\pi/\omega$ is given simply by 
\begin{equation}
\langle A(t)B(t)\rangle_{T}=\frac{1}{2}{\rm Re}\left\{ \tilde{A}\tilde{B}^{*}\right\} \label{eq:time average}
\end{equation}
where 
\begin{align*}
A(t) & ={\rm Re}\left\{ A_{0}e^{-i\omega t}\right\} ={\rm Re}\left\{ \tilde{A}e^{-i\omega t}\right\} \\
B(t) & ={\rm Re}\left\{ B_{0}e^{-i\phi}e^{-i\omega t}\right\} ={\rm Re}\left\{ \tilde{B}e^{-i\omega t}\right\} \,.
\end{align*}
As a rule, one works with the complex fields and takes the real part
at the end of the calculation. In an abuse of notation, the tilde
notation is dropped. We will use Eq.~\ref{eq:time average} to obtain
a time average of the Poynting theorem given in Eq.~\ref{eq:Poynting Theo}.
For that we write Maxwell equations in frequency space, in particular
\begin{align}
\nabla\times\mathbf{H} & =-i\omega\mathbf{D}+\mathbf{j}_{{\rm F}}\label{eq:curl H w}\\
\nabla\times\mathbf{E} & =i\omega\mathbf{B}\,.\label{eq:curl E w}
\end{align}
To obtain the Poynting theorem in complex form, we take the complex
conjugate of Eq.~\ref{eq:curl H w} and perform the scalar product
with $\mathbf{E}$, and take the scalar product of Eq.~\ref{eq:curl E w}
with $\mathbf{H}^{*}$. Subtracting the resulting equations and using
again the vector identity \ref{eq:vector identity} we obtain 
\[
\nabla\cdot\left(\mathbf{E}\times\mathbf{H}^{*}\right)+i\omega\left(\mathbf{E}\cdot\mathbf{D}^{*}-\mathbf{H}^{*}\cdot\mathbf{B}\right)+\mathbf{E}\cdot\mathbf{j}_{{\rm F}}^{*}=0\,,
\]
form where the time average is easily obtained as 
\begin{equation}
{\rm Re}\left\{ \nabla\cdot\left(\mathbf{E}\times\mathbf{H}^{*}\right)+i\omega\left(\mathbf{E}\cdot\mathbf{D}^{*}-\mathbf{H}^{*}\cdot\mathbf{B}\right)+\mathbf{E}\cdot\mathbf{j}_{{\rm F}}^{*}\right\} =0\,.\label{eq:Poynting complex}
\end{equation}
For lossless media, $\langle\nabla\cdot\left(\mathbf{E}\times\mathbf{H}^{*}\right)\rangle_{T}$
must vanish, since all power that enters a volume must leave within
one cycle. Moreover, if there are no free currents, the dissipated
power $\mathbf{E}\cdot\mathbf{j}_{{\rm F}}^{*}$ is also zero. Therefore,
in lossless media 
\begin{equation}
{\rm Re}\left\{ i\omega\left(\mathbf{E}\cdot\mathbf{D}^{*}-\mathbf{H}^{*}\cdot\mathbf{B}\right)\right\} =0\,.\label{eq:Poynting average}
\end{equation}
Moreover, in the frequency domain dispersive effects are included
in a simple way by frequency-dependent permittivity and permeability
tensors 
\begin{align*}
\mathbf{D}(\omega) & =\bar{\varepsilon}(\omega)\cdot\mathbf{E}(\omega)\\
\mathbf{B}(\omega) & =\bar{\mu}(\omega)\cdot\mathbf{H}(\omega)
\end{align*}
where the bar indicates that $\bar{\varepsilon}(\omega)$, $\bar{\mu}(\omega)$
are matrices. Eq.~\ref{eq:Poynting average} then can be written
as \cite{LandauLifschitz,stancil_spin_2009} 
\begin{equation}
\frac{i\omega}{2}\left[\mathbf{E}^{*}\cdot\left(\bar{\varepsilon}^{\dagger}-\bar{\varepsilon}\right)\cdot\mathbf{E}+\mathbf{H}^{*}\cdot\left(\bar{\mu}^{\dagger}-\bar{\mu}\right)\cdot\mathbf{H}\right]=0\label{eq:Poynting final}
\end{equation}
where $\dagger$ indicates complex conjugate and transpose: $(\varepsilon_{ij})^{\dagger}=\varepsilon_{ji}^{*}$
(note that the same expression can be obtained directly from Eq.~\ref{eq:Poynting Theo}
by replacing the real fields using ${\rm Re}\left\{ z\right\} =(z+z^{*})/2$
and noting that $\langle zz\rangle_{T}=\langle z^{*}z^{*}\rangle_{T}=0$).
We deduce therefore that for lossless media 
\begin{align}
\bar{\varepsilon}^{\dagger} & =\bar{\varepsilon}\label{eq:lossless cond}\\
\bar{\mu}^{\dagger} & =\bar{\mu\,,}\nonumber 
\end{align}
that is, the permittivity and permeability must be Hermitian matrices.
Note that if the material is isotropic and $\bar{\varepsilon}(\omega)$,
$\bar{\mu}(\omega)$ can be written as in principle complex scalars
$\varepsilon(\omega)=\varepsilon'(\omega)+i\varepsilon''(\omega)$,
$\mu(\omega)=\mu'(\omega)+i\mu''(\omega)$, the zero-loss condition
implies that the imaginary parts $\varepsilon''(\omega)$ and $\mu''(\omega)$
must vanish.

To define the average electromagnetic energy density in the presence
of dispersion is a little bit more subtle \cite{LandauLifschitz}.
We give here for completeness the corresponding expression without
proof 
\begin{equation}
\langle W\rangle_{T}=\frac{1}{4}\left[\mathbf{E}^{*}\cdot\frac{\partial\left(\omega\bar{\varepsilon}\right)}{\partial\omega}\cdot\mathbf{E}+\mathbf{H}^{*}\cdot\frac{\partial\left(\omega\bar{\mu}\right)}{\partial\omega}\cdot\mathbf{H}\right]\,.\label{eq:energy dispersive}
\end{equation}
If $\bar{\varepsilon}$ and $\bar{\mu}$ are independent of frequency,
this expression reduces to 
\begin{equation}
\langle W\rangle_{T}=\frac{1}{4}\left[\mathbf{E}^{*}\cdot\bar{\varepsilon}\cdot\mathbf{E}+\mathbf{H}^{*}\cdot\bar{\mu}\cdot\mathbf{H}\right]\label{eq:energy non dispersive}
\end{equation}
as expected. 
\begin{enumerate}
\item \textbf{\emph{Exercise: Prove Eq.~\ref{eq:Poynting final} starting
from \ref{eq:Poynting average}.}} 
\end{enumerate}

\subsection*{Check points}
\begin{itemize}
\item Obtain Eq.~\ref{eq:Poynting complex} from Maxwell equations 
\item What is the zero loss condition? 
\item What does it tell us about the symmetries of the permittivity tensor? 
\end{itemize}

\section{Permittivity tensor and magnetization\label{sec:Permittivity-tensor-and}}

In the following section we will use the permittivity tensor $\bar{\varepsilon}$
to describe the Faraday effect in a magnetized sample. For that we
will use the symmetry properties of $\bar{\varepsilon}$ in the presence
of a permanent magnetization, which breaks time reversal invariance.
If we write the permittivity tensor explicitly separating the real
and imaginary parts 
\[
\varepsilon_{ij}=\varepsilon'_{ij}+i\varepsilon''_{ij}\,,
\]
the zero-loss condition \ref{eq:lossless cond} tells us that real
and imaginary parts are respectively symmetric and antisymmetric matrices:
\begin{align}
\varepsilon'_{ij}(\mathbf{M}) & =\varepsilon'_{ji}(\mathbf{M})\label{eq:zero loss eps}\\
\varepsilon''_{ij}(\mathbf{M}) & =-\varepsilon''_{ji}(\mathbf{M})\,,\nonumber 
\end{align}
where we have made explicit a possible dependence on the magnetization
$\mathbf{M}$. On the other hand, Onsager reciprocity relations for
response functions dictate how the permittivity transforms under time
reversal symmetry 
\begin{align}
\varepsilon'_{ij}(\mathbf{M}) & =\varepsilon'_{ji}(-\mathbf{M})\label{eq:Onsager}\\
\varepsilon''_{ij}(\mathbf{M}) & =\varepsilon''_{ji}(-\mathbf{M})\,,\nonumber 
\end{align}
where the time reversed form of $\bar{\varepsilon}$ consists in transposing
the matrix and at the same time inverting the magnetization vector.
We see therefore that the real and imaginary parts are also symmetric
and antisymmetric in the magnetization. Putting all together we obtain
\begin{align}
\varepsilon'_{ij}(\mathbf{M}) & =\varepsilon'_{ji}(\mathbf{M})=\varepsilon'_{ji}(-\mathbf{M})\label{eq:conds eps}\\
\varepsilon''_{ij}(\mathbf{M}) & =-\varepsilon''_{ji}(\mathbf{M})=\varepsilon''_{ji}(-\mathbf{M})\,.\nonumber 
\end{align}
In linear response, the permittivity depends linearly on the magnetization.
This is valid as long as the effect of the magnetization on the permittivity
is small. To fulfill conditions \ref{eq:conds eps} to first order
in the magnetization we write \cite{LandauLifschitz} 
\begin{equation}
\varepsilon_{ij}(\mathbf{M})=\varepsilon_{0}\left(\varepsilon_{r}\delta_{ij}-if\epsilon_{ijk}M_{k}\right)\label{eq:eps M}
\end{equation}
where we have assumed the material is isotropic, and $f$ is a small
material-dependent parameter related to the Faraday rotation as we
show below. A material for which the permittivity takes this form
is denominated \emph{gyrotropic}. 
\begin{enumerate}
\item \textbf{\emph{Exercise: Prove that Eq.~\ref{eq:eps M} fulfills \ref{eq:conds eps}.
}} 
\end{enumerate}

\subsection*{Check points}
\begin{itemize}
\item Explain how Eq.~\ref{eq:eps M} is obtained 
\end{itemize}

\section{Faraday effect\label{sec:Faraday-effect}}

For optical frequencies one can usually safely take the permeability
of a dielectric as the vacuum permeability $\mu_{0}$, even for a
magnetic material. This amounts to neglecting the interaction of the
small oscillating magnetic field part of the optical electromagnetic
field with the material. In turn, the interaction between the electric
part of the optical field and the magnetization in the sample is modeled
by the permittivity given in Eq.~\ref{eq:eps M}. The magnetization
$\mathbf{M}$, even if it has a time dependence, it it much slower
than the optical fields and therefore it is well defined.

To understand how the magnetization dependent permittivity in Eq.
\ref{eq:eps M} encapsulates the Faraday effect we will first derive
the Fresnel equation for the optical field, starting from the Maxwell
equations \ref{eq:curl H w} and \ref{eq:curl E w} in the absence
of free currents, $\mathbf{j}_{{\rm F}}=0$. We are interested in
light propagating through a material, we therefore write the electric
and magnetic fields as plane waves of frequency $\omega$ and wavevector
$\mathbf{k}$ 
\begin{align*}
\mathbf{E}(t,\mathbf{r}) & =\mathbf{E}e^{-i(\omega t-\mathbf{k}\cdot\mathbf{r})}\\
\mathbf{H}(t,\mathbf{r}) & =\mathbf{H}e^{-i(\omega t-\mathbf{k}\cdot\mathbf{r})}\,.
\end{align*}
In momentum and frequency representation, Eqs. \ref{eq:curl H w}
and \ref{eq:curl E w} read 
\begin{align}
\mathbf{k}\times\mathbf{E} & =\mu_{0}\omega\mathbf{H}\label{eq:k E H}\\
\mathbf{k}\times\mathbf{H} & =-\omega\mathbf{D}\,.\label{eq:k H D}
\end{align}
Inserting Eq.~\ref{eq:k E H} into \ref{eq:k H D} , using $\mathbf{D}=\bar{\varepsilon}\cdot\mathbf{E}$
and the product rule $\mathbf{a}\times(\mathbf{b}\times\mathbf{c})=\mathbf{b}(\mathbf{a}\cdot\mathbf{c})-\mathbf{c}(\mathbf{a}\cdot\mathbf{b})$
one obtains 
\begin{equation}
\frac{k^{2}}{\mu_{0}\omega^{2}}\left[\mathbf{E}-\frac{\mathbf{k}(\mathbf{k}\cdot\mathbf{E})}{k^{2}}\right]=\bar{\varepsilon}\cdot\mathbf{E}\,.\label{eq:Fresnel E}
\end{equation}
This is a form of the Fresnel equation, and it determines the dispersion
relation of the electromagnetic wave (that is, $\omega(\mathbf{k})$)
by imposing the determinant of its coefficients to be zero. In components
\[
\frac{k^{2}}{\mu_{0}\omega^{2}}\left[\delta_{ij}E_{j}-\frac{k_{i}k_{j}E_{j}}{k^{2}}\right]=\varepsilon_{ij}E_{j}
\]
and therefore 
\[
{\rm det}\left\{ \frac{k^{2}}{\mu_{0}\omega^{2}}\left[\delta_{ij}-\frac{k_{i}k_{j}}{k^{2}}\right]-\varepsilon_{ij}\right\} =0\,.
\]

The term $\mathbf{k}(\mathbf{k}\cdot\mathbf{E})/k^{2}$ gives us simply
the projection of the electric field along the propagation direction
$\mathbf{k}/k$ . Whereas in vacuum the electric field\textbf{ $\mathbf{E}$}
is purely transverse, in a medium the purely transverse field is actually
$\mathbf{D}$, see Eq.~\ref{eq:k H D}. We will work now however
with the particular form of the permittivity given in Eq.~\ref{eq:eps M},
and consider the direction of propagation $\mathbf{k}/k$ to coincide
with the magnetization axis, $\mathbf{M}=M\hat{z}$, in which case
the $\mathbf{E}$ field is also transverse as one can easily verify
by using the resulting form of the permittivity tensor 
\begin{equation}
\bar{\varepsilon}=\varepsilon_{0}\left(\begin{array}{ccc}
\varepsilon_{r} & -ifM & 0\\
ifM & \varepsilon_{r} & 0\\
0 & 0 & \varepsilon_{r}
\end{array}\right)\,.\label{eq:eps Mz}
\end{equation}
The Fresnel equation reduces therefore to 
\[
\left|\left(\begin{array}{cc}
\frac{k^{2}}{\mu_{0}\omega^{2}}-\varepsilon_{0}\varepsilon_{r} & ifM\\
-ifM & \frac{k^{2}}{\mu_{0}\omega^{2}}-\varepsilon_{0}\varepsilon_{r}
\end{array}\right)\right|=0
\]
with solutions 
\begin{equation}
k_{\pm}^{2}=\left(\frac{\omega}{c}\right)^{2}\left(\varepsilon_{r}\mp fM\right)\label{eq:k pm}
\end{equation}
wehere we have used that $c=1/\sqrt{\varepsilon_{0}\mu_{0}}$. Inserting
these solutions back into Eq.~\ref{eq:Fresnel E} with $\bar{\varepsilon}$
given by Eq.~\ref{eq:eps Mz} (in this case only two dimensional,
since $E_{z}=0$) we obtain 
\[
E_{x}=\mp iE_{y}\,,
\]
with $\mp$ corresponding to $k_{\pm}^{2}$. We have therefore obtained
two solutions for the propagating wave along $\hat{z}$, with the
same amplitude and cicularly polarized in the $xy$ plane, but with
opposite polarizations for $k_{+}$ and $k_{-}$: 
\begin{equation}
\mathbf{E}_{\pm}(z,t)=E_{0}{\rm Re}\left\{ \left(\hat{\mathbf{e}}_{x}\pm i\hat{\mathbf{e}}_{y}\right)e^{i\left(k_{\pm}z-\omega t\right)}\right\} \,.\label{eq:E circ pol}
\end{equation}

\begin{figure}
\centering{}\includegraphics[width=0.8\textwidth]{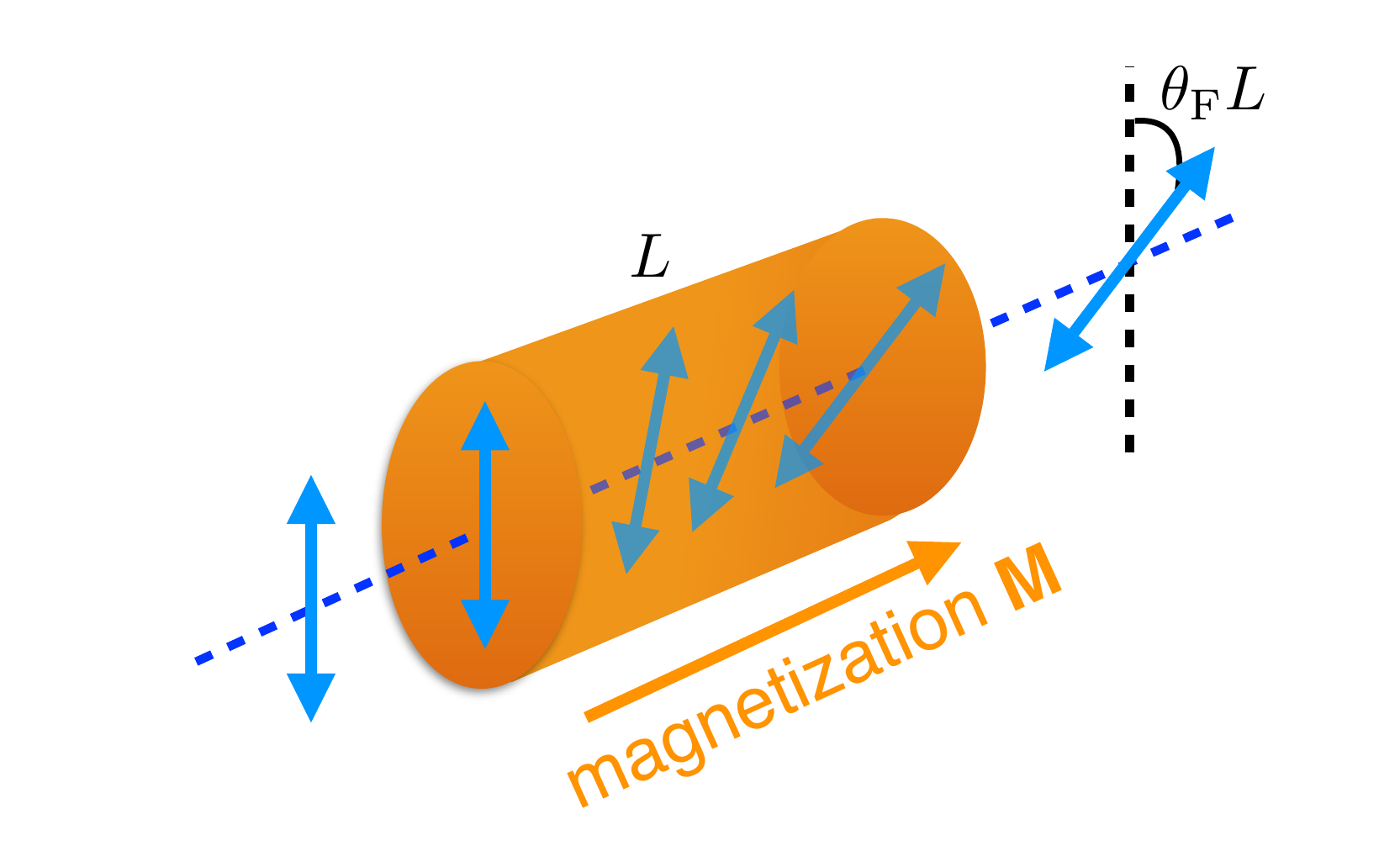}\caption{Vertically polarized light rotates its angle of polarization as it
goes through a magnetized material.}
\label{Fig:FaradayRot} 
\end{figure}

To derive the Faraday rotation we consider now an EM wave propagating
in the medium such that at $z=0$ it is linearly polarized along $\hat{x}$
with amplitude $E_{0}$, 
\begin{equation}
\mathbf{E}(z=0,t)=E_{0}\hat{\mathbf{e}}_{x}e^{-i\omega t}\,.\label{eq:E init cond}
\end{equation}
The linear combination of Eqs. \ref{eq:E circ pol} 
\[
\mathbf{E}(z,t)=\frac{1}{2}\left[\mathbf{E}_{+}(z,t)+\mathbf{E}_{-}(z,t)\right]
\]
fulfills the condition \ref{eq:E init cond}. We therefore have as
solution for the propagating wave $\mathbf{E}(z,t)={\rm Re}\left\{ E_{x}\hat{\mathbf{e}}_{x}+E_{y}\hat{\mathbf{e}}_{y}\right\} $
\begin{align*}
E_{x} & =\frac{E_{0}}{2}\left(e^{ik_{+}z}+e^{ik_{-}z}\right)\\
E_{y} & =i\frac{E_{0}}{2}\left(e^{ik_{+}z}-e^{ik_{-}z}\right)\,.
\end{align*}
After the wave propagated a distance $L$ through the material 
\[
\frac{E_{y}}{E_{x}}=\tan\left[\left(\frac{k_{-}-k_{+}}{2}\right)L\right]\,,
\]
which indicates that the plane of polarization of the light rotated
by an amount $\theta_{{\rm F}}L$, where 
\[
\theta_{{\rm F}}=\frac{k_{-}-k_{+}}{2}
\]
is the \emph{Faraday rotation per unit length}. This is depicted schematically
in Fig. \ref{Fig:FaradayRot}. 

Using Eq.~\ref{eq:k pm} and $fM\ll\varepsilon_{r}$ we obtain 
\begin{equation}
\theta_{{\rm F}}=\frac{\omega}{2c\sqrt{\epsilon_{r}}}fM\,.\label{eq:theta F}
\end{equation}

\subsection*{Check points}
\begin{itemize}
\item What is the Faraday effect? 
\end{itemize}

\section{Magneto-optical energy\label{sec:Magneto-optical-energy}}

In the previous sections we saw that the magnetization in a medium
modifies the permittivity tensor, which acquires an antisymmetric
imaginary term due to the breaking of time reversal symmetry. As light
goes through the material, it experiences this effective permittivity
which we saw leads to the Faraday effect. Light and magnetization
in the sample are therefore coupled. To obtain this coupling, we look
at the electromagnetic energy obtained from Eq.~\ref{eq:energy non dispersive}
by using the permittivity Eq.~\ref{eq:eps M}. We see that the magnetization-dependent
part of the permittivity introduces a correction to the usual electromagnetic
energy expression, given by 
\begin{equation}
U_{{\rm MO}}=-\frac{i}{4}f\varepsilon_{0}\int{\rm d}^{3}r\mathbf{M}(\mathbf{r})\cdot\left[\mathbf{E}^{*}(\mathbf{r})\times\mathbf{E}(\mathbf{r})\right]\,.\label{eq:MO energy}
\end{equation}
One can easily prove that this term is real. In terms of the Faraday
rotation per unit length this can be rewritten as 
\begin{equation}
U_{{\rm MO}}=\theta_{{\rm F}}\sqrt{\frac{\varepsilon_{r}}{\varepsilon_{0}}}\int{\rm d}^{3}r\frac{\mathbf{M}(\mathbf{r})}{M}\cdot\frac{\varepsilon_{0}}{2i\omega}\left[\mathbf{E}^{*}(\mathbf{r})\times\mathbf{E}(\mathbf{r})\right]\,.\label{eq:MO en OSD}
\end{equation}
The term 
\begin{equation}
\mathbf{S}_{{\rm light}}(\mathbf{r})=\frac{\varepsilon_{0}}{2i\omega}\left[\mathbf{E}^{*}(\mathbf{r})\times\mathbf{E}(\mathbf{r})\right]\label{eq:OSD}
\end{equation}
is called the \emph{optical spin density and }it is related to the\emph{
helicity }of light. For example, for circularly polarized light $\mathbf{S}_{{\rm light}}$
points perpendicular to the plane of polarization with a direction
given by the right-hand-rule.

We know from the previous section that the magnetization causes the
plane of polarization of the light to rotate. From Eq.~\ref{eq:MO energy}
we see that the light itself acts as an effective magnetic field on
the magnetization (compare with the usual expression for the Zeeman
energy). This gives rise to the \emph{inverse Faraday effect}, which
takes into account the effect of the light on the magnetization dynamics
of the sample. This effect is usually small, one can show that the
effective light-induced magnetic field for YIG (Yttrium Iron Garnet,
a magnetic insulator, widely used in both technical applications and
current experiments) is of the order of $10^{-11}$T per photon/$\mu^{3}$.
For comparison, the earth's magnetic field is of the order of $10^{-6}$T!
We will see however in the following chapter that this value can be
enhanced by using an optical cavity, effectively " trapping"
photons. 
\begin{enumerate}
\item \textbf{\emph{Exercise: Derive Eq.~\ref{eq:MO en OSD} starting from
\ref{eq:energy non dispersive}.}} 
\end{enumerate}

\subsection*{Check points}
\begin{itemize}
\item Obtain the correction to the electromagnetic energy if a medium is
magnetized. 
\end{itemize}

\chapter{Modern Topics: Cavity Optomagnonics}

In this last chapter we will put together all concepts we learned
so far to derive the basics of a topic of current research: Cavity
Optomagnonics. In these systems, a magnetic insulator material forms
a cavity for the light, which is used to enhance the magnon-photon
coupling. We call this the \emph{optomagnonic coupling}. As we pointed
out in the previous chapter, the Faraday effect is usually a small
effect, which depends on the Faraday rotation constant of the material
and also on the path's length of the light inside of the material.
Since in a cavity the light is " trapped" ,
this effectively enhances the path's length and therefore the coupling.
This is an intuitive way of seeing the enhancement, we will see this
more formally in the next sections. For that we need first to learn
about optical cavities and the quantization of the electromagnetic
field.

\section{Quantization of the electromagnetic field\label{sec:Quantization-of-the}}

We start by quantizing a single-mode field in a cavity formed by perfectly
conducting walls at $z=0$ and $z=L$. We summarize here the most
important concepts, a more thorough discussion can be found for example
in Refs. \cite{cohen-tannoudji_photons_1997,gerry_introductory_2004}.
We assume further the electric field to be polarized along $x$, $\mathbf{E}=E_{x}(z,t)\hat{\mathbf{e}}_{x}$.
The boundary condition therefore implies 
\begin{equation}
E_{x}(z=0,t)=E_{x}(z=L,t)=0\,.\label{eq:BC ex}
\end{equation}
From Maxwell equations in vacuum and no sources, 
\begin{align}
\nabla\times\mathbf{E} & =-\frac{\partial\mathbf{B}}{\partial t} & \nabla\cdot\mathbf{B} & =0\label{eq:rot E}\\
\nabla\times\mathbf{B} & =\mu_{0}\varepsilon_{0}\frac{\partial\mathbf{E}}{\partial t} & \nabla\cdot\mathbf{E} & =0\label{eq:rot B}
\end{align}
we obtain a wave equation for the electric field 
\begin{equation}
\nabla^{2}\mathbf{E}=\mu_{0}\varepsilon_{0}\frac{\partial^{2}\mathbf{E}}{\partial t^{2}}\label{eq:wave eq E}
\end{equation}
which in terms of $\mathbf{E}=E_{x}(z,t)\hat{\mathbf{e}}_{x}$ simplifies
to 
\begin{equation}
\frac{\partial^{2}E_{x}(z,t)}{\partial z^{2}}-\frac{1}{c^{2}}\frac{\partial^{2}E_{x}(z,t)}{\partial t^{2}}=0\,.\label{eq:wave eq Ex}
\end{equation}
The solution of Eq.~\ref{eq:wave eq Ex} satisfying the boundary
conditions in Eq.~\ref{eq:BC ex} is simply 
\begin{equation}
E_{x}(z,t)=\sqrt{\left(\frac{2\omega_{n}^{2}}{V\varepsilon_{0}}\right)}q(t)\sin\left(k_{n}z\right)\label{eq:Ex sol}
\end{equation}
with 
\begin{equation}
\frac{\omega_{n}}{c}=k_{n}=\frac{\pi n}{L}\,.\label{eq:kn}
\end{equation}
In Eq.~\ref{eq:Ex sol}, $q(t)$ has units of length and $V=LS$
is the volume of the cavity, where $S$ is its cross-section. From
the left Eq. in \ref{eq:rot B} one obtains $\mathbf{B}=B_{y}(z,t)\hat{\mathbf{e}}_{y}$
with 
\begin{equation}
B_{y}(z,t)=\frac{\mu_{0}\varepsilon_{0}}{k_{n}}\sqrt{\left(\frac{2\omega_{n}^{2}}{V\varepsilon_{0}}\right)}\dot{q}(t)\cos\left(k_{n}z\right)\,.\label{eq:By sol}
\end{equation}
Inserting Eqs. \ref{eq:Ex sol} and \ref{eq:By sol} into the electromagnetic
energy 
\[
E_{{\rm EM}}=\frac{1}{2}\int{\rm d}V\left(\varepsilon_{0}\mathbf{E}^{2}+\frac{1}{\mu_{0}}\mathbf{B}^{2}\right)
\]
one obtains 
\begin{equation}
E_{{\rm EM}}=\frac{1}{2}\left(\omega_{n}^{2}q^{2}+p^{2}\right)\label{eq:E EM osc}
\end{equation}
where we have defined $p=\dot{q}$ the canonical momentum of a "
particle" of unit mass. Eq.~\ref{eq:E EM osc} is the
energy of a harmonic oscillator of unit mass. We can now proceed to
quantize the theory by 
\begin{align*}
q & \rightarrow\hat{q}\\
p & \rightarrow\hat{p}
\end{align*}
and imposing the commutator $\left[\hat{q},\hat{p}\right]=i\hbar$.
It is convenient to introduce the bosonic creation and annihilation
operators 
\begin{align*}
\hat{a} & =\frac{1}{\sqrt{2\hbar\omega_{n}}}\left(\omega_{n}\hat{q}+i\hat{p}\right)\\
\hat{a}^{\dagger} & =\frac{1}{\sqrt{2\hbar\omega_{n}}}\left(\omega_{n}\hat{q}-i\hat{p}\right)
\end{align*}
which satisfy $\left[\hat{a},\hat{a}^{\dagger}\right]$ in terms of
which the electric and magnetic field can be expressed as 
\begin{align}
\hat{E}_{x}(z,t) & =\sqrt{\left(\frac{\hbar\omega_{n}}{V\varepsilon_{0}}\right)}\left(\hat{a}+\hat{a}^{\dagger}\right)\sin\left(k_{n}z\right)\label{eq:hat E}\\
\hat{B}_{y}(z,t) & =-i\sqrt{\left(\frac{\hbar\omega_{n}}{V\varepsilon_{0}}\right)}\left(\hat{a}-\hat{a}^{\dagger}\right)\cos\left(k_{n}z\right)\,.\label{eq:hat B}
\end{align}
The energy Eq.~\ref{eq:E EM osc} gives rise to the usual harmonic-oscillator
Hamiltonian 
\[
\hat{H}_{{\rm EM}}=\hbar\omega_{n}\left(\hat{a}^{\dagger}\hat{a}+\frac{1}{2}\right)\,.
\]
The time dependence of the ladder operators can be obtained from the
Heisenberg equation of motion and is given by 
\begin{align*}
\hat{a}(t) & =\hat{a}(0)e^{-i\omega t}\\
\hat{a}^{\dagger}(t) & =\hat{a}^{\dagger}(0)e^{i\omega t}\,.
\end{align*}
Note that an eigenstate of the number operator $\hat{n}=\hat{a}^{\dagger}\hat{a}$,
$\hat{n}|n\rangle=n|n\rangle$, is an energy eigenstate but the electric
field operator's expectation value vanishes 
\[
\langle n|\hat{E}_{x}|n\rangle=0
\]
and therefore it is not well defined. The expectation value squared
field $\langle n|\hat{E}_{x}^{2}|n\rangle$ is however finite, as
one can easily prove. This reflects the uncertainty in the phase of
the electric field, which is conjugate to the number operator. We
call the excitation with energy $\hbar\omega_{n}$ a \emph{photon}.

We now proceed to the quantization of multimode fields in a 3D cavity.
For that, it is convenient to use the Coulomb gauge 
\begin{equation}
\nabla\cdot\mathbf{A}(\mathbf{r},t)=0\,,\label{eq:Coulomb gauge}
\end{equation}
in which both electric and magnetic field can be expressed in terms
of the vector potential $\mathbf{A}(\mathbf{r},t)$ 
\begin{align}
\mathbf{E}(r,t) & =-\frac{\partial\mathbf{A}(\mathbf{r},t)}{\partial t}\label{eq:EdA}\\
\mathbf{B}(r,t) & =\nabla\times\mathbf{A}(\mathbf{r},t)\,.\label{eq:BrA}
\end{align}
From Maxwell's equations, one obtains a wave equation for the vector
potential 
\[
\nabla^{2}\mathbf{A}-\frac{1}{c^{2}}\frac{\partial^{2}\mathbf{A}}{\partial t}=0
\]
which can be solved by separating time end space variables. It is
customary to split the time dependence 
\[
\mathbf{A}(\mathbf{r},t)=\mathbf{A}^{+}(\mathbf{r},t)+\mathbf{A}^{-}(\mathbf{r},t)
\]
such that 
\begin{align*}
\mathbf{A}^{+}(\mathbf{r},t) & =\sum_{k}c_{k}\mathbf{u}_{k}(\mathbf{r})e^{-i\omega_{k}t}\\
\mathbf{A}^{-}(\mathbf{r},t) & =\sum_{k}c_{k}^{*}\mathbf{u}_{k}^{*}(\mathbf{r})e^{i\omega_{k}t}
\end{align*}
where $\omega_{k}\ge0$. The mode-functions $\mathbf{u}_{k}$ are
solutions of 
\[
\left(\nabla^{2}+\frac{\omega_{k^{2}}}{c^{2}}\right)\mathbf{u}_{k}(\mathbf{r})=0
\]
satisfying orthonormality 
\[
\int{\rm d}V\mathbf{u}_{k}^{*}(\mathbf{r})\mathbf{u}_{k'}(\mathbf{r})=\delta_{k,k'}
\]
and the appropriate boundary conditions. The index $k$ indicates
both the mode and the polarization vector. Moreover, due to the Coulomb
gauge 
\begin{equation}
\nabla\cdot\mathbf{u}_{k}(\mathbf{r})=0\,.\label{eq:div u}
\end{equation}
Guided by our one-mode example, we quantize replacing the amplitudes
in the sum over modes by creation and annihilation operators 
\[
\hat{\mathbf{A}}(\mathbf{r},t)=\sum_{k}\sqrt{\frac{\hbar}{2\omega_{k}\varepsilon_{0}}}\left[\hat{a}_{k}\mathbf{u}_{k}(\mathbf{r})e^{-i\omega_{k}t}+\hat{a}_{k}^{\dagger}\mathbf{u}_{k}\text{\^{ }*}(\mathbf{r})e^{i\omega_{k}t}\right]
\]
from which, using Eq.~\ref{eq:EdA}, we obtain the electric field
operator 
\begin{equation}
\hat{\mathbf{E}}(\mathbf{r},t)=\hat{\mathbf{E}}^{+}(\mathbf{r},t)+\hat{\mathbf{E}}^{-}(\mathbf{r},t)=i\sum_{k}\sqrt{\frac{\hbar\omega_{k}}{2\varepsilon_{0}}}\left[\hat{a}_{k}\mathbf{u}_{k}(\mathbf{r})e^{-i\omega_{k}t}-\hat{a}_{k}^{\dagger}\mathbf{u}_{k}\text{\^{ }*}(\mathbf{r})e^{i\omega_{k}t}\right]\label{eq:E op}
\end{equation}
with the same convention for $\hat{\mathbf{E}}^{\pm}(\mathbf{r},t)$
as for $\hat{\mathbf{A}}^{\pm}(\mathbf{r},t)$. Using Eq.~\ref{eq:BrA}
we can obtain the corresponding expression for the magnetic field.
The bosonic ladder operators $\hat{a}_{k}$, $\hat{a}_{k}^{\dagger}$
as defined are dimensionless and satisfy the usual commutation relations
\begin{align*}
\left[\hat{a}_{k},\hat{a}_{k'}\right] & =\left[\hat{a}_{k}^{\dagger},\hat{a}_{k'}^{\dagger}\right]=0\\
\left[\hat{a}_{k},\hat{a}_{k'}^{\dagger}\right] & =\delta_{k,k'}
\end{align*}
and the corresponding Hamiltonian is that of a collection of non-interacting
harmonic oscillators 
\[
H_{{\rm EM}}=\sum_{k}\hbar\omega_{k}\left(\hat{a}_{k}^{\dagger}\hat{a}_{k}+\frac{1}{2}\right)\,.
\]
The sum over modes has no cutoff and therefore the factor $1/2$ leads
to a divergence. In these notes we will not worry about that, since
we will always work with differences of energies, where this factor
cancels out. We will therefore simply omit this term in the following.
The energy eigenstates are also eigenstates of the number operator
$\hat{n}_{k}=\hat{a}_{k}^{\dagger}\hat{a}_{k}$ 
\begin{align*}
\hat{n}_{k}|n_{k}\rangle & =n_{k}|n_{k}\rangle\\
E_{n_{k}} & =\hbar\omega_{k}\left(n_{k}+\frac{1}{2}\right)\,.
\end{align*}
The states $|n_{k}\rangle$ form an orthonormal basis of the Hilbert
space and are called number or \emph{Fock} states, where $n_{k}$
gives the number of photons in state $k$. A general multi-mode state
can be written as 
\[
|\psi\rangle=\sum_{n_{1},n_{2},...}c_{n_{1},n_{2},...}|n_{1},n_{2},...\rangle\,.
\]

In the simplest example, one uses periodic boundary conditions in
a cubic box of side length $L$. In this case, 
\[
\mathbf{u}_{k}(\mathbf{r})=\frac{1}{L^{3/2}}\hat{\mathbf{e}}^{\lambda}e^{i\mathbf{k}\cdot\mathbf{r}}
\]
with $\mathbf{k}=2\pi/L\left(n_{x},n_{y},n_{z}\right)$, $n_{i}=0,\pm1,\pm2,...$
and $\lambda=1,2$ indicating the polarization, which from Eq.~\ref{eq:div u}
must fulfill 
\[
\hat{\mathbf{e}}^{\lambda}\cdot\mathbf{k}=0\,.
\]
One can analogously use reflecting boundary conditions, where the
solutions are standing waves as in the single mode we studied above.
Note that in this case the normalization factor of $\hat{\mathbf{E}}(\mathbf{r},t)$
and the quantization of $\mathbf{k}$ will be different. 
\begin{enumerate}
\item \textbf{\emph{Exercise: Derive Eq.~\ref{eq:wave eq E}}} 
\item \textbf{\emph{Exercise: Derive Eq.~\ref{eq:E EM osc}}} 
\end{enumerate}

\subsection*{Check points}
\begin{itemize}
\item Write a general expression for the quantized electric field 
\end{itemize}

\section{Optical cavities as open quantum systems\label{sec:Optical-Cavities-as}}

\label{sec:OCOQS}

Cavities are usually open systems, in contact with an environment
both because, for example, the mirrors are not perfect, allowing contact
to a bath of photons or/and phonons, and because we want to have access
to the cavity by means of an external probe. The environment (also
called bath, or reservoir) is assumed to be very large and in thermal
equilibrium, and it is modeled as a collection of harmonic oscillators.
The simplest Hamiltonian of the cavity plus bath is written as 
\[
\hat{H}=\hat{H}_{S}+\hat{H}_{R}+\hat{H}_{I}
\]
where $\hat{H}_{S}$ , $\hat{H}_{R}$, and $\hat{H}_{I}$ are the
cavity, reservoir, and interaction Hamiltonians respectively 
\begin{align*}
\hat{H}_{S} & =\hbar\Omega\hat{a}^{\dagger}\hat{a}\\
\hat{H}_{R} & =\sum_{k}\hbar\omega_{k}\hat{b}_{k}^{\dagger}\hat{b}_{k}\\
\hat{H}_{I} & =\hbar\sum_{k}\left(g_{k}\hat{a}^{\dagger}\hat{b}_{k}+g_{k}^{*}\hat{b}_{k}^{\dagger}\hat{a}\right)\,,
\end{align*}
where we have taken for simplicity only one mode for the cavity $\hat{a}$
with frequency $\Omega$. The interaction Hamiltonian represents an
excitation in the cavity being converted into one in the reservoir
and vice-versa, with coupling constant $g_{k}$.

Our aim is to obtain an effective equation of motion for the cavity
mode $\hat{a}$ which encapsulates the effect of the bath \cite{meystre_elements_2007}.
This procedure is denominated to \emph{integrate out }the bath, and
it means that, since we are not interested in the dynamics of the
bath per se, we want to eliminate these degrees of freedom and just
retain the ones we are interested in, in this case, the single cavity
mode. We begin by writing the Heisenberg equations of motion for both
the cavity mode and the reservoir modes 
\begin{align}
\dot{\hat{a}}(t) & =-i\Omega\hat{a}(t)-i\sum_{k}g_{k}\hat{b}_{k}(t)\label{eq:a dot}\\
\dot{\hat{b}}_{k}(t) & =-i\omega_{k}\hat{b}(t)-ig_{k}^{*}\hat{a}(t)\,.\label{eq:b dot}
\end{align}
We can integrate formally Eq.~\ref{eq:b dot} to obtain 
\begin{equation}
\hat{b}_{k}(t)=\hat{b}_{k}(0)e^{-i\omega_{k}t}-ig_{k}^{*}\int_{0}^{t}{\rm d}t'\hat{a}(t')e^{-i\omega_{k}\left(t-t'\right)}\,,\label{eq:bk int out}
\end{equation}
where the first term corresponds to the free evolution of $\hat{b}_{k}$
and the second one is due to the interaction with the cavity. Substituting
Eq.~\ref{eq:bk int out} into \ref{eq:a dot} we obtain 
\begin{equation}
\dot{\hat{a}}(t)=-i\Omega\hat{a}(t)-i\sum_{k}g_{k}\hat{b}_{k}(0)e^{-i\omega_{k}t}-\sum_{k}\left|g_{k}\right|^{2}\int_{0}^{t}{\rm d}t'\hat{a}(t')e^{-i\omega_{k}\left(t-t'\right)}\,.\label{eq:a dot int}
\end{equation}
We now transform the cavity operators to a rotating frame with frequency
$\Omega$ 
\[
\hat{A}(t)=\hat{a}(t)e^{i\Omega t}\,.
\]
We see that this transformation preserves the bosonic commutation
relations 
\begin{align*}
\left[\hat{A}(t),\hat{A}(t)\right] & =\left[\hat{A}^{\dagger}(t),\hat{A}^{\dagger}(t)\right]=0\\
\left[\hat{A}(t),\hat{A}^{\dagger}(t)\right] & =1
\end{align*}
and removes the free, fast rotating term from the equation of motion:
\begin{equation}
\dot{\hat{A}}(t)=-i\sum_{k}g_{k}\hat{b}_{k}(0)e^{i\left(\Omega-\omega_{k}\right)t}-\sum_{k}\left|g_{k}\right|^{2}\int_{0}^{t}{\rm d}t'\hat{A}(t')e^{i\left(\Omega-\omega_{k}\right)\left(t-t'\right)}\,.\label{eq:A rot frame}
\end{equation}

The first term in Eq.~\ref{eq:A rot frame} is denominated the \emph{noise
operator } 
\[
\hat{F}(t)=-i\sum_{k}g_{k}\hat{b}_{k}(0)e^{i\left(\Omega-\omega_{k}\right)t}\,.
\]
We see that this operator is composed of many different frequencies
and therefore oscillates rapidly in time. Its effect on the cavity
mode is that of exerting random " quantum kicks"
. Its expectation value for a reservoir in thermal equilibrium is
easily shown to be zero 
\[
\langle\hat{F}(t)\rangle_{R}=0
\]
and therefore this operator is the quantum analog to the noise due
to the environment responsible for the Brownian motion of a classical
particle. The second term in Eq.~\ref{eq:A rot frame} 
\begin{equation}
\hat{B}_{ba}=-\sum_{k}\left|g_{k}\right|^{2}\int_{0}^{t}{\rm d}t'\hat{A}(t')e^{i\left(\Omega-\omega_{k}\right)\left(t-t'\right)}\label{eq:B aux}
\end{equation}
is due to \emph{backaction}: changes in the cavity mode affect slightly
the bath, which in turn acts back onto the cavity. We will see in
the following that this term leads to decay of the cavity mode, which
corresponds to dissipation of energy from the cavity into the environment.

To analyze the second term in Eq.~\ref{eq:A rot frame} we use that
the environment volume is large, which allows us to take the continuum
limit for the bath modes. We write the sum over modes directly in
terms of a \emph{density of states} (DOS) which we do not specify,
this will depend on the details of the bath. The DOS $\mathcal{D}(\omega_{k})$
gives the number of modes with frequency between $\omega_{k}$ and
$\omega_{k}+{\rm d\omega_{k}}$. In terms of the DOS, the second term
in Eq.~\ref{eq:A rot frame} is 
\begin{equation}
\hat{B}_{ba}=-\int_{0}^{\infty}{\rm d}\omega_{k}\mathcal{D}(\omega_{k})\left|g(\omega_{k})\right|^{2}\int_{0}^{t}{\rm d}t'\hat{A}(t')e^{i\left(\Omega-\omega_{k}\right)\left(t-t'\right)}\,.\label{eq:aux 1}
\end{equation}
To perform this integral we have to resort to approximations that
rely on the physics of the system. We will work with what is known
as the Weisskopf-Wigner approximation, which is at its core a Markovian
approximation: the evolution of the system of interest is local in
time. This implies a separation of time scales between the bath, which
we assume to be the fast, and the system (in our case, the cavity
mode) which is slow. The information from the system that goes into
the reservoir is lost, since the bath fluctuates very rapidly. The
system therefore is said to have \emph{no memory}. The timescale of
the bath is defined by the inverse bandwidth $1/W$. If the rate of
variation $\hat{A}(t)$ is slow compared to this timescale, we can
replace $\hat{A}(t')\rightarrow\hat{A}(t)$ in Eq.~\ref{eq:aux 1},
and extend the integral from $t$ to $\infty$ 
\[
\hat{B}_{ba}\approx-\int_{0}^{\infty}{\rm d}\omega_{k}\mathcal{D}(\omega_{k})\left|g(\omega_{k})\right|^{2}\hat{A}(t)\int_{0}^{\infty}{\rm d}\tau e^{i\left(\Omega-\omega_{k}\right)\tau}\,
\]
where we defined $\tau=t-t'$. We use now that 
\[
\int_{0}^{\infty}{\rm d}\tau e^{i\left(\Omega-\omega_{k}\right)\tau}=\pi\delta\left(\omega_{k}-\Omega\right)-i\mathcal{P}\left(\frac{1}{\omega_{k}-\Omega}\right)\,,
\]
where the last term indicates the principal part. We neglect this
term for the moment, since it leads to a frequency shift (note that
its contribution is proportional to the cavity operator, and has an
$i$ in front). Evaluating the Delta function we obtain 
\[
\hat{B}_{ba}\approx-\hat{A}(t)\pi\mathcal{D}(\Omega)\left|g(\Omega)\right|^{2}
\]
and therefore the effective equation of motion for the cavity mode
in the rotating frame is 
\begin{equation}
\dot{\hat{A}}(t)=-\frac{\gamma}{2}\hat{A}(t)+\hat{F}(t)\,,\label{eq:EOM A}
\end{equation}
with 
\[
\gamma=\pi\mathcal{D}(\Omega)\left|g(\Omega)\right|^{2}
\]
the \emph{cavity decay rate}. Eq.~\ref{eq:EOM A} is a \emph{quantum
Langevin equation }and the decay rate $\gamma$ and the noise operator
$\hat{F}(t)$ can be shown to fulfill the \emph{fluctuation-dissipation
theorem } 
\[
\gamma=\frac{1}{\bar{n}}\int_{-\infty}^{\infty}{\rm d}\tau\langle\hat{F}^{\dagger}(\tau)\hat{F}(0)\rangle_{R}
\]
in equilibrium, with 
\[
\bar{n}=\langle\hat{b}^{\dagger}(\Omega)\hat{b}(\Omega)\rangle_{R}=\frac{1}{e^{\hbar\beta\Omega}-1}
\]
the thermal occupation of the bath at the cavity frequency. In the
Markov approximation the noise correlators fulfill 
\begin{align*}
\langle\hat{F}^{\dagger}(t')\hat{F}(t'')\rangle_{R} & =\gamma\bar{n}\delta\left(t'-t''\right)\\
\langle\hat{F}(t')\hat{F}^{\dagger}(t'')\rangle_{R} & =\gamma\left(\bar{n}+1\right)\delta\left(t'-t''\right)\,.
\end{align*}
In particular for the vacuum one obtains 
\[
\langle0|\hat{F}(t')\hat{F}^{\dagger}(t'')|0\rangle_{R}=\gamma\delta\left(t'-t''\right)\,.
\]
This delta-correlated noise shows clearly that the dynamics of the
bath is fast compared to that of the system of interest, and that
it has no memory since every " quantum kick"
is uncorrelated with the previous one. Usually the noise operator
is normalized to an operator $\hat{A}_{{\rm in}}$ such that 
\[
\langle0|\hat{A}_{{\rm in}}(t')\hat{A}_{{\rm in}}^{\dagger}(t'')_{{\rm }}|0\rangle_{R}=\delta\left(t'-t''\right)
\]
and the equation of motion is written as 
\[
\dot{\hat{A}}(t)=-\frac{\gamma}{2}\hat{A}(t)+\sqrt{\gamma}\hat{A}_{{\rm in}}(t)
\]
or, in the original frame, 
\[
\dot{\hat{a}}(t)=-i\Omega\hat{a}(t)-\frac{\gamma}{2}\hat{a}(t)+\sqrt{\gamma}\hat{a}_{{\rm in}}(t)\,.
\]
For the decay rate sometimes $\kappa=\gamma/2$ is used.

\subsection*{Check points}
\begin{itemize}
\item Write the total Hamiltonian of a cavity coupled to an environment,
and explain each term 
\item How does one obtain an effective equation of motion for the cavity
mode? What approximations are involved? 
\item Write the effective equation of motion for the cavity mode 
\item What is the meaning of $\kappa$ (or $\gamma$)? 
\end{itemize}

\section{\label{sec:The-optomagnonic-Hamiltonian}The optomagnonic Hamiltonian}

We will now put together all the elements from the previous sections
to derive the optomagnonic Hamiltonian, that is, the Hamiltonian for
a system in which optical photons couple to magnons. For that, we
will quantize the interaction term given by the Faraday effect, Eq.
\ref{eq:MO energy}. This section and the next follow the recent work
in Ref. \cite{viola_kusminskiy_coupled_2016}.

We can quantize the electric field following Sec. \ref{sec:Quantization-of-the},
$\mathbf{\hat{E}^{+}}(\mathbf{r},t)=\sum_{\beta}\mathbf{E}_{\beta}(\mathbf{r})\hat{a}_{\beta}(t)$
and correspondingly $\mathbf{\hat{E}^{-}}(\mathbf{r},t)=\sum_{\beta}\mathbf{E}_{\beta}^{*}(\mathbf{r})\hat{a}_{\beta}^{\dagger}(t)$,
where $\mathbf{E}_{\beta}(\mathbf{r})$ indicates the $\beta^{{\rm th}}$
eigenmode of the electric field (eigenmodes are indicated with greek
letters in what follows). The magnetization requires more careful
consideration, since $\mathbf{M}(\mathbf{r})$ depends on the local
spin operator which, in general, cannot be written as a linear combination
of bosonic modes. There are however two simple cases: (i) the spin-wave
approximation, which is valid for small deviations of the spins from
equilibrium and, as we saw in Sec. \ref{sec:Spin-wave-approximation},
the Holstein-Primakoff representation can be truncated to linear order
in the bosonic magnon operators, and (ii) considering the homogeneous
Kittel mode \footnote{The Kittel mode is a spin wave with $\mathbf{k}=0$, so that all spins
precess in phase and can be replaced by a precessing macrospin.} $\mathbf{M}(\mathbf{r})=\mathbf{M}$, for which we can work simply
with the resulting macrospin $\mathbf{S}$. In the following we treat
this second case. Although it is valid only for the homogeneous magnon
mode, it allows us to capture the nonlinear dynamics of the spin.

From Eq.~\ref{eq:MO energy} we obtain the coupling Hamiltonian 
\begin{equation}
\hat{H}_{MO}=\hbar\sum_{j\beta\gamma}\hat{S}_{j}G_{\beta\gamma}^{j}\hat{a}_{\beta}^{\dagger}\hat{a}_{\gamma}\label{eq:OM Ham coupling}
\end{equation}
with coupling constants 
\begin{equation}
G_{\beta\gamma}^{j}=-i\frac{\varepsilon_{0}f\,M_{s}}{4\hbar S}\epsilon_{jmn}\int{\rm d}\mathbf{r}E_{\beta m}^{*}(\mathbf{r)}E_{\gamma n}(\mathbf{r})\,,\label{eq:G_jbg-1}
\end{equation}
where we replaced $M_{j}/M_{s}=\hat{S}_{j}/S$, with $S$ the extensive
total spin (scaling like the magnetic mode volume). $G^{j}$ are hermitian
matrices which in general cannot be simultaneously diagonalized. For
simplicity, in the following we treat the case of a strictly diagonal
coupling to some optical eigenmodes ($G_{\beta\beta}^{j}\neq0$ but
$G_{\alpha\beta}^{j}=0$).

\begin{figure}
\centering{}\includegraphics[width=0.7\textwidth]{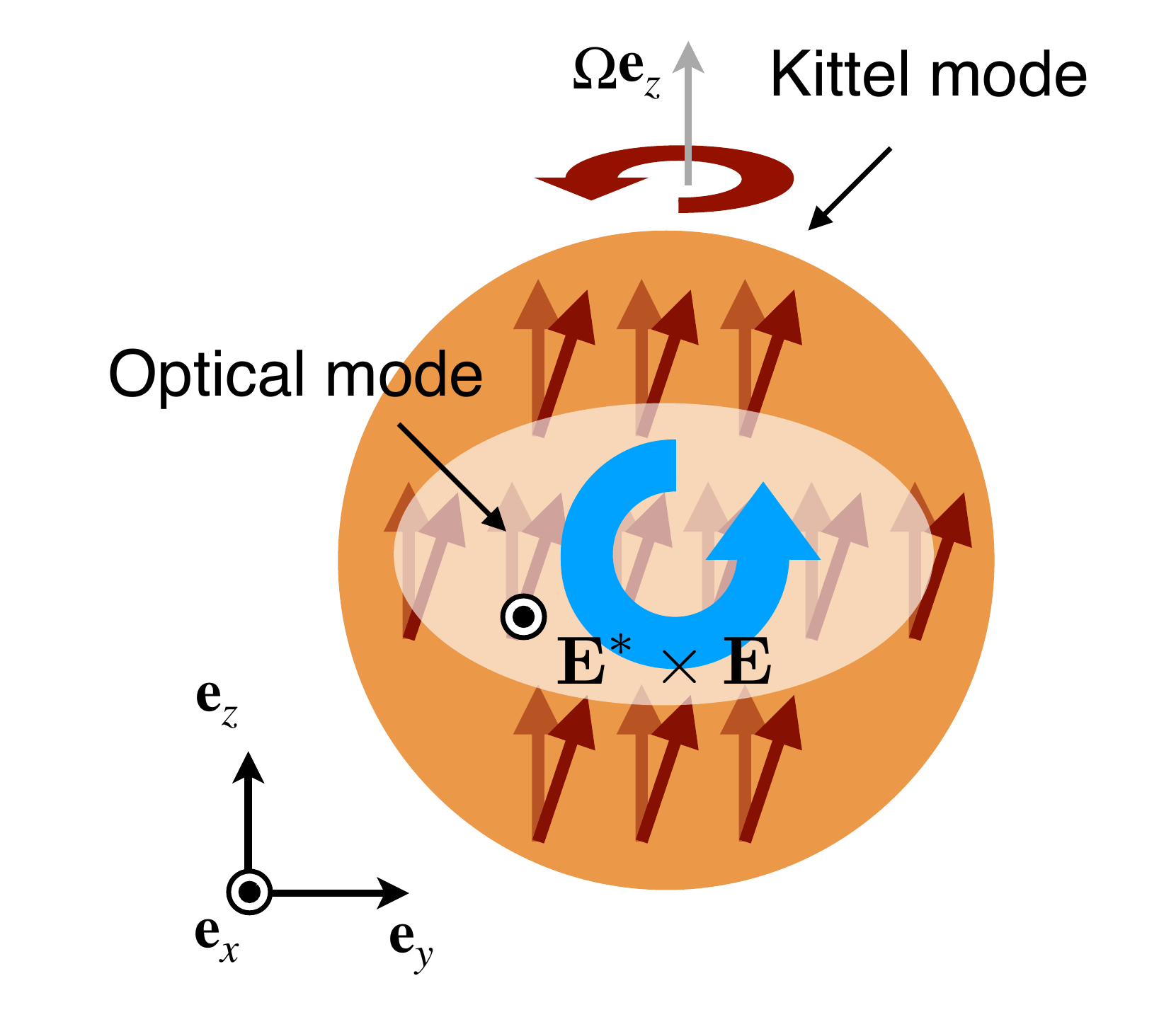}\caption{The Kittel mode with frequency $\Omega$ is excited on top of a homogeneous
magnetic ground state with magnetization along the $\mathbf{e}_{z}$
axis. The same material serves as the optical cavity. The optical
mode is right circularly polarized in the $y-z$ plane and the optical
spin density points along the $\mathbf{e}_{z}$ axis.}
\label{Fig:setupOM} 
\end{figure}

As an example, we consider circular polarization (R/L) in the $y-z$-plane.
In this case, the optical spin density is perpendicular to this plane,
and therefore $G^{x}$ is diagonal while $G^{y}=G^{z}=0$. The Hamiltonian
$H_{MO}$ is then diagonal in the the basis of circularly polarized
waves, $\mathbf{e}_{R/L}=\frac{1}{\sqrt{2}}\left(\mathbf{e}_{y}\mp i\mathbf{e}_{z}\right)$.
We choose moreover the magnetization axis along the $\mathbf{\hat{z}}$
axis. This setup is shown schematically in Fig. \ref{Fig:setupOM}.

The rationale behind choosing the coupling direction \emph{perpendicular}
to the magnetization axis, is to maximize the coupling to the magnon
mode, that is to the\emph{ deviations} of the magnetization with respect
to the magnetization axis. The light field hence only couples to the
$x$ component of the spin operator, $\hat{S}_{x}$.

Again for simplicity, we consider the case of plane waves for quantizing
the electric field. Therefore 
\[
\mathbf{\hat{E}^{+(-)}}(\mathbf{r},t)=+(-)i\sum_{_{j}}\mathbf{e}_{j}\sqrt{\frac{\hbar\omega_{j}}{2\varepsilon_{0}\varepsilon V}}\hat{a}_{j}^{(\dagger)}(t)e^{+(-)i\mathbf{k_{j}\cdot r}}\,,
\]
where $V$ is the volume of the cavity, $\mathbf{k}_{j}$ the wave
vector of mode $j$ and we have identified the positive and negative
frequency components of the field as $\mathbf{E}\rightarrow\hat{\mathbf{E}}^{+}$,
$\mathbf{E^{*}}\rightarrow\hat{\mathbf{E}}^{-}$. In the normalization
of the fields we have used the relative permittivity $\varepsilon$
of the magnetic insulator, since the electric fields considered live
in the cavity formed by the material. The factor of $\varepsilon_{0}\varepsilon$
in the denominator ensures the normalization $\hbar\omega_{j}=\varepsilon_{0}\varepsilon\langle j|\int d^{3}\mathbf{r}|\mathbf{E}(\mathbf{r})|^{2}|j\rangle-\varepsilon_{0}\varepsilon\langle0|\int d^{3}\mathbf{r}|\mathbf{E}(\mathbf{r})|^{2}|0\rangle$,
which corresponds to the energy of a photon in state $|j\rangle$
above the vacuum $|0\rangle$. For two degenerate (R/L) modes at frequency
$\omega$, using Eq.~\ref{eq:theta F} we see that the frequency
dependence cancels out and we obtain the simple form for the optomagnonic
Hamiltonian 
\[
H_{MO}=\hbar G\hat{S}_{x}(\hat{a}_{L}^{\dagger}\hat{a}_{L}-\hat{a}_{R}^{\dagger}\hat{a}_{R})
\]
with 
\[
G=\frac{1}{S}\frac{c\,\theta_{F}}{4\sqrt{\varepsilon}}.
\]
In general however an overlap factor $\xi\le1$ appears, which takes
into account that there is a mismatch between the optical and magnonic
mode volumes. For example, current experiments couple optical whispering
gallery modes (WGM) in a YIG sphere to the magnetic Kittel mode \cite{osadaCavityOptomagnonicsSpinOrbit2016,zhangOptomagnonicWhisperingGallery2016,haighTripleResonantBrillouinLight2016}.
The Kittel mode is a bulk mode, and lives on the whole sphere, that
is, the magnetic mode volume equals the volume of the sphere. The
WGMs live however very close to the surface, and therefore its volume
is smaller than the magnetic one, leading to and overlap factor $\xi<1$,

\begin{equation}
G_{LL}^{x}=-G_{RR}^{x}=G=\frac{1}{S}\frac{c\,\theta_{F}}{4\sqrt{\varepsilon}}\xi\,.\label{eq:GMO}
\end{equation}

We now consider an incoming laser, which drives only one of the two
circular polarizations in the cavity. The total Hamiltonian of the
cavity optomagnonic system is therefore given simply by 
\begin{equation}
H=-\hbar\Delta\hat{a}^{\dagger}\hat{a}-\hbar\Omega\hat{S}_{z}+\hbar G\hat{S}_{x}\hat{a}^{\dagger}\hat{a}\,,\label{eq:OM Ham one mode}
\end{equation}
where $\hat{a}^{\dagger}$ ($\hat{a}$) is the creation (annihilation)
operator for the cavity mode photon which is being driven. We work
in a frame rotating at the laser frequency $\omega_{las}$, and $\Delta=\omega_{las}-\omega_{cav}$
is the detuning vs. the optical cavity frequency $\omega_{cav}$.
In Eq.~\ref{eq:OM Ham one mode} we included that the dimensionless
macrospin ${\bf S}=(S_{x},S_{y},S_{z})$ has a magnetization axis
along $\mathbf{\hat{z}}$, and a Larmor precession frequency $\Omega$
which can be controlled by an external magnetic field. 
\begin{enumerate}
\item \textbf{\emph{Exercise: show that in the rotating frame the free Hamiltonian
for a cavity driven mode is given by $\hbar\Delta\hat{a}^{\dagger}\hat{a}$
.}} 
\end{enumerate}

\subsection*{Check points}
\begin{itemize}
\item Derive the optomagnonic coupling Hamiltonian 
\end{itemize}

\section{\label{sec:Coupled-Equations-of}Coupled equations of motion and
fast cavity limit}

The coupled Heisenberg equations of motion are obtained by using $\left[\hat{a},\hat{a}^{\dagger}\right]=1$,
$\left[\hat{S}_{i},\hat{S}_{j}\right]=i\epsilon_{ijk}\hat{S}_{k}$.
We next focus on the classical limit, where we replace the operators
by their expectation values: 
\begin{eqnarray}
\dot{a} & = & -i\left(GS_{x}-\Delta\right)a-\frac{\kappa}{2}\left(a-\alpha_{{\rm max}}\right)\nonumber \\
\dot{\mathbf{S}} & = & \left(Ga^{*}a\mathbf{\,e}_{x}-\Omega\,\mathbf{e}_{z}\right)\times\mathbf{S}+\frac{\eta_{{\rm G}}}{S}(\mathbf{\dot{S}}\times\mathbf{S})\,.\label{eq:EOM COM}
\end{eqnarray}
From Sec. \ref{sec:OCOQS}, we know our optical cavity is an open
system and the optical fields are subject to a decay rate. Here we
introduced the cavity decay rate $\kappa$ phenomenologically, its
value is in general determined by the scpecific experimental setup.
We also included the driving laser amplitude $\alpha_{{\rm max}}$
for the optical mode. This gives the steady state amplitude of the
light field when it is not coupled to the magnetics and for zero detuning
of the driving laser. We also added an intrinsic damping for the spin
$\eta_{{\rm G}}$, which can be due to phonons and defects and it
is material dependent. This coefficient is denominated \emph{Gilbert
damping} and, whereas it does not change the magnitude of the spin
vector, it causes a decay of the Larmor precession to the stable equilibrium
of the spin. The equation of motion for the spin without coupling
to the light reduces to 
\[
\dot{\mathbf{S}}=-\Omega\,\mathbf{e}_{z}\times\mathbf{S}+\frac{\eta_{{\rm G}}}{S}(\mathbf{\dot{S}}\times\mathbf{S})\,,
\]
which is known as the\emph{ Landau-Lifschitz-Gilbert equation}. We
have encountered this equation before, abeit without the damping term.

We see hence that the light acts as a kind of effective magnetic field
on the spin. Actually, since the field $a$ depends on time, and the
spin-light dynamics is coupled, retardation effects cause also dissipation
for the spin, in a similar way in which an environment causes the
dissipation term $\kappa$ for the optical field. In the following
we will obtain the effective equation of motion for the spin induced
by the light, " integrating out" the light
field. For that we have to resort to an approximation, which is denominated
the \emph{fast cavity limit}, where the dynamics of the light is much
faster than that of the spin (sometimes it is also called the bad
cavity limit, since it implies that $\kappa$ is large). That means
that the photons spend in average a very short time in the cavity,
during which they " see" the spin almost
as static.

\begin{figure}
\centering{}\includegraphics[width=0.7\textwidth]{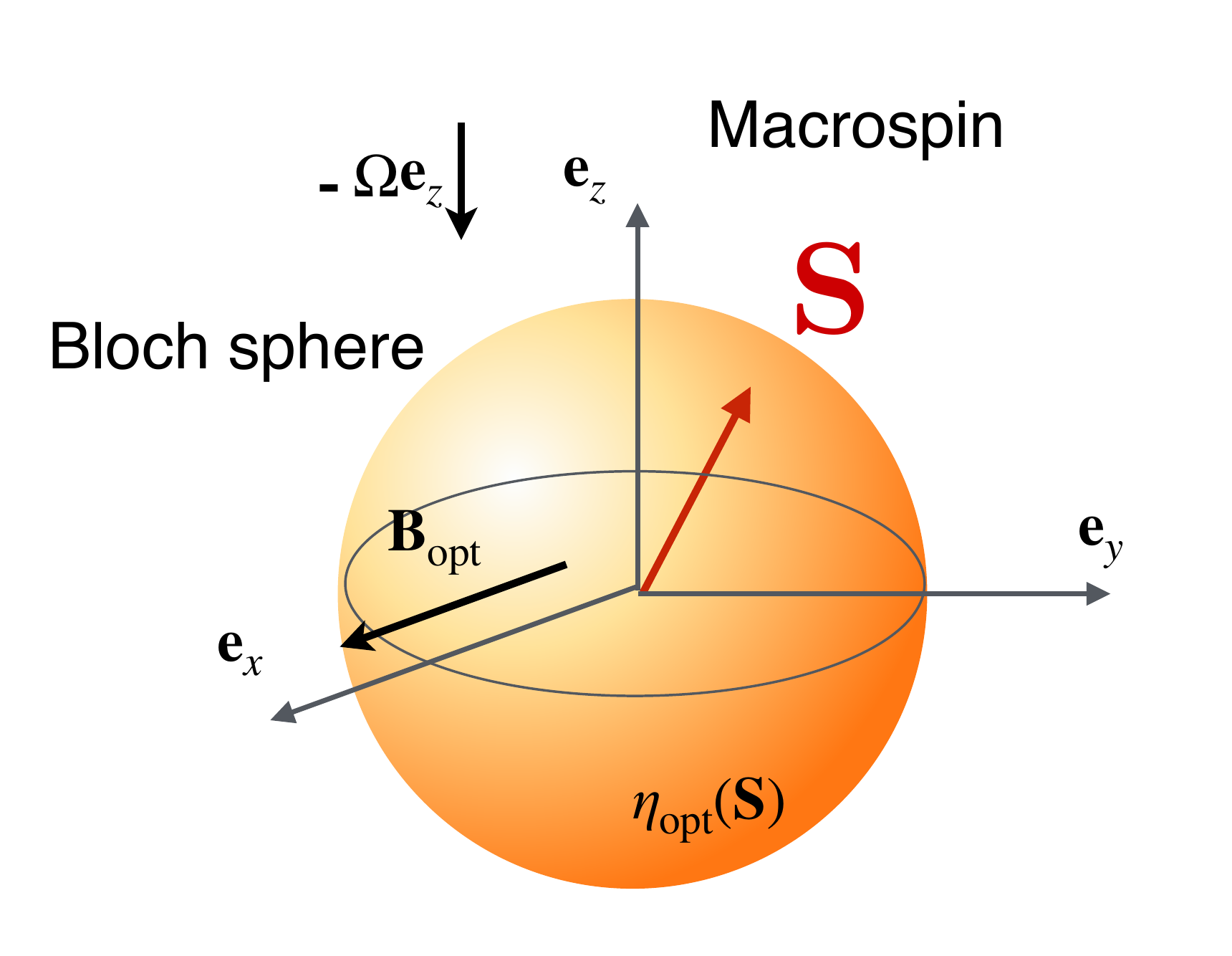}\caption{Optically induced effective magnetic field $\mathbf{B}_{{\rm opt}}$
and dissipation coefficient $\eta_{{\rm opt}}$. Together with the
external applied magnetic field, which controls the Kittel mode frequency
$\Omega$, they govern the nonlinear dynamics of the macrospin $\mathbf{S}$
on the Bloch sphere.}
\label{Fig:optind} 
\end{figure}

The condition for the fast cavity limit to be valid is $G\dot{S}_{x}\ll\kappa^{2}$.
In that case we can expand the field $a(t)$ in powers of $\dot{S}_{x}$.
We write $a(t)=a_{0}(t)+a_{1}(t)+\ldots$, where the subscript indicates
the order in $\dot{S}_{x}$. From the equation for $a(t)$, we find
that $a_{0}$ fulfills the instantaneous equilibrium condition 
\begin{equation}
a_{0}(t)=\frac{\kappa}{2}\alpha_{{\rm max}}\frac{1}{\frac{\kappa}{2}-i\left(\Delta-GS_{x}(t)\right)}\,,\label{eq:a0}
\end{equation}
from which we obtain the correction $a_{1}$: 
\begin{equation}
a_{1}(t)=-\frac{1}{\frac{\kappa}{2}-i\left(\Delta-GS_{x}\right)}\frac{\partial a_{0}}{\partial S_{x}}\dot{S}_{x}\,.\label{eq:a1}
\end{equation}
To derive the effective equation of motion for the spin, we replace
$|a|^{2}\approx|a_{0}|^{2}+a_{1}^{*}a_{0}+a_{0}^{*}a_{1}$ in Eq.~\ref{eq:EOM COM}
which leads to 
\begin{equation}
\dot{\mathbf{S}}=\mathbf{B}_{{\rm eff}}\times\mathbf{S}+\frac{\eta_{{\rm opt}}}{S}(\dot{S}_{x}\,\mathbf{e}_{x}\times\mathbf{S})+\frac{\eta_{{\rm G}}}{S}(\mathbf{\dot{S}}\times\mathbf{S})\,.\label{eq:EOM_eff}
\end{equation}
Here $\mathbf{B}_{{\rm eff}}=-\Omega\mathbf{e}_{z}+\mathbf{B}_{{\rm opt}}$,
where $\mathbf{B}_{{\rm opt}}(S_{x})=G|a_{0}|^{2}\,\mathbf{e}_{x}$
is the purely static contribution and acts as an optically induced
magnetic field. The second term is due to retardation effects, and
it reminiscent of Gilbert damping, albeit with spin-velocity component
only along $\mathbf{\hat{x}}$ due to the chosen geometry. These are
depicted in Fig. \ref{Fig:optind}. 

Both the induced field $\mathbf{B}_{{\rm opt}}$ and dissipation coefficient
$\eta_{{\rm opt}}$ depend explicitly on the instantaneous value of
$S_{x}(t)$: 
\begin{eqnarray}
\mathbf{B}_{{\rm opt}} & = & \frac{G}{[(\frac{\kappa}{2})^{2}+(\Delta-GS_{x})^{2}]}\left(\frac{\kappa}{2}\alpha_{{\rm max}}\right)^{2}\mathbf{e}_{x}\label{eq:B_ind}\\
\eta_{{\rm opt}} & = & -2G\kappa S\,|\mathbf{B}_{{\rm opt}}|\,\frac{(\Delta-GS_{x})}{[(\frac{\kappa}{2})^{2}+(\Delta-GS_{x})^{2}]^{2}}\,.\label{eq:G_ind}
\end{eqnarray}
These fields are highly non-linear functions of the spin. Note that
the optically induced dissipation can change sign! This leads to very
interesting dynamics. Two distinct solutions can be found: generation
of new stable fixed points (switching) and optomagnonic limit cycles
\cite{viola_kusminskiy_coupled_2016}. .
\begin{enumerate}
\item \textbf{\emph{Exercise: fill in the steps of the derivation above.}} 
\end{enumerate}

\subsection*{Check points}
\begin{itemize}
\item What is the fast cavity limit? 
\item How do you obtain an effective equation of motion for the spin, and
what is the meaning of each term? 
\end{itemize}

\section{\label{sec:Linearized-Hamiltonian}Linearized optomagnonic Hamiltonian}

As we mentioned in Sec. \ref{sec:The-optomagnonic-Hamiltonian}, one
can also treat the optomagnonic Hamiltonian in the limit of small
oscillations of the spins, by using a truncated Holstein Primakoff
expansion so that the spin ladder operators are replaced by linear
bosonic operators. This allows to study the behavior of the system
beyond the homogeneous magnetic Kittel mode studied in the previous
two sections, but restricts the analysis to small displacements of
the spins with respect of their equilibrium positions.

In order to proceed with the linearization, we consider spin wave
excitations on top of a possibly nonuniform static ground state $\mathbf{m}_{0}(\mathbf{r})$,
\begin{equation}
\delta\mathbf{m}(\mathbf{r},t)=\mathbf{m}(\mathbf{r},t)-\mathbf{m}_{0}(\mathbf{r})\,.
\end{equation}
For small deviations $|\delta\mathbf{m}|\ll1$ we can express these
in terms of harmonic oscillators, which correspond to the magnon modes.
This is equivalent to a local Holstein Primakoff approximation. We
can quantize the spin wave as 
\begin{equation}
\delta\mathbf{m}(\mathbf{r},t)\rightarrow\frac{1}{2}\sum_{\gamma}\left[\delta\mathbf{m}_{\gamma}(\mathbf{r})\hat{b}_{\gamma}e^{-i\omega_{\gamma}t}+\delta\mathbf{m}_{\gamma}^{*}(\mathbf{r})\hat{b}_{\gamma}^{\dagger}e^{i\omega_{\gamma}t}\right]\,.
\end{equation}
In turn, the quantization of the optical fields can be written as
\begin{eqnarray}
\mathbf{E}(\mathbf{r},t)\rightarrow & \sum_{\beta}\mathbf{E}_{\beta}(\mathbf{r})\hat{a}_{\beta}e^{-i\omega_{\beta}t}\\
\mathbf{E}^{*}(\mathbf{r},t)\rightarrow & \sum_{\beta}\mathbf{E}_{\beta}^{*}(\mathbf{r})\hat{a}_{\beta}^{\dagger}e^{i\omega_{\beta}t}
\end{eqnarray}
From Eq.~\ref{eq:MO energy} we obtain the coupling Hamiltonian linearized
in the spin fluctuations \cite{grafCavityOptomagnonicsMagnetic2018}
\begin{equation}
\hat{H}_{MO}=\sum_{\alpha\beta\gamma}G_{\alpha\beta\gamma}^{{\rm }}\hat{a}_{\alpha}^{\dagger}\hat{a}_{\beta}\hat{b}_{\gamma}+{\rm h.c.}\label{eq:linSHMO}
\end{equation}
where 
\begin{equation}
G_{\alpha\beta\gamma}^{{\rm }}=-i\frac{\theta_{{\rm F}}\lambda_{n}}{4\pi}\frac{\varepsilon_{0}\varepsilon}{2}\int{\rm d}\mathbf{r}\,\delta\mathbf{m}_{\gamma}(\mathbf{r})\cdot[\mathbf{E_{\alpha}^{*}}\left(\mathbf{r}\right)\times\mathbf{E}_{\beta}\left(\mathbf{r}\right)]\label{eq:coupling}
\end{equation}
is the optomagnonic coupling. Note that $\hat{a}$ correspond to photon
operators, while $\hat{b}$ correspond to magnonic ones. The Greek
subindices indicate the respective magnon and photon modes which are
coupled. The information on the specific shape and normalization of
the magnon and optical modes is encoded in the respective mode functions
$\delta\mathbf{m}_{\gamma}(\mathbf{r})$ and $\mathbf{E}_{\alpha}(\mathbf{r})$.
Eqs. \ref{eq:linSHMO} and \ref{eq:coupling} allow to treat arbitrary
geometries for the optical cavity, arbitrary magnetic ground states,
and arbitrary spin wave modes.

The Hamiltonian in Eq. \ref{eq:linSHMO} is still nonlinear, since
it involves products of three bosonic operators. This Hamiltonian
therefore still contains " interacting"
three-particle terms. In particular, it describes scattering processes
in which a photon in mode $\beta$ and a magnon in mode $\gamma$
are annihilated creating a photon in the mode $\alpha$, and the complementary
process in which a photon $\alpha$ is annihilated creating a photon
$\beta$ and a magnon $\gamma$. To bring this Hamiltonian into a
solvable, quadratic form, we linearize now the Hamiltonian Eq. \ref{eq:linSHMO}
in the optical fields. For that we consider fluctuations of the optical
fields around a steady state solution $\langle\hat{a}_{\alpha}\rangle$,
$\langle\hat{a}_{\beta}\rangle$ 
\begin{eqnarray}
\hat{a}_{\alpha}= & \langle\hat{a}_{\alpha}\rangle+\delta\hat{a}_{\alpha}\nonumber \\
\hat{a}_{\beta}= & \langle\hat{a}_{\beta}\rangle+\delta\hat{a}_{\beta}\,.
\end{eqnarray}
The steady state solutions $\langle\hat{a}_{\alpha}\rangle$, $\langle\hat{a}_{\beta}\rangle$
satisfy $\langle\dot{\hat{a}}_{\alpha}\rangle=0$, $\langle\dot{\hat{a}}_{\beta}\rangle=0$,
where the time evolution is given by the coupled equations of motion
dictated by the interaction Hamiltonian Eq. \ref{eq:linSHMO} plus
driving and free terms, obtained by generalizing Eq. \ref{eq:OM Ham one mode}
to multiple modes. The average number of photons circulating in cavity
mode $\alpha$ in steady state is simply given by $n_{\alpha}=|\langle\hat{a}_{\alpha}\rangle|^{2}$
and it is related to the input laser power. To linear order in the
fluctuations defined by Eqs. \ref{eq:fluct_a}, the Hamiltonian Eq.
\ref{eq:linSHMO} reduces to 
\begin{equation}
\hat{H}_{lin}=\sum_{\alpha\beta\gamma}G_{\alpha\beta\gamma}^{{\rm }}\left(\sqrt{n_{\alpha}}\delta\hat{a}_{\beta}\hat{b}_{\gamma}+\sqrt{n_{\beta}}\delta\hat{a}_{\alpha}^{\dagger}\hat{b}_{\gamma}\right)+{\rm h.c.}\,,\label{eq:linSAHMO}
\end{equation}
which is a Hamiltonian linear both in magnon and photon operators.
It contains two types of terms, denominated \emph{parametric amplifier},
corresponding to those terms which simultaneously create or annihilate
a photon and a magnon ($\delta\hat{a}_{\beta}\hat{b}_{\gamma}$ and
$\hat{b}_{\gamma}^{\dagger}\delta\hat{a}_{\beta}^{\dagger}$), and
\emph{beam splitter}, which converts a photon into a magnon ($\hat{b}_{\gamma}^{\dagger}\delta\hat{a}_{\alpha}$),
and vice-versa ($\delta\hat{a}_{\alpha}^{\dagger}\hat{b}_{\gamma}$).
Which of the two types of processes dominates depends on which of
them is in resonance, and can be tuned by the external laser driving.
We note that the optomagnonic coupling constant $G_{\alpha\beta\gamma}$
is enhanced by the square root of the number of photons circulating
in the corresponding cavity mode. This is similar to \emph{optomechanics},
where light in an optical cavity couple to phonons \cite{aspelmeyerCavityOptomechanics2014}.

\section{\label{sec:Prospects}Prospects in cavity optomagnonics}

Cavity optomagnonic systems are the newest addition to a collection
of platforms being studied nowadays with the aim of manipulating,
processing, and storing quantum information. These systems, usually
of nano and micro scale dimensions, are called \emph{hybrid quantum
systems} \cite{PNASHybrid}, since they combine different degrees
of freedom (such as electronic, mechanical, photonic, or magnetic)
to enhance functionality. For example, whereas optical photons are
good carriers of information, they are not so good for information
processing. Further examples of hybrid systems are nanoelectromechanical
or optomechanical systems. A general underlying property of these
is that they use collective excitations (such as phonons or magnons)
whose properties can be engineered by proper design at the nanoscale.

A challenge in many of these platforms is that the coupling between
the different degrees of freedom is weak, even taking into account
the enhancement obtained by the use of a cavity. Strong coupling,
together with low losses, are required for quantum information applications.
This is because one should be able to process and transfer information
before it is lost to the environment. In particular for magnonic systems,
it has been shown that strong coupling to microwave photons is possible,
by using a microwave cavity \cite{soykal_strong_2010,huebl_high_2013,tabuchiHybridizingFerromagneticMagnons2014,zhangStronglyCoupledMagnons2014}.
Note that in this case the coupling is resonant, meaning that the
frequencies from both microwave and magnonic excitations can be matched,
being both in the GHz range. The magnons in this case couple directly
to the slow, oscillating magnetic field, as in ferromagnetic resonance
experiments. We have not discussed this coupling in these notes, but
it can be shown it has the form 
\begin{equation}
\hat{S}^{+}\hat{a}^{\dagger}+\hat{S}^{-}\hat{a}
\end{equation}
in terms of the spin ladder operators $\hat{S}^{\pm}$ and the microwave
photons $\hat{a}$. This interaction converts a magnon into a photon
and vice-versa. The optomagnonic coupling instead is parametric in
the photon fields (coupling instead to terms of the form $\hat{a}^{\dagger}\hat{a}$).
This is in general the case for non-resonant interactions, where the
frequency mismatch has to be accounted for (note that optical photons
have frequencies of hundreds of THz), and usually results in small
intrinsic coupling values.

The fact that magnons can couple coherently both to microwaves and
to light is however a big incentive to pursue the strong coupling
regime also in the optical domain. That would allow coherent transfer
of information from the microwave regime, where the information is
usually processed (e.g. with superconducting qubits \cite{tabuchiCoherentCouplingFerromagnetic2015}),
to the telecom regime, where information can be communicated through
long distances and at room temperature with the help of optical fibers.
We can expect that the next few years will bring many exciting advances
in this field.

 \bibliographystyle{unsrt}
\bibliography{QMSWL}

\end{document}